\documentclass[twocolumn]{aastex63} 
\usepackage{amsmath} 
\usepackage{amssymb} 
\usepackage{amsfonts}
\usepackage{mathrsfs}
\usepackage{graphicx}
\usepackage{natbib}
\usepackage{color}
\usepackage{hyperref}

\newcommand{\mbh}{\ensuremath{M_{\rm BH}}}
\newcommand{\msun}{\ensuremath{M_\odot}}

\newcommand{\mstar}{\ensuremath{M_\ast}}

\newcommand{\ergspersec}{erg~s\ensuremath{^{-1}}}

\begin{document}
\title{Forward Modeling Populations of Flares from Tidal Disruptions of Stars by Super-massive Black Holes}

\author[0000-0002-6485-2259]{Nathaniel Roth}
\affiliation{Lawrence Livermore National Laboratory, Livemore, CA 94550, USA}

\author[0000-0002-3859-8074]{Sjoert van Velzen}
\affiliation{Leiden Observatory, Leiden University, PO Box 9513, 2300 RA Leiden, The Netherlands}
\affiliation{Center for Cosmology and Particle Physics, New York University, NY 10003, USA}
\affiliation{Department of Astronomy, University of Maryland, College Park, MD 20742, USA}

\author[0000-0003-1673-970X]{S. Bradley Cenko}
\affiliation{Astrophysics Science Division, NASA Goddard Space Flight Center, 8800 Greenbelt Rd, Greenbelt, MD 20771, USA}
\affiliation{Joint Space-Science Institute, University of Maryland, College Park, MD 20742, USA}

\author[0000-0002-7962-5446]{R.F. Mushotzky}
\affiliation{Department of Astronomy, University of Maryland, College Park, MD 20742, USA}
\affiliation{Joint Space-Science Institute, University of Maryland, College Park, MD 20742, USA}

\correspondingauthor{Nathaniel Roth}
\email{roth14@llnl.gov}

\begin{abstract}
Detections of the tidal disruption flares (TDFs) of stars by supermassive black holes (SMBHs) are rapidly accumulating as optical surveys improve. These detections may provide constraints on SMBH demographics, stellar dynamics, and stellar evolution in galaxies. To maximize this scientific impact, we require a better understanding of how astrophysical parameters interact with survey selection effects in setting the properties of detected flares. We develop a framework for modeling the distributions of optical TDF detections in surveys across attributes of the host galaxies and the flares themselves. This model folds in effects of the stellar disruption rate in each galaxy, the flare luminosity and temperature distributions, the effects of obscuration and reddening by dust in the host galaxy, and survey selection criteria. We directly apply this model to the sample of TDFs detected by the Zwicky Transient Facility and find that the overall flare detection rate is in line with simple theoretical expectation. The model can also reproduce the distribution of total stellar mass and redshift of the host galaxies, but fails to match all details of the detected flares, such as their luminosity and temperature distributions. We also find that dust obscuration likely plays an important role in suppressing the TDF detection rate in star-forming galaxies. While we do not find that the unusual preference of TDFs to have hosts in post-starburst galaxies in the ``green valley'' can be entirely explained by selection effects, our model can help to quantify the true rate enhancement in those galaxies.     
\end{abstract}

\section{Introduction}
As wide-field surveys have advanced, the study of tidal disruptions of stars by super-massive black holes (SMBHs) has progressed beyond the detailed follow-up of events in isolation, and has now entered the era of science enabled by the study of samples. This opens up exciting new possibilities to learn about the cosmic demographics of SMBHs and the host galaxies in which they reside. For example, \citet{Stone2016-1} (hereafter SM16) proposed how TDF\footnote{Throughout this paper we will follow the terminology suggested by \citet{van-Velzen2018-1}, in which detected flares are termed ``tidal disruption flares'' (TDFs), to distinguish them from the more general phenomenon of ``tidal disruption events'' (TDEs). Every TDF is associated with a TDE, but not every TDE leads to a detectable TDF.} detection rates can be used to constrain the galaxy occupation fraction of SMBHs. \citet{Kochanek2016-2}, \citet{Stone2018-2}, and \citet{DOrazio2019} have proposed how detection rates can be used to constrain the star formation history and stellar mass function in galaxies.

Aside from increasing the detection rate, additional pre-requisites remain before these ambitious science goals can be reached. Many of these relate to better characterizing survey detection efficiencies and selection effects. After the first generation of searches for X-ray and optical surveys of flares from tidal disruption events, much attention was paid to the so-called ``rate problem'', relating to the fact that the rate of TDF detections in most galaxies, $\sim 10^{-5}$ yr$^{-1}$ \citep{Donley2002,van-Velzen2014}, seemed roughly an order of magnitude too low compared to theoretical predictions of TDE rates based on stellar dynamics \citep{Wang2004}. However, \citet{Gezari2008} determined that the detection rate of ultraviolet (UV) TDFs found by the Galaxy Evolution Explorer (GALEX) was consistent with theoretical predictions, albeit with substantial uncertainty.  More recently, \citet{van-Velzen2018-1} made use of TDF detections from previous optical and UV surveys to constrain the distribution of peak $g$-band luminosities of TDFs, using an inverse-volume weighting \citep{Schmidt1968} to account for flares that were likely missed because they were too faint. After taking into the account the higher rate of low-luminosity TDFs, the observed TDF rate was found to be in better agreement with the theoretical TDE rate.

A separate puzzle has emerged regarding the host galaxies of TDFs. \citet{Arcavi2014} pointed out that TDFs detected by the Palomar Transient Factory showed an unusual preference for a class of post-starburst galaxies known as E+A galaxies. This preference was further investigated and confirmed by \citet{French2017-0} and by \citet{Graur2018-0} for a larger sample of TDF hosts. Other peculiarities of TDF host galaxies have also been identified. \citet{Law-Smith2017} compared TDF hosts to a sample of galaxies matched by the mass of their super-massive black holes, \mbh. They found that this reduced, but did not entirely explain, the preference for E+A galaxies, and identified that TDF hosts tend to be more centrally concentrated, with higher Sersic indices. This last effect was also explored in a mass-matched galaxy sample by \citet{French2020-0}.

Recently, a TDF survey was conducted with the Zwicky Transient Facility (ZTF), and this has roughly doubled the number of optically selected TDFs \citep{van-Velzen2020}. Once again, the TDF hosts were found to cluster in an unusual region of galaxy parameter space, the so-called ``green valley'' which separates the peak populations of blue, star-forming galaxies from red, quiescent galaxies in galaxy color-magnitude space. This preference had also been identified in \citet{Law-Smith2017}. It is not yet clear whether the more specific preference for E+A galaxies remains as strong in the new sample, and the issue is confounded by the fact that the population of E+A galaxies significantly overlaps in color-magnitude space with the more general class of post-starburst galaxies that reside in the green valley.

More work is needed to understand the role of selection effects in setting both the rate of TDF detections and the host galaxies in which they are found. In particular, an effect that has often been acknowledged, but rarely quantified, is the role of obscuration by dust and neutral gas in the host galaxy. While neutral gas will be the dominant source of obscuration for X-ray flares, dust is the most important obscurer for optical and UV detections. As we will see, dust obscuration is likely to be a crucial ingredient in flare selection effects, leading us to predict that roughly twice as many TDFs than those that are detected might be missed solely because they are dust obscured.

One route toward understanding these selection effects is to perform end-to-end forward modeling, beginning with the processes that govern the rate of stellar disruptions, then accounting for the luminosity distribution of the resulting flares, how these flares might be obscured by material in their host galaxies, and then applying survey selection criteria to determine which of these disruptions will actually been detected. Work along these lines has been done by \citet{Kochanek2016-2}, SM16, \citet{van-Velzen2018-1}, and \citet{DOrazio2019}. The aim of the present paper is to bring together a number of effects that have been separately accounted for in these previous models. Here we make a special application to the ZTF TDF sample from \citet{van-Velzen2020}, because it provides the largest sample of flares with detections that can be modeled uniformly using a single set of survey selection criteria. We also make predictions for TDFs that may be detected with the upcoming Vera C. Rubin Observatory (VRO/LSST). The models presented here are tailored for optical surveys, but future work could use a similar framework applied to X-ray surveys such as those enabled by eROSITA. The code used to generate the results in this paper is publicly available at \href{url}{https://github.com/nroth/tdegb}.

In section~\ref{sec:preliminary}, we present a preliminary calculation that indicates that the vast majority of SMBHs at masses and redshifts favorable for visible disruptions should be in ``blue'' galaxies, which contrasts with the host population of observed TDFs. In section~\ref{sec:Method} we discuss the methods and assumptions that go into our forward modeling. In section~\ref{sec:Results} we present our results in detail. Finally, in section~\ref{sec:Discussion} we recap our primary conclusions and highlight areas that could be improved with future work. Throughout we assume a flat cosmology with $H_0 = 70$ km s$^{-1}$ Mpc$^{-1}$, $\Omega_M = 0.3$ and $\Omega_\Lambda = 0.7$.

\section{Preliminary calculation}
\label{sec:preliminary}
The main result of the paper is from forward modeling using a synthetic galaxy catalog and survey selection criteria. But first, we discuss a simpler calculation that helps to motivate our more detailed calculation. 

We start with galaxy luminosity functions for two galaxy populations, ``red'' and ``blue'', separated by a color-magnitude cut, as in \citet{Cool2012} Figure~3. This uses the Schechter function fits from their table~4, for a redshift bin between $z = 0.05$ and $z = 0.15$. 

\begin{figure*}[htb!]
    \plottwo{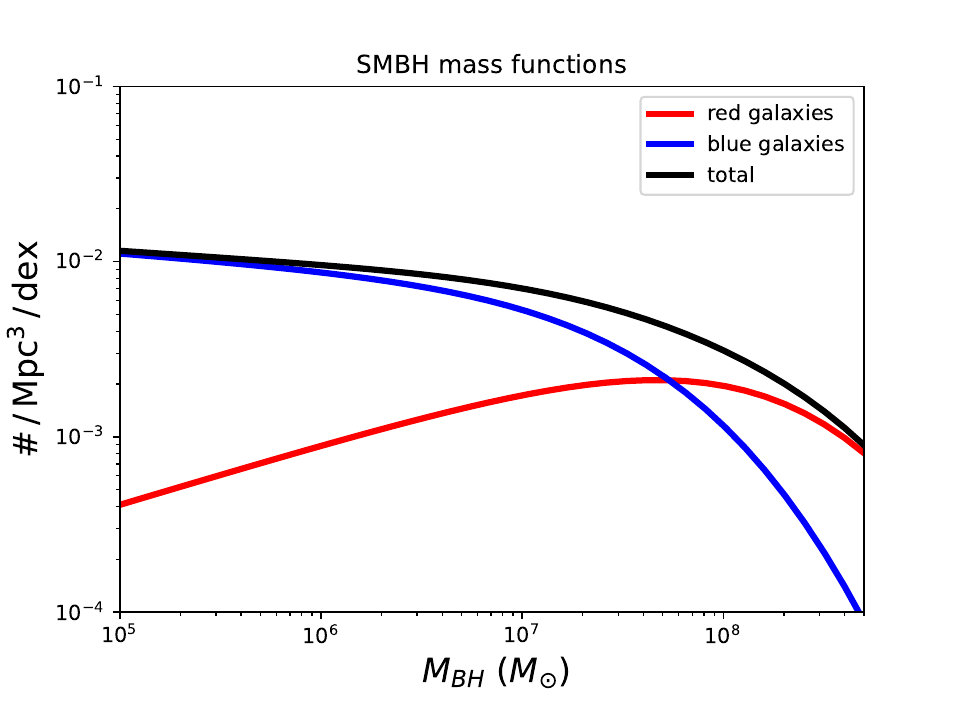}{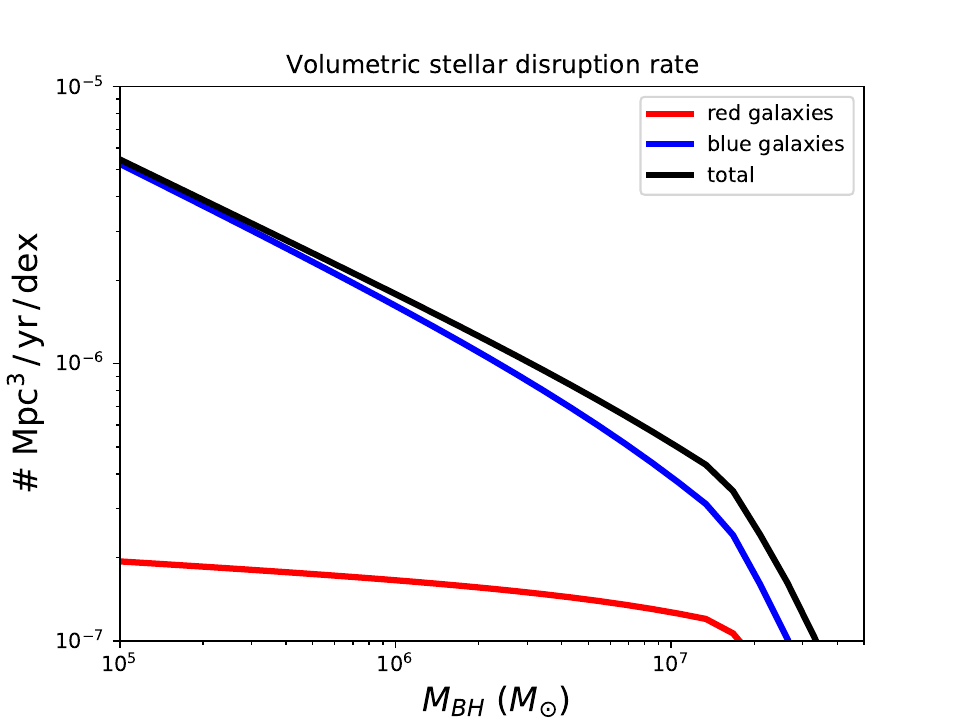}
    \caption{\emph{First panel}: This shows the SMBH mass function corresponding to to galaxies within $z = 0.15$ separated by a color-magnitude cut as in \citet{Cool2012}. \emph{Second panel}: This shows the volumetric rate of tidal disruptions of stars outside of the BH event horizon, again separated between the two galaxy populations, for a disruption rate that is proportional to \mbh$^{-0.4}$, and with a distribution of stellar masses related to a Kroupa IMF. For these assumptions, the majority of the stellar disruptions oustide the BH event horizon in nearby galaxies should be in ``blue'' galaxies. This is not seen in present survey data, which helps to motivate our more detailed forward modeling. } 
    \label{fig:SmbhMassFuncVolDisrupt}
\end{figure*}

To convert these luminosity functions to SMBH mass functions, we make use of the \mbh --$L_{\rm bulge}$ relation from \citet{Tundo2007} for early-type galaxies. We account for the scatter in this relation to convert between a given galaxy bulge magnitude and a distribution of likely SMBH masses. Following \citet{Shankar2004} we assume a bulge-to-total luminosity ratio of 0.85 for all early-type galaxies (which we apply to the ``red'' galaxies) and 0.35 for all late-type galaxies (which we apply to the ``blue'' galaxies), and we apply the same \mbh --$L_{\rm bulge}$ relation to all bulge components. We assume a SMBH halo occupation fraction of unity for all the galaxies that might produce flares we can see, for all galaxy types. The resulting SMBH mass functions, shown in the first panel of Figure~\ref{fig:SmbhMassFuncVolDisrupt}, look very similar to Figure~5 of \citet{Shankar2004}. 

The volumetric disruption rate, as a function of black hole mass, depends both on the SMBH mass function and the per-galaxy disruption rate as a function of \mbh. Here we follow a procedure identical to the one described in SM16 to produce the volumetric disruption rate as a function of \mbh, but now we have it for the red and blue galaxies separately. Here we assume the per-galaxy disruption rates to be the same in the red and blue galaxies, both for the overall rate normalization and the dependence on \mbh, proportional to \mbh$^{-0.4}$, as suggested by SM16 and similar to equation~\eqref{eq:MbhRate}, one of the prescriptions we will use in the more detailed forward modeling that follows. The result, shown in Figure~\ref{fig:SmbhMassFuncVolDisrupt}, is very similar to Figure~8 of SM16, but it is now divided between the two galaxy populations.

As in SM16, to generate these curves we needed to integrate the disruption rates over a present-day stellar mass function, since for each stellar mass (and radius) there is a maximum \mbh\ that can disrupt it without hiding the result within the event horizon (the Hills mass, see equation~\ref{eq:Hills}). Here we're using the same present-day stellar mass function in both the blue and red galaxies, which is a Kroupa initial mass function (IMF) but truncated to only include stars with mass $< 1 \msun$. In principle we could use a different stellar mass function for the blue and red galaxies, but the results are not very sensitive to it. 

The fact that the volumetric disruption rate is so much higher for blue galaxies than for red galaxies is simply a reflection of the fact that most galaxies in the local universe with \mbh ~in the right range to disrupt stars are in blue galaxies (this can be seen in the galaxy luminosity functions). The sharp downturn at in the rate for masses above a few times $10^7 \msun$ is because SMBHs at those high masses start to exceed the Hills mass for the stars they might disrupt (SM16).

For this plot, we have not yet accounted for the flare luminosity function and how that might effect flare visibility from these disruptions. In particular, this means that any possible effect of the Eddington limit has been ignored. Since lower mass SMBHs in the local universe tend to be found more often in blue galaxies (again, see left panel of Figure~\ref{fig:SmbhMassFuncVolDisrupt}), and since the Eddington limit would prevent these flares from becoming as bright as those from higher mass black holes, this is one reason that the above plot of volumetric disruption rate might cause one to over-estimate the fraction of flares found in blue galaxies. Nor have we accounted for dust in the host galaxies or how the flares might contrast with their host light. Galaxies with more star formation tend to be dustier, while also tending to have more young stars that move the galaxy to the blue side of a $u-r$ cut. This is another reason that the above plot might cause one to over-predict the number of flares found in blue galaxies. All of those details will be included in the more detailed forward modeling. 

The take-away here is that, if the disruption rates were the same in the red and blue galaxies modulo \mbh, and the flares produced by these disruptions were equally visible at these redshifts, we should expect survey detections to be dominated by blue galaxy hosts. The fact that surveys such as ZTF do not find this suggests significant differences in the rates of disruption in these galaxies (which could in turn depend on stellar dynamics as well as the SMBH halo occupation fraction), a dependence of the flare luminosity function on the galaxy type (e.g. via the Eddington limit or the stellar population), other survey selection effects, or a combination of all of the above. Being able to quantify which of these effects dominate for different physical assumptions about the disruption process and the flare production process is a major goal of this study. 

\section{Method}
\label{sec:Method}
\subsection{Galaxy catalog}
\label{sec:catalog}
We use the synthetic catalog from \citet{van-Velzen2018-1}. We use this, rather than a direct query of the Sloan Digital Sky Survey (SDSS), for the same reason as in \citet{van-Velzen2018-1}: since TDFs are currently found by scanning nuclear transients in host galaxies from archival flux-limited surveys, we want to represent a complete flux-limited survey of galaxies in a cosmological volume. However, at redshifts $\gtrsim 0.1$, SDSS galaxies do not generally have all the information we need for forward modeling (e.g., spectroscopic redshifts, velocity dispersion to measure \mbh, total stellar masses, star formation rates, surface brightness profiles, etc). Therefore the synthetic catalog contains galaxies whose properties are drawn from distributions of a smaller number of measured galaxy properties, mostly from the NYU Value-Added Galaxy Catalog \citep{Blanton2005}, which does draw on data from SDSS \citep{Adelman-McCarthy2008-0,Padmanabhan2008-0} and is intended to preserve empirical galaxy correlations and their redshift dependencies. The catalog was populated with galaxy down to an apparent magnitude of $m_r=22$.

The total stellar mass of the galaxies in the synthetic catalog was obtained using the result of \citet{Mendel2014-0}, who applied the Flexible Stellar Population Synthesis code \citep{Conroy2010Code} to the SDSS photometry. The same code and with identical model assumptions were used by \citet{van-Velzen2020} to estimate the stellar mass of the TDE host galaxies. 

The black hole masses in the catalog are derived from an empirical \mbh--$\sigma$ relation --- specifically, the \citet{Gultekin2009} relation for ``all'' galaxies (both early- and late- type). The stellar velocity dispersions are estimated from the galaxy stellar mass and effective radius using the virial theorem. A 0.4 dex scatter about this \mbh--$\sigma$ relation was included when generating \mbh\ values for the catalog galaxies \citep{van-Velzen2018-1}. This process has also assumed that the aforementioned \mbh--$\sigma$ relation can be extrapolated down to black hole masses as low as $\sim 10^5 \msun$. Future work can consider how the relation might need to be modified at the low-mass end. 

The catalog does not contain a morphological classification for each galaxy. However, it is populated with galaxies that represent a wide range of morphologies, and this is reflected in their attributes including stellar mass, color, surface brightness profile, star formation rate, etc, all of which correlate with morphology. Regarding surface brightness, for each galaxy the catalog stores an $r$-band Sersic index and half-light radius, representing a pure Sersic profile fit, with no attempt to include a separate disk contribution. Based on these two quantities, the Sersic profile can be use to find the power-law slope of the surface brightness profile at a resolution limit of 0.04''. 
We refer to this inner power-law value as $\gamma^\prime$, related to the so-called Nuker $\gamma$ used to fit the inner power-law of galaxy surface brightness \citet{Lauer2007}. This is an imperfect procedure --- in particular, two identical galaxies located at different redshifts would have different values of $\gamma^\prime$ computed this way. However, we chose to use the 0.04'' resolution limit for measuring $\gamma^\prime$ because this is the resolution limit of the HST WFPC2 camera which was used to measure $\gamma^\prime$ for the majority of the galaxies in \citet{Lauer2007}, and those galaxies were in turn used to calibrate per-galaxy disruption rates based on $\gamma^\prime$, as discussed below. While the Sersic index and effective radius measurements that went into the galaxy catalog were based on SDSS imaging data with a point-spread funtion (PSF) of approximately 1'' full-width at half maximum (FWHM), we are extrapolating these fits down to the resolution at which the rate measurements were calibrated. We hope to improve the treatment of galaxy surface brightness profiles in future work. 

\subsection{Per-galaxy TDE rates}
\label{sec:PerGalaxyRates}
Accurately modeling the stellar disruption rate in a galaxy is a detailed task. The steps for doing so can be found in \citet{Wang2004} and SM16. One begins with the stellar surface brightness profile of the galaxy, then de-projects this to obtain the three-dimensional stellar density profile, gravitational potential, and phase space distribution function. This process is sensitive to the properties of the stellar velocity distribution and the overall symmetry, or lack thereof, of the spatial distribution of the stars. Once these steps are complete, the flux of stars into the loss cone and finally the stellar disruption rate can be computed. These final steps depend on the mass distribution of the stars and the properties of the black hole including its mass and spin, and may be different for stars of different masses and radii. 

For this paper, we will gloss over most of these details, choosing instead from simplified parameterizations of the disruption rate in each galaxy taken from SM16. These authors computed disruption rates in a sample of 219 galaxies, with surface brightnesses measured from \citet{Lauer2007} and \citet{Trujillo2004}, assuming isotropic distributions for the stellar velocities. In addition to performing the detailed calculation described above, they offered summary relations for how these rates correlated with \mbh\ and $\gamma^\prime$. In all cases, it must be remarked that reducing the disruption rate to a single parameter dependence results in a large amount of scatter around that relation, which is not presently included in the forward modeling.

The first parameterization we use is one based on $\gamma^\prime$. Here there is a distinction drawn between “cored” galaxies ($\gamma^\prime < 0.3$) and non-cored galaxies (larger $\gamma^\prime$). In line with previous work, SM16 found that the disruption rate in cored galaxies is systematically lower than in non-cored galaxies. This in turn causes the $\gamma^\prime$ rate parameterization to change for high-mass galaxies compared to lower-mass galaxies, because cores are more often found in higher-mass galaxies. The rate parameterization based on gamma-prime, calibrated for galaxies in which the black hole mass is below the Hills mass for a solar-type star, is 
\begin{equation}
\label{eq:GammaPrimeRate}
\dot{N}_{\rm TDE} = 10^{-3.79} {\gamma^\prime}^{0.852} \,\, {\rm yr}^{-1} \, .
\end{equation}
(N. Stone, private communication). We also consider a parameterization based on \mbh\ only. For galaxies without cores, the result from SM16 is 
\begin{equation}
\label{eq:MbhRate}
\dot{N}_{\rm TDE} = 10^{-4.19} \left(\frac{M_{\rm BH}}{10^8 M_\odot}\right) ^{-0.223} \,\, {\rm yr}^{-1} \, .
\end{equation}
Since this second rate based on \mbh\ is ignoring cored galaxies entirely, we will be over-estimating our disruption rates. However, results from the \mbox{ATLAS$^{3{\rm D}}$} survey indicate that cores are rare in galaxies with total $M_\ast < 8 \times 10^{10} \msun$. Galaxies more massive than that tend to harbor black holes more massive than the Hills mass for a solar type star. Indeed, our results will indicate that the expected rate of flare detections from such massive galaxies is vanishingly small for this reason, even if we are overestimating their disruption rates by ignoring that a substantial fraction of them are cored. 

\begin{figure}[htb!]
\gridline{\fig{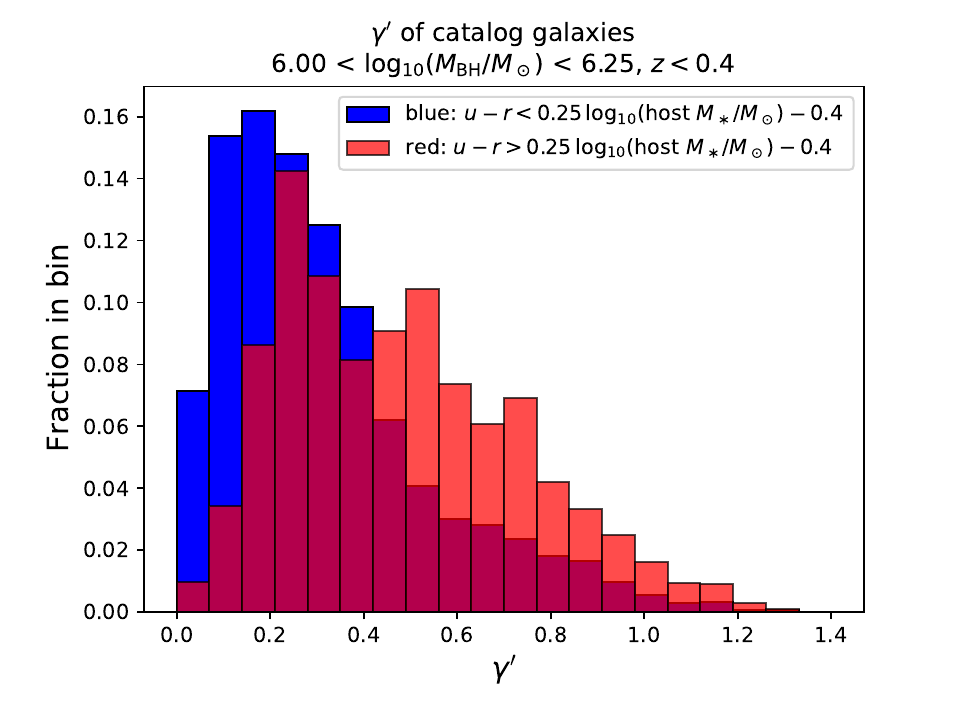}{0.5 \textwidth}{(a)}}
\gridline{\fig{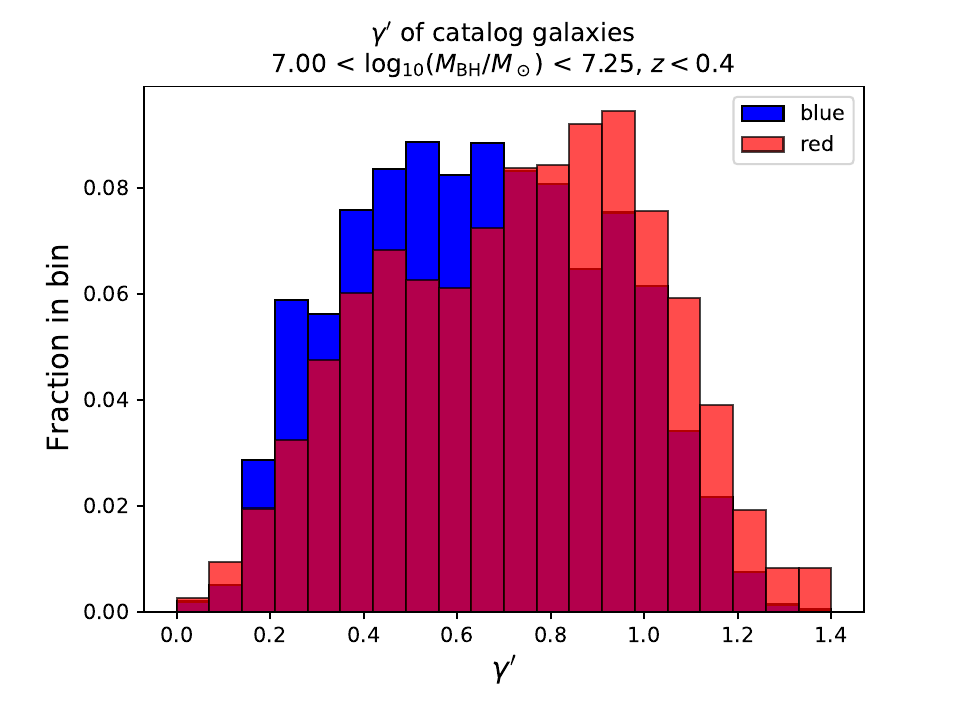}{0.5\textwidth}{(b)}}
\caption{Distributions of $\gamma^\prime$ for mock catalog galaxies. Panel (a) is for an \mbh\ bin near $10^6 \,M_\odot$ while panel (b) is for a bin near $10^7 \,M_\odot$. The division between `red' vs. `blue' galaxies is made via the color-magnitude cut indicated in the legend. Red galaxies tend to have higher $\gamma^\prime$, and galaxies with \mbh\ closer to $10^7$ tend to have higher $\gamma^\prime$ than those with \mbh\ closer to $10^6$. In panel (a) there are 147,828 and 17,657 red and blue galaxies, respectively. In panel (b) there are 10,578 and 42,988 such galaxies. }
\label{fig:GammaprimeBins}
\end{figure}

These parameterized rates are based on the more detailed calculations which were performed \emph{only for elliptical galaxies}, which made up the \citet{Lauer2007} and \citet{Trujillo2004} samples. There is no comprehensive published study on stellar disruption rates in spiral galaxies, and this remains an important detail that must be improved in future modeling. Galaxy morphology does end up affecting the disruption rate with our simple parameterizations to the extent that it correlates with \mbh\ and $\gamma^\prime$. In particular, the $\gamma^\prime$ inferred for catalog entries representing spiral galaxies, based on their Sersic index and half-light radii, tend to be smaller than for elliptical galaxies. Some distributions of $\gamma^\prime$ for catalog galaxies are shown in Figure~\ref{fig:GammaprimeBins}.

We are assuming that the SMBH occupation fraction is unity in all galaxies. Future work can consider different guesses for how the occupation fraction might drop in lower-mass galaxies to see if this improves matches to observed flare distributions. 

As noted in SM16, these rate estimates do not account for nuclear star clusters.  The disruption rate will likely be enhanced in galaxies that contain these \citep[e.g.][]{Pfister2020-0}, and this is an effect that should be included in future modeling efforts along these lines. 

We assume that the stellar disruption rate is the same for all masses of the stars being disrupted. Past theoretical work suggests that this is true for main sequence stars to a high degree of accuracy. In the following, $M_\ast$ and $R_\ast$ will denote the mass and radius, respectively, of the star being disrupted. Let $r_t \equiv R_\ast (\mbh / M_\ast)^{1/3}$ denote the tidal radius. \citet{MacLeod2012} found $\dot{N}_{\rm TDE} \propto r_t^{1/4} \propto M_\ast^{-1/12}$ for a galaxy with a given \mbh. While this weak dependence could be included, we are ignoring it. As in SM16, we are not considering disruptions of giant stars. 

Equations~\eqref{eq:GammaPrimeRate} and \eqref{eq:MbhRate} represent the rate at which stars in the galaxy are directed into the ``loss cone'', where they will be ripped apart by tidal forces. However, these equations do not account for whether the resulting TDE produces material that is entirely swallowed by the SMBH. For a given stellar mass and stellar radius, all the stellar material will be swallowed if \mbh\ exceeds the Hills mass, which depends on both \mstar\ and $R_\ast$:
\begin{align}
\label{eq:Hills}
M_{\rm Hills} &\equiv \sqrt{\frac{5}{8}} \left(\frac{R_\ast c^2}{2 G M_\ast^{1/3}} \right)^{3/2} \nonumber \\
&= 9.0 \times 10^7 \, \msun \left(\frac{M_\ast}{\msun}\right)^{-1/2}  \left(\frac{R_\ast}{R_\odot}\right)^{3/2} \, .
\end{align}
\citep{Hills1975,Beloborodov1992,Leloudas2016}, where $G$ is Newton's gravitational constant and $c$ is the speed of light. For this reason, when we randomly generate disruptions, we must sample from a distribution of stellar masses to determine whether the star in question can be disrupted outside the event horizon of the black hole in question. Here we are ignoring how black hole spin affects the distance from the black hole at which stars are disrupted - such a dependence could be included in future work.

Following SM16, we assume that the present-day stellar mass function can be approximated by adapting a Kroupa IMF, and truncating the distribution at high stellar mass, to account for the deaths of massive stars over time. Specifically, we truncate the stellar mass at 1 \msun\ in all galaxies. For added detail we could adopt different truncation masses for galaxies with different star-formation histories. However, given all our other assumptions, this effect on the disruption rate is expected to be small; given how weighted the IMF is toward low-mass stars, the fraction of disruptions corresponding to stars more massive than 1 solar mass would always be small. However, we acknowledge that if disruptions of massive stars give rise to systematically brighter flares than we have accounted for, then our neglect of these stars could become more problematic. 

\subsection{Optical/UV flare luminosity and temperature distributions}
\label{sec:FlareLF}
This paper is focused on flares detected in optical surveys such as ZTF and the upcoming VRO/LSST. While these optically-selected flares are sometimes found to be X-ray or radio sources in follow-up, we will not be making use of emission in any band outside of the optical surveys for determining the flare detection rates. 

In practice, the optical/UV emission is fit to a blackbody. Therefore, we also use blackbody emission in our forward modeling for the optical flux, regardless of whether this accurately describes emission in other bands. We will use the label $L_{\rm bb}$ to refer to the peak, frequency-integrated luminosity in the blackbody for UV/optical detections, and $T_{\rm bb}$ to refer to the associated temperature of this blackbody.

A complete theoretical understanding of TDF emission would allow us to derive $L_{\rm bb}$ and $T_{\rm bb}$ from first principles for a given set of disruption parameters. Several theoretical frameworks have been proposed to accomplish this, such as those discussed in \citet{Mockler2019} and \citet{Ryu2020-0}, However, no consensus for this procedure currently exists, and the mechanism that produces the optical/UV light in TDFs remains incompletely characterized. For this reason, we turn to the empirical luminosity function (LF) from \citet{van-Velzen2018-1}, who found that the underlying distribution of peak rest-frame $g$-band luminosity $L_g$ follows a distribution approximately described by $d N / (d L_g) \propto L_g^{-2.5}$, or equivalently  $dN / (d \log_{10} L_g) \propto L_g^{-1.5}$. 

 Rather than implementing the empirical $g$-band LF directly in the forward modeling, we do the following: For each disruption generated in the model, we randomly draw $L_{\rm bb}$ from a probability distribution $\psi$ where $\psi(L_{\rm bb})$ follows the same power-law in terms of $L_{\rm bb}$ as the empirical $g$-band LF follows for $L_g$. Meanwhile, $T_{\rm bb}$ is set independently and drawn from a uniform distribution confined to the range 10,000 to 50,000 Kelvin, which brackets the fitted temperatures in real events. The procedure for using these rest-frame, unextincted values for flare luminosity and temperature to produce observer-frame, host-extincted measures of flux in a given band will be described in Sections~\ref{sec:HostDust} and \ref{sec:Kcorrection}, and in Appendices~\ref{sec:ImplementDustKcorrection} and \ref{sec:Tfit}. 
 
An advantage of working in terms of $L_{\rm bb}$ instead of $L_{g}$ is that we may naturally impose physically motivated constraints on how large it may be. For example, one of these constraints comes from the fact that for the TDF hosts where \mbh\ has been estimated via the \mbh--$\sigma$ relation, $L_{\rm bb}$ is consistent with being Eddington-limited \citet{Wevers2019-1}. While the bolometric luminosity that comes from adding the UV/optical blackbody component, a separate thermal X-ray contribution, and possibly an additional excess in the EUV, might result in luminosity that is super-Eddington by a small factor, we believe it is reasonable to impose the Eddington-limit constraint on $L_{\rm bb}$.

Another luminosity constraint becomes important at sufficiently high black hole masses, and this is set by the rate at which stellar material is returning to the pericenter of the original stellar orbit following disruption. We do not allow the optical peak luminosity to exceed $0.1 \, \dot{M}_{\rm peak} c^2$. To determine $\dot{M}_{\rm peak}$ we make use of the fitting functions from \citet{Guillochon2013} for the peak mass fallback rate as a function of black hole mass and stellar mass, for disruptions with penetration factors $\beta$ = 1, where $\beta$ is defined as $r_t$ divided by the pericenter radius of the initial orbit of the disrupted star. These functions were calibrated using hydrodynamic simulations of polytrope stellar models, in Newtonian gravity and neglecting effects of black hole spin. Since we are only considering disruptions of stars with mass $\le 1 \msun$, we only use the fitting function for polytropes with  $\gamma$-polytrope $= 5/3$. Future work can sample from distributions of the disruption penetration factors, and make use of fitting functions based on more recent hydrodynamic simulations of disruptions of more realistic stellar models \citep{Ryu2020-2, Golightly2019-2,Law-Smith2019,Goicovic2019-0}. However, we point out that in most cases, the Eddington limit provides the operational luminosity constraint, rather than the peak fallback rate (see Figure~\ref{fig:LFitMbh2D}).

A final consideration is how faint we allow $\psi$ to extend with non-zero probability. The fainter flares can be, the smaller the flare detection rate will be -- surveys will not be able to detect as many disruptions because flares from a larger number of disruptions will be too dim. Since $\psi$ is so steep, a small change in this lower luminosity limit in our models can have a large effect in the flare detection rate expected from those models. For most of the models discussed below we assume that $\psi = 0$ for $L_{\rm bb} < 10^{43}$ \ergspersec. This is partially motivated by the fact that iPTF16fnl, the faintest event seen so far, had $L_{\rm bb}$ of approximately that value. We will refer to this universal minimum value of $L_{\rm bb}$ as $L_{\rm bb, min}$. To further understand the consequences of our choice of value for $L_{\rm bb, min}$, in section~\ref{sec:ChangingLmin} we consider one model where $L_{\rm bb, min} = 10^{42}$ \ergspersec . For the precise description of $\psi(L_{\rm bb})$, how it makes use of $L_{\rm bb,min}$, and how we sample from this distribution, please consult Appendix~\ref{sec:ImplementLF}.

The choice of $L_{\rm bb, min}$, along with the stipulation that $L_{\rm bb}$ respect the Eddington limit, effectively imposes a cut on the black hole masses that produce detectable flares in our model survey. As \mbh\ is lowered, eventually $L_{\rm Edd}$ will fall below $L_{\rm bb, min}$. TDFs for those black holes cannot simultaneously satisfy the criteria that $L_{\rm bb} \ge L_{\rm bb, min}$ and $L_{\rm bb} \le L_{\rm Edd}$, and so we do not consider flares produced in these systems to be visible. For $L_{\rm bb, min} = 10^{43}$ \ergspersec, this effectively limits detections to black holes with masses above $10^{4.9} \msun$. More generally, the effective lower limit on \mbh\ that comes from requiring $L_{\rm Edd} > L_{\rm bb,min}$ is
\begin{equation}
    \mbh > 10^{4.9} \msun \left(\frac{ L_{\rm bb,min}}{10^{43} \,\,{\rm \ergspersec}} \right) \, \, ,
\end{equation}
where we are using the definition $L_{\rm Edd} \equiv 4 \pi G \mbh m_p c / \sigma_{T}$, and where $m_p$ is the proton mass and $\sigma_T$ is the Thomson electron scattering cross section. 

The combined effects of all of these constraints from the flare luminosity probability distribution, and their dependencies on \mbh\ and $M_\ast$, can be seen in Figure~\ref{fig:LFitMbh2D} as part of the description of our results. 

The procedure for handling the flare luminosity and temperature distributions discussed in this section is one of the most uncertain ingredients in our modeling. Throughout section~\ref{sec:Results} we will discuss the potential drawbacks of this approach and how it might be improved in future work. In particular, in section~\ref{sec:ResultsLuminosity}, we show how the expected distribution of $L_{g,\rm fit}$ for detected flares produced by the model (which includes an estimate of the effect of obscuration by host dust; see Appendix~\ref{sec:Tfit} for a precise definition) compares to the observed distribution of $L_g$ of detected flares in ZTF, and we also show how the \emph{volumetric} distribution of $L_{g,\rm fit}$ compares to the measured volumetric LF produced from past survey data as analyzed by \citet{van-Velzen2018-1}. In section~\ref{sec:TemperatureResults}, we discuss some of the consequences of the decision to sample $T_{\rm bb}$ uniformly and independently of $L_{\rm bb}$, and we explore an alternative possibility for how to set $T_{\rm bb}$.

In addition to the peak luminosity and temperature of the flare, other aspects of the light-curve, especially how rapidly the light curve rises to peak light, can play an important role in a survey's detection efficiency of these flares. However, in the case where the survey cadence is short compared to the flare rise-times, details of the light-curve aside from the peak luminosity and temperature have little effect on the survey detection efficiency. The cadence for ZTF was usually no longer than three days, which is short compared to the expected and observed light-curve rise-times for the TDFs produced by the disruption of main-sequence stars that we are considering. We therefore feel confident that our restriction of attention to peak flare quantities is not contributing to errors in our expected detection rates for ZTF. Such a restriction is more likely to affect our predictions for VRO/LSST, as we will discuss in Section~\ref{sec:VROLSST}. 

\subsection{Host dust extinction and reddening}
\label{sec:HostDust}
While most surveys correct for Milky Way extinction when fitting for flare temperatures and luminosities, traditionally extinction from the host galaxy has been ignored. Often this can be justified if the spectra of the flare and the galactic nucleus show no evidence of significant extinction. However, a major objective of this work is to quantify how many flares surveys might be missing because they happen to take place in highly obscured galactic nuclei, so we will need a way to model this effect.

For optical surveys, the primary contributor to extinction is dust (neutral gas has a subdominant effect, although for X-ray surveys it would be dominant). For highest accuracy, a modeler would require detailed maps of dust distributions in all galaxies that might host detectable disruption flares. Here, we instead attempt to quantify this effect using a more approximate approach. We refer to \citet{Garn2010-1}, who measured the typical extinction in star-forming galaxies by measuring Balmer decrements. They point out that while the dust extinction they measure correlates with galaxy total stellar mass, galaxy metallicity, and star formation rate, the most ``fundamental'' of these correlations in star-forming galaxies is the total stellar mass of the galaxy. We adopt their parameterization of this relationship: 
\begin{align}
\label{eq:GarnBestLaw}
 A_{{\rm H} \alpha,\rm median} &= 0.91 + 0.77 x + 0.11 x^2 -0.09 x^3 \, ,\; \; {\rm where}  \nonumber \\
 x &\equiv \log_{10}\left(\frac{{\rm host \, \,} M_\ast}{10^{10} M_\odot}\right) \, ,
\end{align}
and the distribution for these extinction values around the median can be treated as Gaussian (with a floor at zero), with standard deviation of 0.28 mags. This relationship was only calibrated for star-forming galaxies with total stellar mass between $10^{8.5}$ and $10^{11.5}$ \msun. For galaxies that fall outside of that range, we just use the edge values of the relationship (evaluated either at $10^{8.5}$~\msun\ or $10^{11.5}$~\msun). However, the vast majority of hosts of detected flares in our model surveys ($\approx 97.6\%$ in the fiducial model presented in section~\ref{sec:SixPanelDiscussion}) fall within the calibration range. 

We set a specific star-formation rate (sSFR) of at least $10^{-11.3}$ ~yr$^{-1}$ as our criterion for treating a galaxy as “star-forming.” This separates the two galaxy populations in the mock catalog, as can be seen in Figures~\ref{fig:MstarsSFR_Fiducial} and \ref{fig:MstarsSFR_MbhRate} below where galaxies are binned by sSFR. For galaxies with sSFRs lower than this, we assume a median $A_V$ of 0.2 mags, with a distribution about the median given by a Gaussian of 0.06 mags, with a floor at $A_V = 0$. This choice of median extinction at the galaxy center is guided by spatially resolved observations of $A_V$ in early-type galaxies from the CALIFA survey \citep{Gonzalez-Delgado2015}. 

To convert from the extinction at H$\alpha$, or from $A_V$, to the extinction at all observed bands shifted to the host rest-frame, we use the \citet{Calzetti2000} law. We use $R_V = 4.2$ for all galaxies.

This treatment of dust extinction in our modeling comes with important caveats. For TDF detections, we are interested in the extinction toward the center of the galaxy, which is a column with a tiny cross section (namely the TDF photosphere). Thus, a relevant question is how many dusty regions around stars are sampled by the line of sight towards the SMBH in each galaxy producing TDFs. The Balmer decrement measurements from \citet{Garn2010-1} tend to be dominated by the brightest stars in star-forming regions, since these contribute the most light to the SDSS spectra. Since we are applying these extinction measurements to our observations of the narrow, central columns of galaxies, we therefore run the risk of incorrectly estimating the extinction toward the black hole at the galaxy center. This concern becomes especially pertinent if the dust in the galaxies is concentrated into narrow ``dust lanes'', the prevalence of  which depends on galaxy mass \citep{Dalcanton2004-0}. Since we lack a reliable means of modeling the 3D dust distribution in all the catalog galaxies, we have chosen the approximate approach described above,  but the limitations of this approach must be kept in mind for future improvements.

\subsection{K-correction}
\label{sec:Kcorrection}
For each randomly generated disruption, once we have sampled the flare luminosity, temperature, and dust extinction in the galaxy rest-frame, we follow the procedure from \citet{Hogg1999} to K-correct this extincted and reddened emission and to determine the band flux in the observer frame, treating the observed band as infinitely narrow for this purpose. No additional correction is made for Milky Way extinction. For more details, consult Appendix~\ref{sec:ImplementDustKcorrection}.

\subsection{Survey-selection criteria}
To count as visible, the flare must pass a number of cuts which can be set based on the survey being modeled. We approximate ZTF's selection criteria as follows: first, we require the peak flux from the flare to be brighter than 19 apparent mag in both the $g$-band and $r$-band. While in principle ZTF can detect transients fainter than this, it is also important that the rise-to-peak in the light curve be detected, which translates into a requirement that the peak flux be a few mags brighter than the faintest transients the instrument can detect. At this point we also mention that in all cases when converting from model spectra to band magnitudes, we treat the band filters as delta functions peaked at the mean wavelength of the filter

A cut that tends to separate TDFs from other nuclear transients such as supernovae is that they are persistently ``blue'', meaning the flare $g - r$ is sufficiently low before, during, and after peak light \citep{Hung2018, van-Velzen2020}. In our models we account for this by requiring $g - r < 0$ for the peak emission of the flare.

Another requirement is that the flare contrast sufficiently against the light from the galaxy host contained within the PSF. In other words, we must have 
\begin{equation}
\label{eq:DeltaMPrimeDefinition}
2.5 \log_{10}\left(1 + \frac{F_{\rm TDF}}{F_{\rm host,PSF}}\right) > \lvert \Delta m^{\prime} \rvert \, \, ,
\end{equation}
where $F_{\rm TDF}$ is the peak specific flux of the transient in the band of interest (not counting contributions from the host), and $F_{\rm host,PSF}$ is the background flux of starlight contained within the PSF around the center of the host in the band of interest. The quantity $|\Delta m^{\prime}|$ is survey-dependent, and given our definition in Equation~\eqref{eq:DeltaMPrimeDefinition}, values closer to zero correspond to surveys that have an easier time picking out faint flares against the host light. The above condition can be rearranged to create a condition on the TDF flux, expressed in magnitudes, yielding 
\begin{equation}
\label{eq:HostContrast1}
m_{\rm TDF} < m_{\rm PSF} - 2.5 \log_{\rm 10}\left( 10^{\frac{\lvert \Delta m ^\prime \rvert}{2.5}} - 1 \right) \, \, .
\end{equation}
However, it is sometimes more natural to think of the survey contrast in another way, in which the requirement is expressed as
\begin{equation}
\label{eq:HostContrast2}
m_{\rm TDF} < m_{\rm PSF} + \Delta m \, \, ,  
\end{equation}
where again $m_{\rm TDF}$ refers to the flux from the TDF alone (not including the host background). Given this definition of $\Delta m$, which can take on both positive and negative values, a larger value corresponds to a survey that has an easier time picking out faint flares against the host light. In principle, for ZTF, detections may correspond to $\Delta m$ as high as 3.0, although a more typical practical limit is to use $\Delta m$ of 1, which is what we use whenever we apply a cut on host contrast in the forward-modeling. A $\Delta m$ of 1 corresponds to a $|\Delta m^\prime|$ of approximately 0.36. For comparison, the value of $|\Delta m^\prime|$ specified for the intermediate Palomar Transient Factory (iPTF) TDF survey was 0.5 \citep{Hung2018}. We require that the flux from the TDF surpass these limits in both the $r$- and the $g$-bands. We use 2.0'' and 2.1'' for the FWHM of the $r$-band and $g$-band PSFs, respectively \citep{Bellm2019}. To determine the host light contained in the PSF, we integrate the stellar surface brightness in a disk with radius equal to the band FWHM, using the Sersic profile information in the synthetic galaxy catalog (see section~\ref{sec:catalog}).

\subsection{Procedure}

For each galaxy in the catalog we generate 100 disruptions, and for each of these we randomly sample \mstar, $L_{\rm bb}$, $T_{\rm bb}$, and $A_V$ from the relevant distributions. We use the fraction of these flares that are detected, multiplied by the galaxy's total disruption rate (set either by $\gamma^\prime$ or \mbh\ depending on the case under consideration), to determine the expected rate of detections from that galaxy. 

For flares that pass the survey selection criteria we perform a least-squares fit to the dust obscured, \emph{galaxy rest-frame} spectrum to determine the peak luminosity and blackbody temperature that would be inferred for the flare. We make use of flux measurements at the the mean filter wavelengths for the ZTF $r$- and $g$-bands, along with the UVW1, UVM2 and UVW2 bands for the UV-optical telescope \citep{Roming2005-0} aboard the Neil Gehrels Swift Observatory. To approximate statistical error for the measurement, we apply a Gaussian random error, with $\sigma$ = 10\% of the original value, to the flux measurement (in magnitudes) in each band before performing the fit. In the results that follow, we use the labels $L_{\rm bb, fit}$, $T_{\rm bb, fit}$ and $L_{g, \rm fit}$ to refer to these fitted values, to distinguish them from the intrinsic (unobscured) flare properties $L_{\rm bb}$, $T_{\rm bb}$, and $L_{g}$ that we have discussed in section~\ref{sec:FlareLF}. For more details about the fitting procedure for $T_{\rm bb}$ and $L_{\rm bb}$ consult Appendix~\ref{sec:Tfit}, and for more details about how we convert randomly generated flares into distributions of detected flare properties, consult Appendix~\ref{sec:MonteCarloExplanation}.

\section{Results}
\label{sec:Results}
\subsection{Effects of combining model ingredients}
\label{sec:SixPanelDiscussion}
\begin{figure*}[htb!]
\gridline{\fig{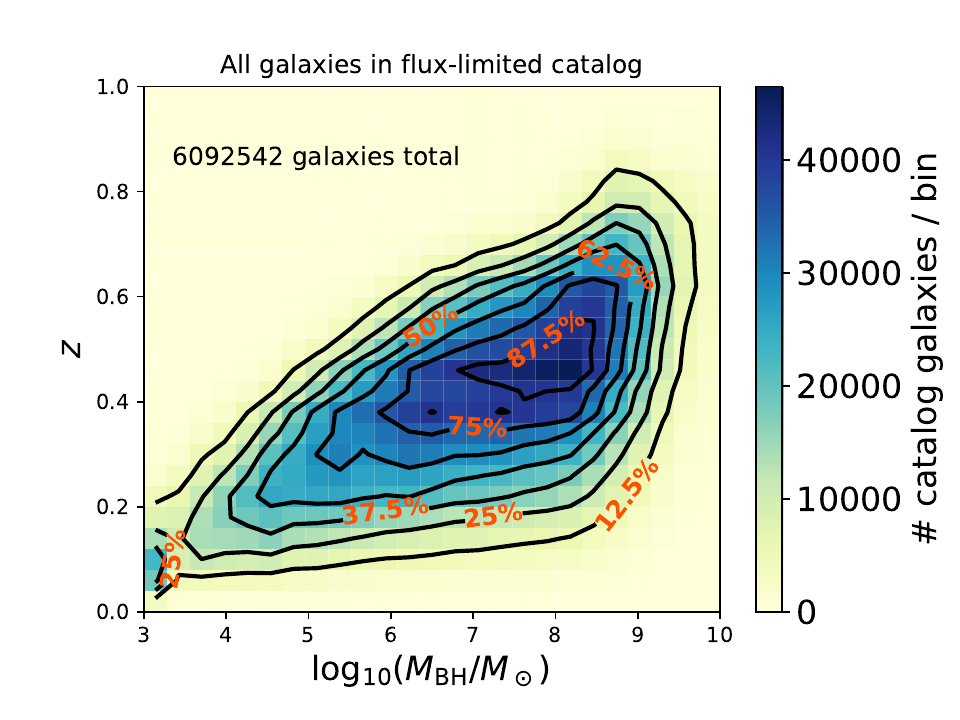}{0.45 \textwidth}{(a)}
          \fig{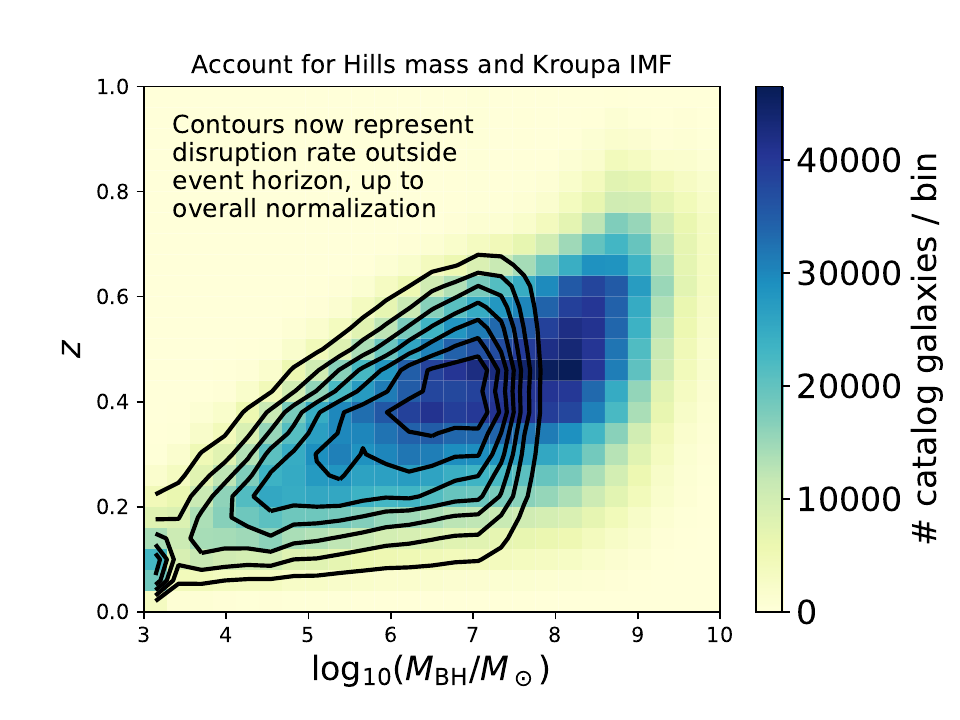}{0.45 \textwidth}{(b)}}
\gridline{\fig{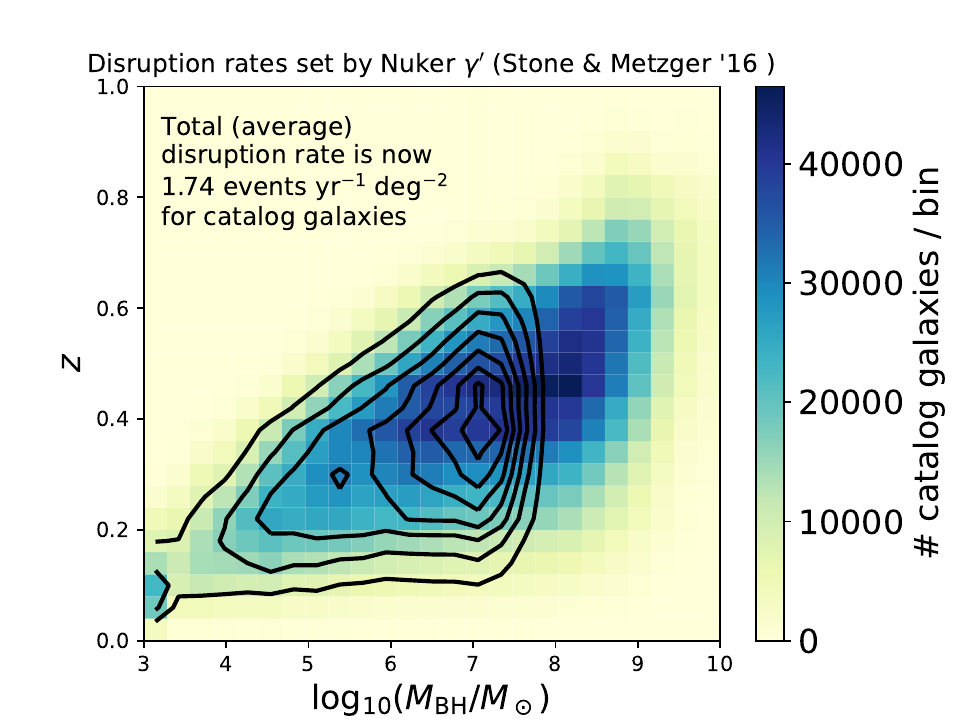}{0.45 \textwidth}{(c)}
          \fig{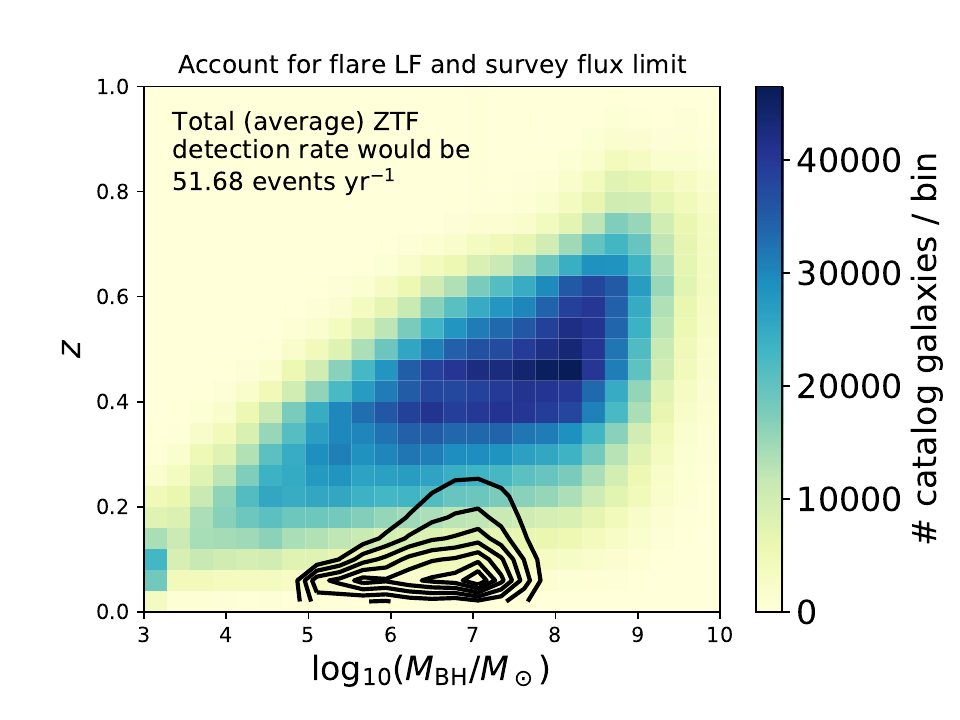}{0.45 \textwidth}{(d)}}
\gridline{\fig{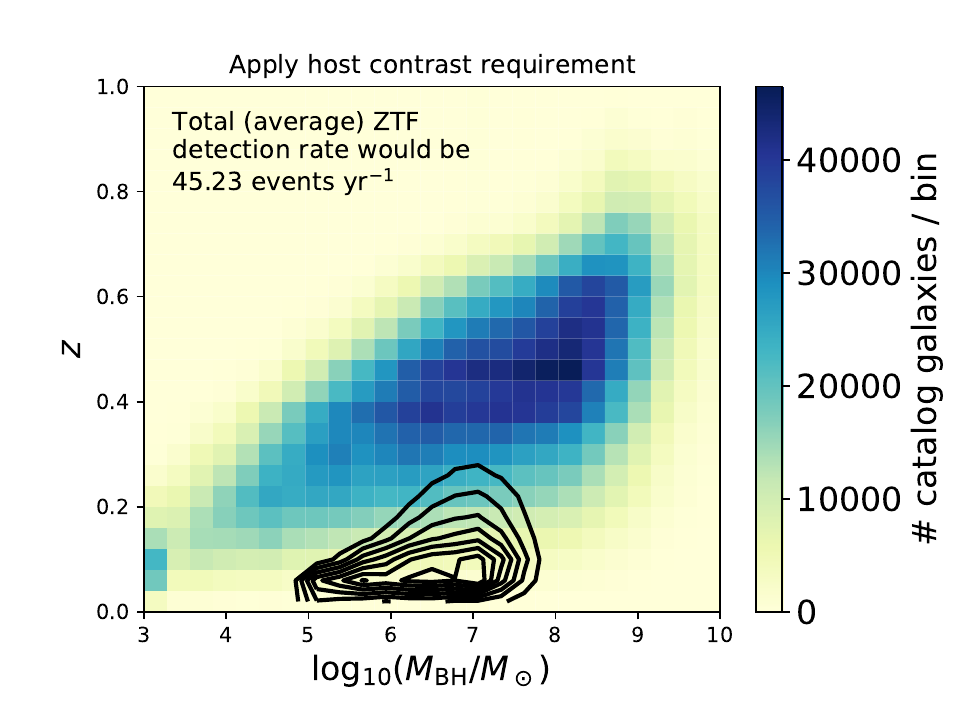}{0.45 \textwidth}{(e)}
          \fig{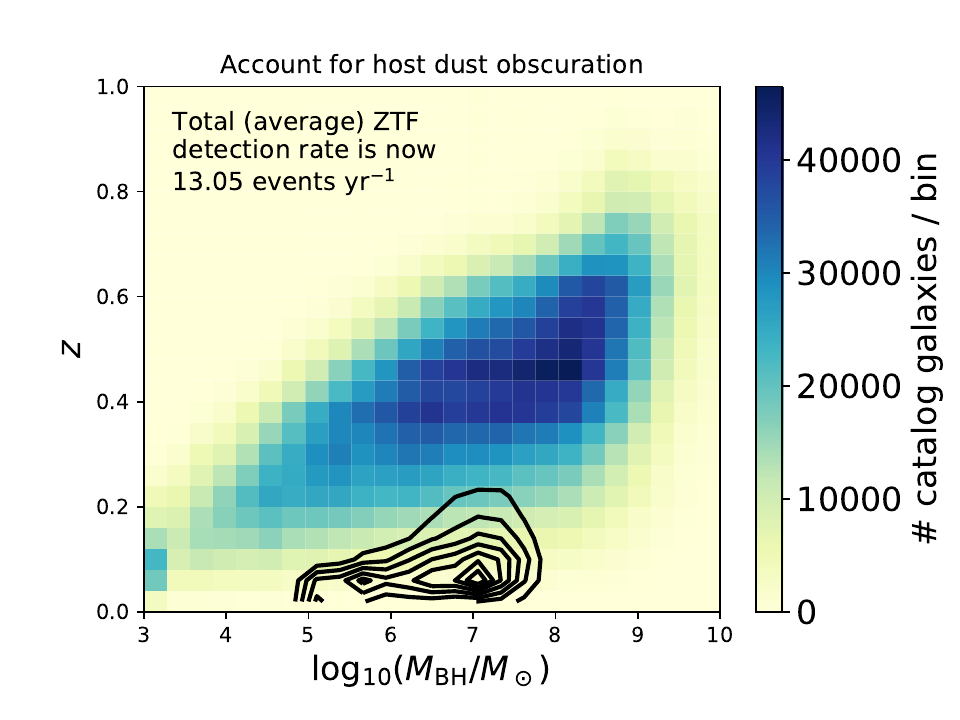}{0.45 \textwidth}{(f)}}
\caption{Two-dimensional histograms comparing populations of host galaxies of TDFs (contours), to all of the galaxies in the mock catalog (colored bins), based on the values of \mbh\ and redshift $z$ for the catalog galaxies. The contours in panel (f) represent the final expected distribution of host galaxies of detected flares in our fiducial model survey, while panels (a) through (e) demonstrate how this population is approached by incrementally adding model details and assumptions. In all panels, galaxies with $\log_{10}( \mbh / \msun) < 3 $ are included in the lowest mass bin. }
\label{fig:SixPanel}
\end{figure*}

First, we will present a sequence of plots to build intuition for how various ingredients in our forward modeling affect the final distribution of properties of observed flares and their hosts. We begin with 2D histograms of host galaxies, with the horizontal axis corresponding to \mbh\ of the host (as found from the mock catalog), and the vertical axis corresponding to the host redshift $z$. The next six subsections are devoted to explaining these histograms, shown as six panels in Figure~\ref{fig:SixPanel}. We will refer to the final result the ``fiducial model''.

\subsubsection{Galaxies in mock flux-limited catalog}
Panel (a) of Figure~\ref{fig:SixPanel} shows the galaxies in the mock catalog binned by \mbh\ and redshift $z$. Notice how distribution of \mbh\ shifts to higher values at higher redshifts. This is because at greater cosmological distances, only more massive galaxies which generally contain more massive black holes pass the flux limit ($m_r \lesssim 22$) to be represented in the catalog, and we are not considering the possibility of ``orphan flares'' that are identified without an association to an archival galaxy. For all panels of Figure~\ref{fig:SixPanel}, the colored bins in the background will remain unchanged, and they represent the number of catalog galaxies in each 2D bin. Contours will be overlaid on this plot, and the contours will always represent a distribution related to tidal disruption rates. To start, if the disruption rate were the same in every galaxy, and if we do not account for which disruptions produce visible flares, then the shape of the disruption rate distribution corresponds exactly to the relative number of catalog galaxies in each bin. The contour levels for this disruption rate distribution are labeled here, and these fractional values for the contours with respect to the peak of the rate distribution will remain the same in the following collection of plots. 

\subsubsection{Accounting for Hills mass and Kroupa IMF}
For panel (b), we have performed the random sampling of 100 disruptions for each galaxy, with stellar masses drawn a Kroupa IMF truncated at 1 \msun, and we have kept track of the fraction of these events such that the star has been disrupted outside the black hole event horizon, simply by comparing the black hole mass to the formula for the Hills mass with respect to the stellar mass in question, given by equation~\eqref{eq:Hills}. Then each galaxy, weighted by this fraction, contributes toward the rate distribution represented by the contours. The effect is that our distribution now drops off at high \mbh, but otherwise tracks the number of galaxies in the catalog. Since we are only computing fractional weights here, the volumetric disruption rate here can be adjusted by any overall normalization.

\subsubsection{Volumetric rates}
 For panel (c) of Figure~\ref{fig:SixPanel} we allow the disruption rate to vary between galaxies based on their surface brightness profiles, specifically their $\gamma^\prime$, using the SM16 rate as expressed by equation~\eqref{eq:GammaPrimeRate}. This effect is cumulative with the effects of the Hills mass and stellar mass function that were discussed in the previous panel. With this rate specification, we can also compute the total, average rate of TDF detections based on the model ingredients included so far. We say ``total'' because we are summing the rate contribution over all the 2D histogram bins. We say ``average'' because this sum over galaxy event rates gives us an overall number of events per unit time, but flares are discrete events that are sampled from this distribution over a specified interval of time, and so the number of events taking place in any time interval is subject to Poisson noise. There is not a big change to the shape of the distribution, but some of the lower mass events are suppressed, so that the distribution is now concentrated more toward $10^7 \msun$ (see Figure~\ref{fig:GammaprimeBins}). 
 
 \subsubsection{Effects of flare luminosity and temperature distributions, and survey flux limit}
 Panel (d) of Figure~\ref{fig:SixPanel} shows the effects of sampling $L_{\rm bb}$ from $\psi(L_{\rm bb})$ and independently sampling $T_{\rm bb}$ from a uniform distribution between $10,000$ and $50,000$ Kelvin, as discussed in section~\ref{sec:FlareLF}. This is combined with the effect of the assumed survey flux limit (peak $m_r$ and $m_g < 19$), and all previous effects discussed so far in these panels. At the high-mass end, this simply squashes the distribution to lower redshift, i.e. more nearby galaxies. On the low-mass end there is also a shift to even lower $z$, and the detection rate also vanishes at the lowest masses, $\log_{10}(\mbh / \msun) < 4.9 $, as discussed at the end of section~\ref{sec:FlareLF}. For black holes more massive than that cutoff, they still only produce flares up to their Eddington limit, which suppresses the detectability of flares from galaxies at the lower end of \mbh. The final expected rate of flare detections in this panel, approximately 50 per year for ZTF, is much lower than the rate of disruptions outside the SMBH event horizon in catalog galaxies represented in panel (c).
 
 \subsubsection{The effect of the cut on host contrast}
In panel (e) of Figure~\ref{fig:SixPanel} we add the requirement that the flare is sufficiently bright compared to the PSF light of its host, as specified by equation~\ref{eq:HostContrast2} with $\Delta m = 1$. About 12\% of the flares detected per unit survey time are lost compared to panel (d). There is not a big change to the shape of the distribution, although we lose more flares at high $z$ than at low $z$, and at high \mbh\ than at low \mbh. This is as expected, because galaxies that are farther away have more light contained in the PSF, and more massive galaxies tend to have brighter centers. 

\subsubsection{The effect of host dust obscuration}
Finally, in panel (f), we account for the effect of host dust obscuration, as described in section~\ref{sec:HostDust}. The majority of the flare detections are lost --- only $\approx 30$\% of the detection rate from panel (e) is retained. The distribution of the hosts of detected flares shifts slightly toward lower redshifts, because overall flares have become fainter. The effect on the \mbh\ distribution is more subtle, but it remains weighted toward $10^7$ \msun.

\subsection{Comparing expected host $M_\ast$ and $z$ distributions of flare hosts to ZTF detections}

\begin{figure}[htb!]
\gridline{\fig{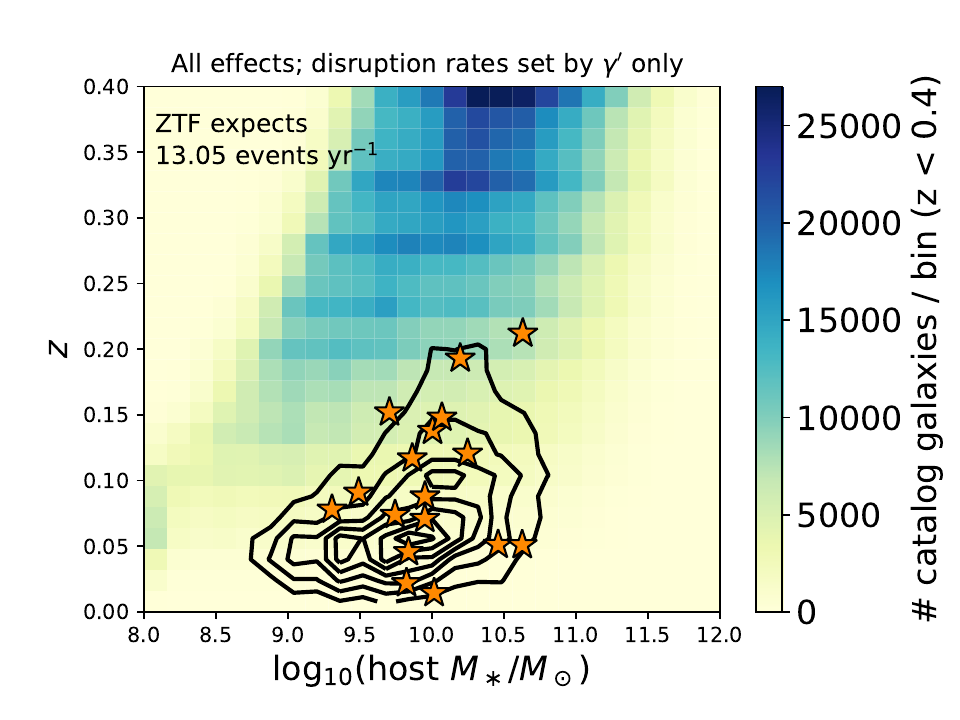}{0.4 \textwidth}{(a)}}
\gridline{\fig{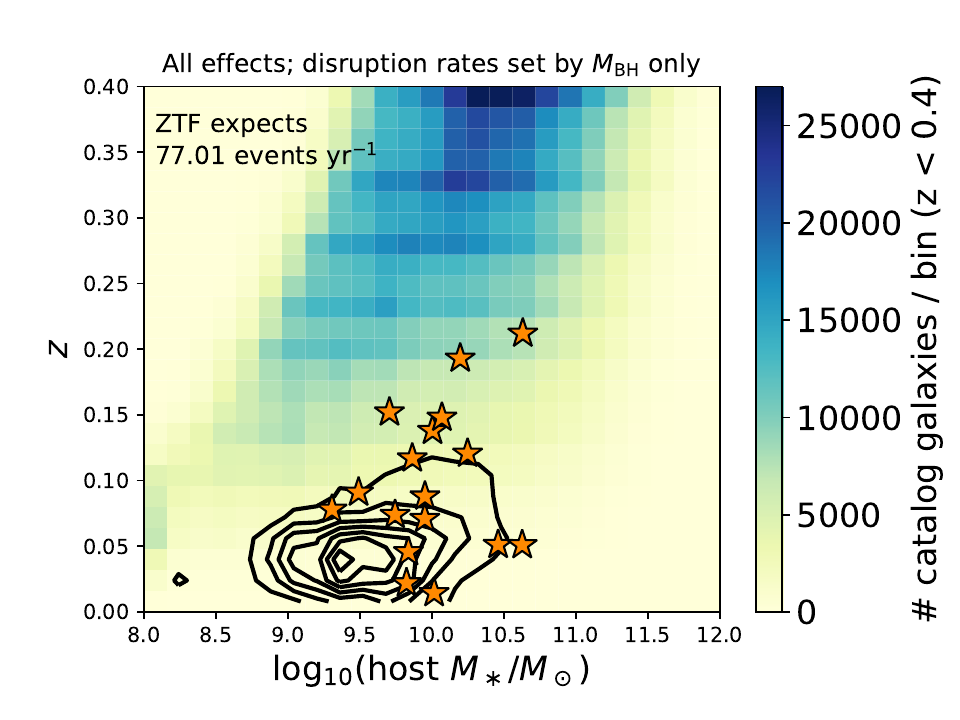}{0.4\textwidth}{(b)}}
\gridline{\fig{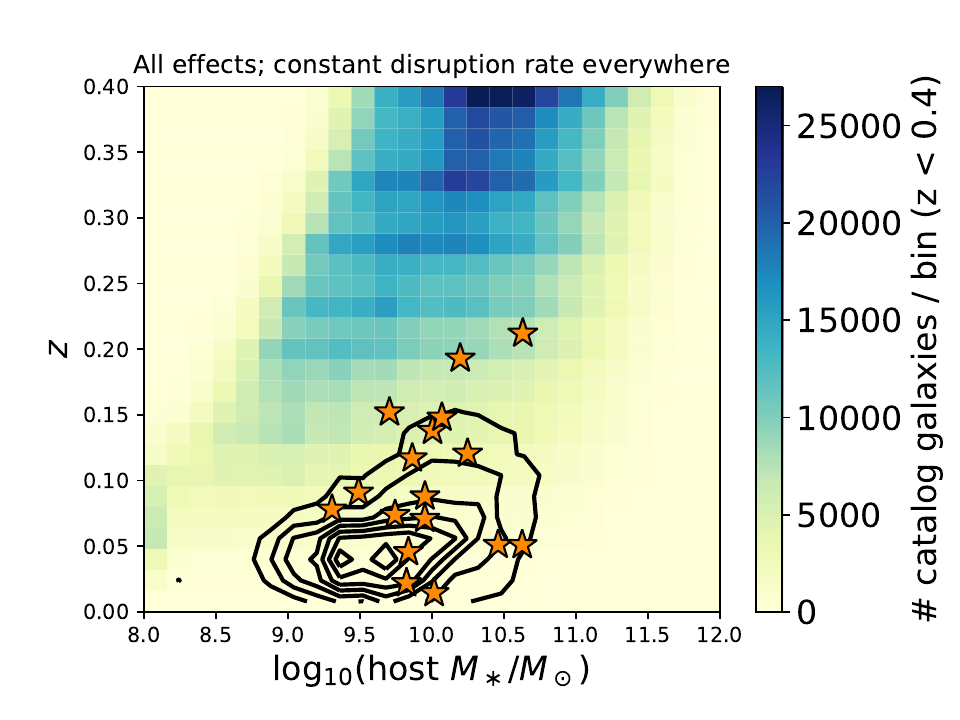}{0.4\textwidth}{(c)}}
\caption{Contours represent expected survey detection rates, which are binned based on host galaxy total \mstar\ and host redshift $z$. The colors of the bins correspond to the number of mock galaxies in those bins. The orange stars represent the ZTF TDF detections presented in \citet{van-Velzen2020}. The top panel corresponds to the fiducial model survey.}
\label{fig:MstarZ3Panel}
\end{figure}

The histograms discussed in the previous section included an axis for \mbh\ of the host. However, in practice it is much easier to obtain an estimate of the host total \mstar\ than \mbh. Therefore, in Figure~ \ref{fig:MstarZ3Panel} we compare the expected distributions of host properties as binned by total host \mstar\ and host $z$, for three separate TDE rate prescriptions used in the model: one using $\gamma^\prime$ (equation~\ref{eq:GammaPrimeRate}), shown in panel (a); one using \mbh\ (equation~\ref{eq:MbhRate}) shown in panel (b), and one where the TDE rate is the same in all galaxies (c). Panel (a) corresponds to the same as the fiducial model represented by panel (f) in Figure~\ref{fig:SixPanel}. Except for the TDE rate prescription, these models are the same and include all of the details that went into constructing panel (f) in Figure~\ref{fig:SixPanel}. For these panels we have restricted the range to galaxies with $z < 0.4$ since nearly all flare detections in the model come from galaxies in that range. If the constant rate were set equal to the rate for $\mbh = 10^6 \msun$ from equation~\eqref{eq:MbhRate}, then the ZTF expectation for this case would be $\approx$ 83 events per year. The orange stars represent the ZTF TDF detections presented in \citet{van-Velzen2020}.

\begin{figure*}[htb!]
    \plottwo{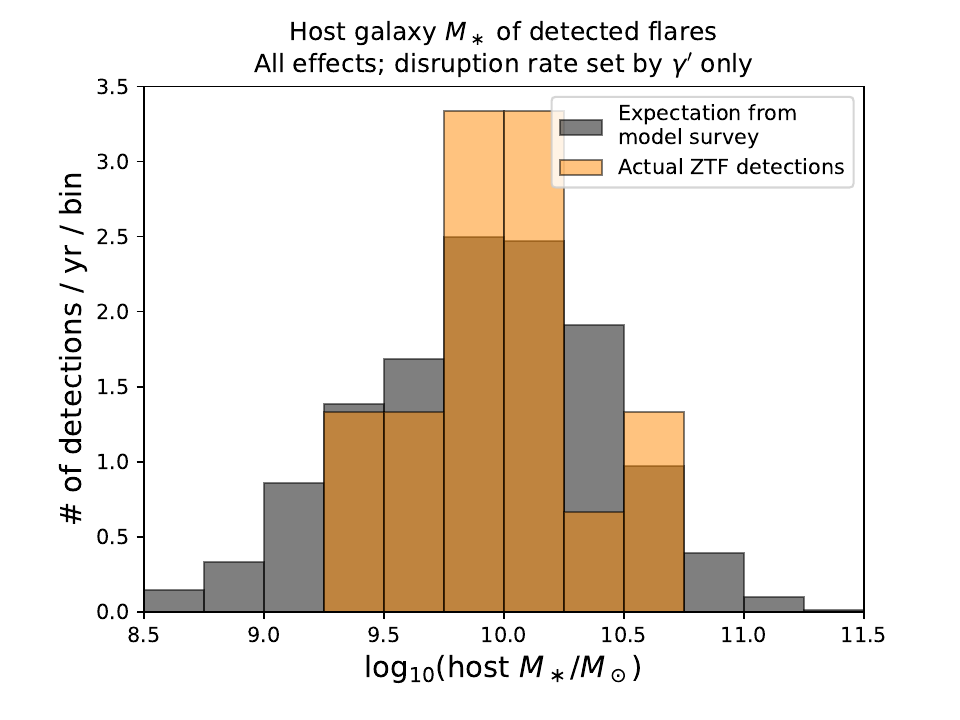}{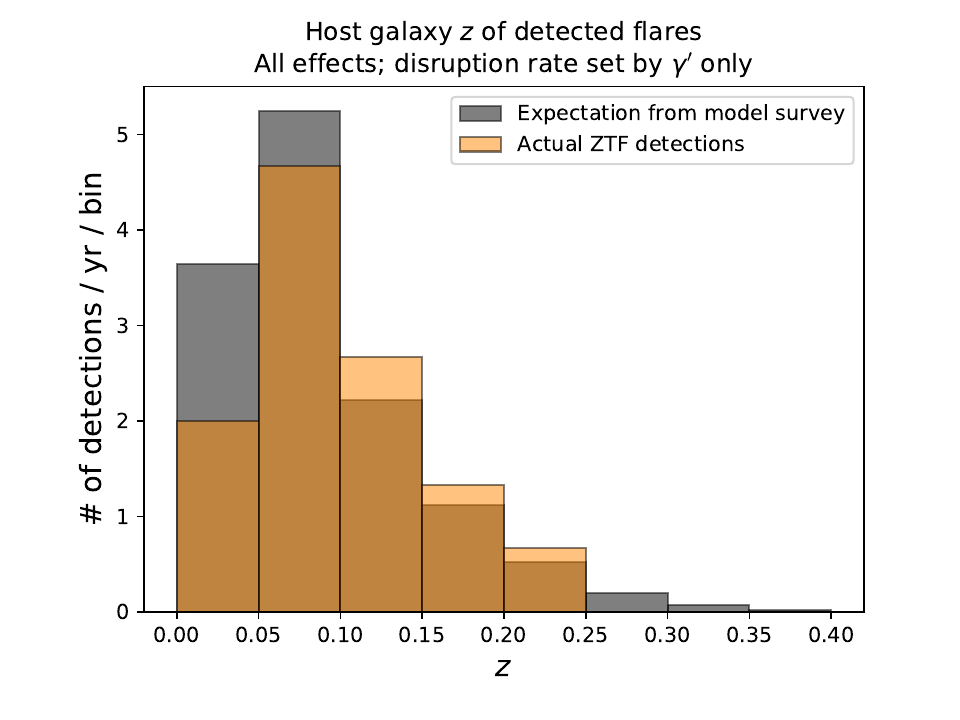}
    \caption{Distributions of host properties expected from the model survey, in gray, compared to the properties of hosts of TDFs discovered by ZTF, in orange. The first panel shows the distribution for host \mstar, while the second panel shows the distribution for host redshift $z$. The binning in these plots is arbitrary. K-S tests, which do not rely on the binning, suggest that these distributions are consistent: the $p$-values for rejecting this hypothesis are 0.36 and 0.50 for the host $M_\ast$ and $z$ distributions, respectively.}
    \label{fig:MstarZ_ZTF_Comp}
\end{figure*}

In Figure~\ref{fig:MstarZ_ZTF_Comp} we project along each axis for panel (a) of Figure \ref{fig:MstarZ3Panel} to compare the distributions for host \mstar\ and $z$ separately. To compute the ZTF detection rate, we are using the fact that the 17 flare detections published in \citet{van-Velzen2020} were made over a span of approximately 1.5 years of surveying time. The distribution of host $z$ from the fiducial model survey has a similar shape to the real ZTF detections, although the expectation is slightly weighted toward more nearby events. When binned by host \mstar, the peak of the model distribution lines up very well with the real distribution, although the model expects a slightly wider spread in host stellar masses than the real detections indicate.

To better quantify how closely the model distributions match the observed distributions, we performed Kolmogorov-Smirnov (K-S) tests. These tests are independent of the binning used to generate the histograms in Figure~\ref{fig:MstarZ_ZTF_Comp}, which was chosen arbitrarily. For the fiducial model, the detected flare distribution is constructed from more than 130,000 randomly generated (weighted) flares which passed the survey criteria. Therefore, we treat the empirical cumulative distribution function for these flares as a continuous distribution, and then perform the K-S test for the ZTF sample compared to this continuous model distribution. All K-S tests performed in this paper were done in this manner. We find that the model distributions for the host $M_\ast$ and $z$ distributions are consistent with the hypothesis that the ZTF flares are drawn from the same distribution; the $p$-values for rejecting this hypothesis are 0.36 and 0.50 for the $M_\ast$ and $z$ distributions, respectively.

To summarize, for the fiducial model survey in which galaxy disruption rates are set entirely by $\gamma^\prime$ as extrapolated from the Sersic profile fits in the catalog, we are successful in matching the overall flare detection rate and the distributions of the total $M_\star$ and $z$ of the host galaxies of the flares identified by ZTF. While this is encouraging, it might also be due, in part, to cancellation of errors in a number of the assumptions that have gone into the model, and will need to be updated and re-tested as these assumptions are improved. This is especially true in light of the fact that there are a number of observables that are not well-reproduced by the model, as will be discussed in later sections. Nevertheless, we can confidently say that the Hills mass is playing an important role in preventing the host total $M_\star$ distribution from being weighted toward higher-mass galaxies.

\subsection{Changing how faint $L_{\rm bb}$ may be, and implications for constraining the true flare LF}
\label{sec:ChangingLmin}

\begin{figure}[htb!]
    \includegraphics[width=0.5\textwidth]{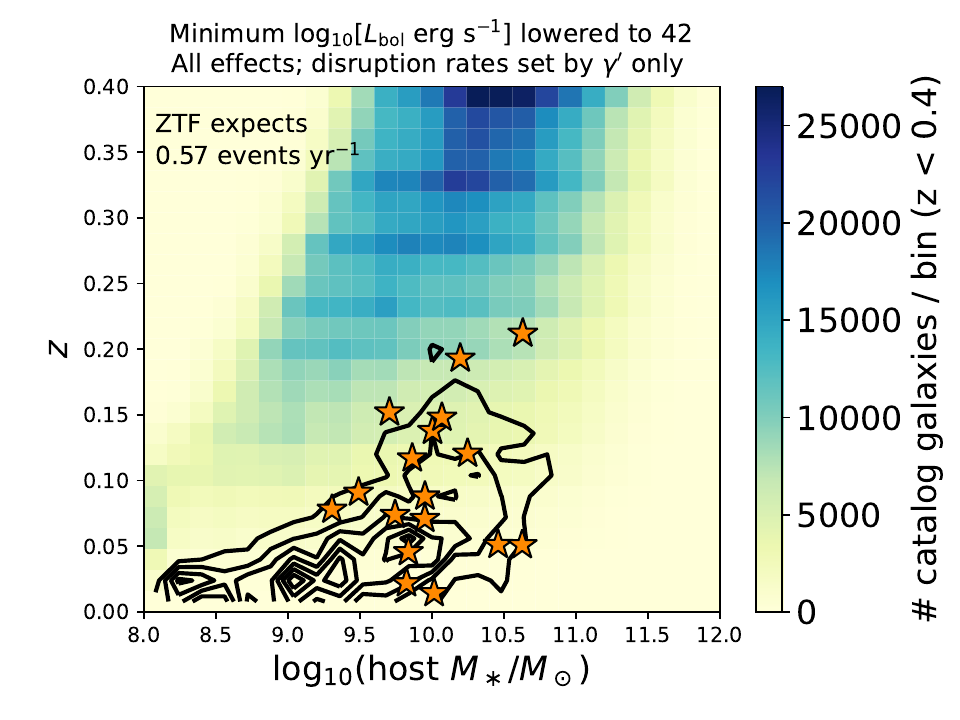}
    \caption{Like the top panel of Figure~\ref{fig:MstarZ3Panel}, but now $L_{\rm bb, min} = 10^{42}$ \ergspersec\ instead of $10^{43}$ \ergspersec. By allowing $L_{\rm bb}$ to be fainter than before, the model survey expectations for the detected rates, represented by the contours, do not agree as well with the ZTF detections as they did in Figure~\ref{fig:MstarZ3Panel}}.
    \label{fig:MinLogL42}
\end{figure}

We can consider what happens if we allow $\psi(L_{\rm bb})$ to remain nonzero down to a luminosity that is a factor of 10 lower than we had before. That is, we set $L_{\rm bb, min}$ to $10^{42}$ \ergspersec\ instead of $10^{43}$ \ergspersec. The 2D histogram of host \mstar\ versus host $z$ for this adjusted model compared to the ZTF detections is shown in Figure~\ref{fig:MinLogL42}. A striking effect of this change is that the expected number of detections in our model survey drops by a factor of about 24, as a larger number of flares become too faint to be detected. This model also expects the distribution of host galaxies of detected flares shifts to lower redshift, and a tail of low-mass galaxies contribute to the observed flares. The shift to lower redshift is a consequence of the flares becoming fainter overall. The tail of low mass galaxies appears in part because now, for a larger number of them than before, the Eddington limit corresponding to their black hole exceeds the allowed value for the peak luminosity. 

This model also expects that about 40\% of detected flares would have $L_{\rm bb, fit }$ below $10^{43}$ \ergspersec, compared to 15\% for the fiducial model. Keep in mind that $L_{\rm bb, fit }$ of just about every TDF detected to date, including all of those detected by ZTF, exceeds $10^{43}$ \ergspersec. The faintest TDF detected by any survey to date, iPTF16fnl, had $L_{\rm bb, fit }$ of almost exactly $10^{43}$ \ergspersec.

To summarize, the number of flares detected in a survey is highly sensitive to $L_{\rm bb, min}$, and our modeling suggests that this cutoff is in fact around $10^{43}$ \ergspersec. When comparing our model survey results to ZTF for the overall detection rate, host stellar masses, host redshifts, and distribution of $L_{\rm bb,fit}$, our fiducial disruption model matches the data far better when $L_{\rm bb, min} = 10^{43}$ \ergspersec than when $L_{\rm bb, min} = 10^{42}$ \ergspersec.

\subsection{Detection mass function, luminosity function, and Eddington ratio distribution}
\subsubsection{\mbh\ distribution of detected flares}
\begin{figure}[htb!]
\gridline{\fig{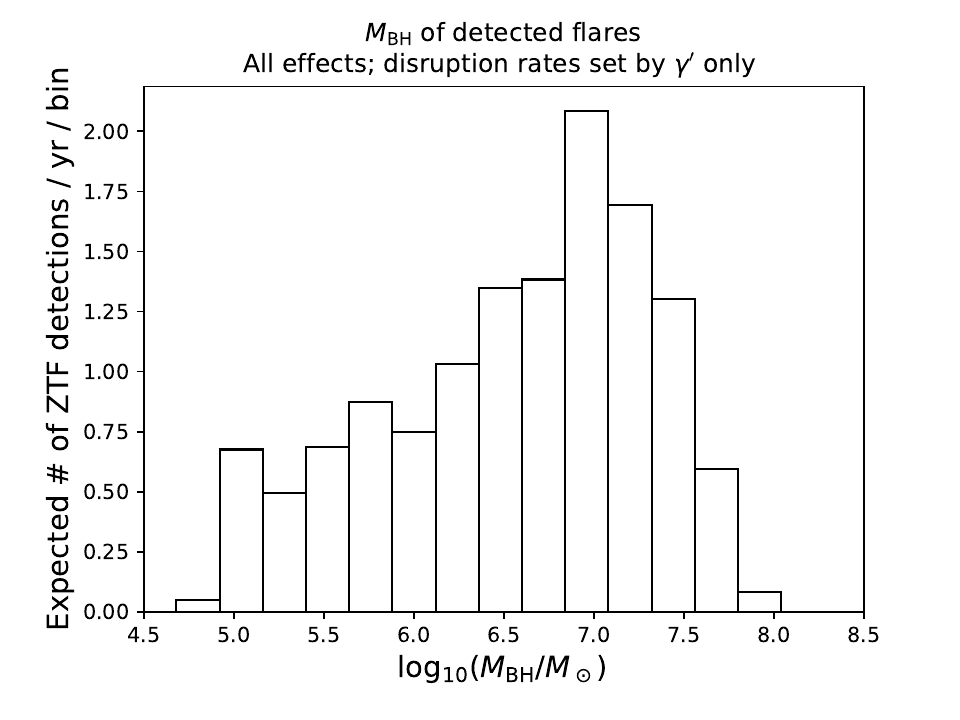}{0.4 \textwidth}{(a)}}
\gridline{\fig{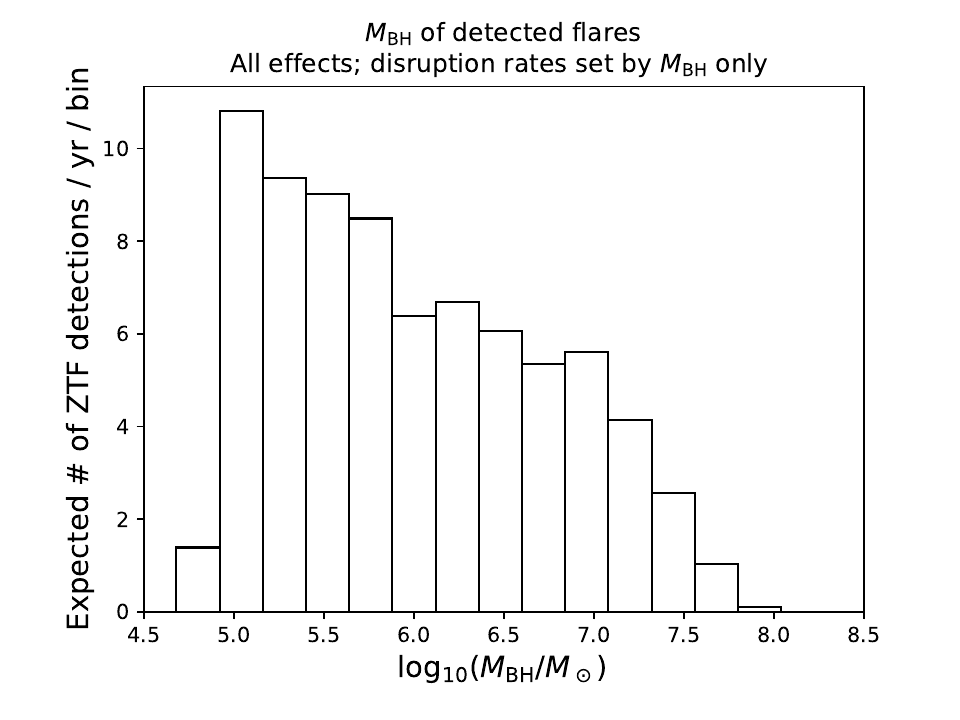}{0.4\textwidth}{(b)}}
\gridline{\fig{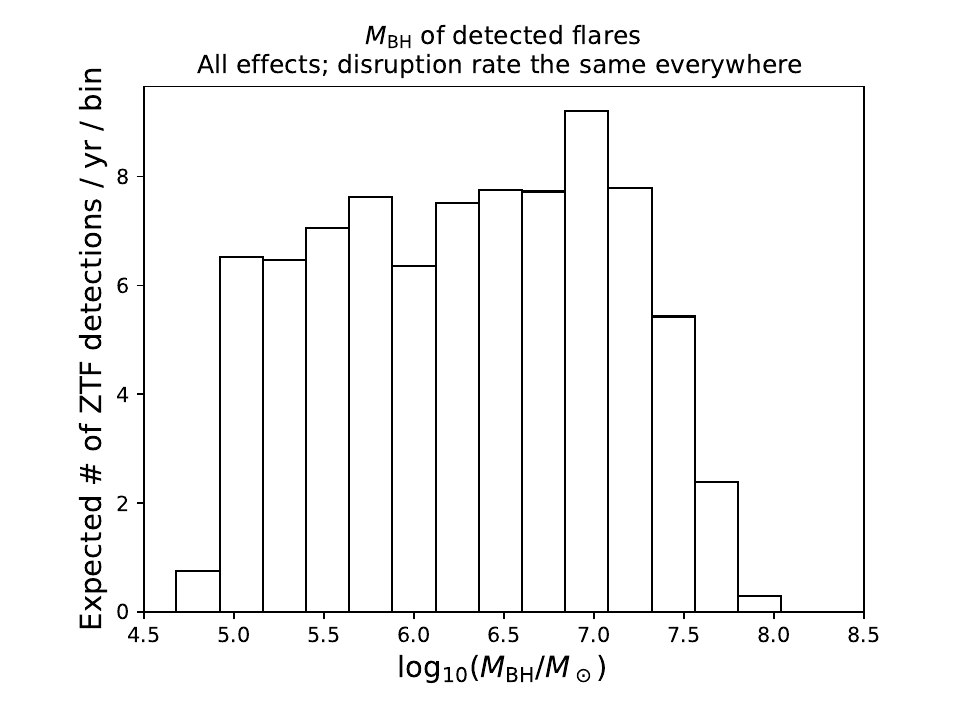}{0.4\textwidth}{(c)}}
\caption{The distributions of \mbh\ corresponding to detected flares, for model surveys where the TDE rate is set in three separate ways. The top-most panel corresponds to the fiducial model.}
\label{fig:Mbh1D3Panel}
\end{figure}

In Figure~\ref{fig:Mbh1D3Panel} we show the distribution of \mbh\ (the catalog value) for detected flares in various model survey scenarios. Panel (a) corresponds to our fiducial model where disruption rates are set by $\gamma^\prime$ and where the overall detection rate matches ZTF very well ($\sim$ 13 detections per year), the peak of the distribution is at around $10^7 \, \msun$. This is high compared to samples of TDFs with \mbh\ measured via the \mbh--$\sigma$ relation \citep{Wevers2017, Wevers2019-1}, where the distribution peaks closer to $10^6\, \msun$. The weighting toward higher \mbh\ in the model is due to a combination of factors. There is the role of the Eddington limit \citep[cf.][]{Kochanek2016-2}, and the role of dust in hiding flares from nearby star forming galaxies, which tend to harbor lower mass SMBHs on average (see Figure~\ref{fig:SmbhMassFuncVolDisrupt} and the arguments in Section~\ref{sec:preliminary} that generated it). But that alone does not explain the peak near $10^7 \msun$, as we can see by comparing to panel (c) where the disruption rate is taken to be the same in every galaxy --- in that case the expected distribution of \mbh\ in detected flares is relatively flat between $10^5$ and $10^7$ \msun. So, for the fiducial case, we are also seeing an effect where, for \mbh\ that is not large enough to exceed the Hills mass for most stars being disrupted, the galaxies with the highest $\gamma^\prime$ tend to be clustered around $\mbh \sim 10^7 \msun$ (refer again to Figure~\ref{fig:GammaprimeBins}, and panel (c) in Figure~\ref{fig:SixPanel}). Again we emphasize that this statement might depend on the the method we have described for computing $\gamma^\prime$ in this catalog.

Panel (b), corresponding to the model in which TDE rates are set by \mbh, expects a distribution weighted even more toward lower-mass SMBHs than real detections seem to suggest, along with generally over-predicting the detection rate. Also recall from Figure~\ref{fig:MstarZ3Panel} earlier that this model expects the detections to be dominated from very low-redshift galaxies compared to real detections.

When considering the observed \mbh\ distribution in the case of constant per-galaxy rate as in panel (c), it is somewhat of a coincidence that the distribution is so flat near peak --- There is pressure toward higher mass because of the Eddington limit and dust, but there is also pressure toward lower mass because of the larger number of lower mass SMBHs nearby, as seen in Figure~\ref{fig:SmbhMassFuncVolDisrupt}.

Note that in none of these panels, not even the middle panel, do we have a rate expectation that rises as steeply toward lower \mbh\ as in the volumetric disruption rate shown in the right panel of Figure~\ref{fig:SmbhMassFuncVolDisrupt}. Once again, the cutoff at $\log_{10}(\mbh / \msun) = 4.9$ is from the combination of our requirements that $L_{\rm bb} > L_{\rm bb, min}$ and $L_{\rm bb} < L_{\rm Edd}$. 

To summarize, the fiducial model survey expects a distribution of \mbh\ for the detected flares that peaks around $10^7$ \msun. These masses seem systematically higher than the \mbh\ distribution inferred from previously detected TDFs, although we do not have \mbh\ measurements for the ZTF black holes that make use of the \mbh--$\sigma$ relation. The model favors higher-mass black holes partly because of the role of the Eddington limit in our construction of $\psi(L_{\rm bb})$, but also because of correlations between $\gamma^\prime$ and the amount of dust obscuration with \mbh\ in the host galaxies in the catalog. As expected, if the model instead assumes that the disruption rate has an inverse dependence on \mbh\, this shifts the \mbh\ distribution for observed flares toward lower masses.
\subsubsection{Luminosity distributions of detected flares}
\label{sec:ResultsLuminosity}
There are several aspects of flare luminosities in the model that we wish to compare to the data. We will separately consider the expected distributions of $L_{g,\rm fit}$ and $L_{\rm bb,fit}$ of detected flares produced from the models. Recall that $L_{g,\rm fit}$ is of special importance to us because we designed $\psi(L_{\rm bb})$ so that the model would match the empirical rest-frame $g$-band LF from \citet{van-Velzen2018-1}. So, for $L_{g, \rm fit}$, we will also examine the inferred volumetric rate distribution, which differs from the distribution of detected flares. To produce the volumetric rate, we apply an inverse-volume weighting to the flare detection rates from the model, just as is done to contstruct the volumetric LF for the real survey detections. Once again, for the details of how $L_{g,\rm fit}$ is computed in the model, please consult Appendix~\ref{sec:Tfit}. 

In this section we are considering the \emph{fitted} luminosities, as opposed to the raw values of $L_{\rm bb}$ and $L_{g}$ generated for each flare, because only the fitted values account for any dust obscuration that might be arise from the host galaxies. In section~\ref{sec:ResultDustEffects} we will discuss how the fitted luminosities compare to the true values of the unobscured flares that are generated in the model. 

\begin{figure}[htb!]
\includegraphics[width=0.5\textwidth]{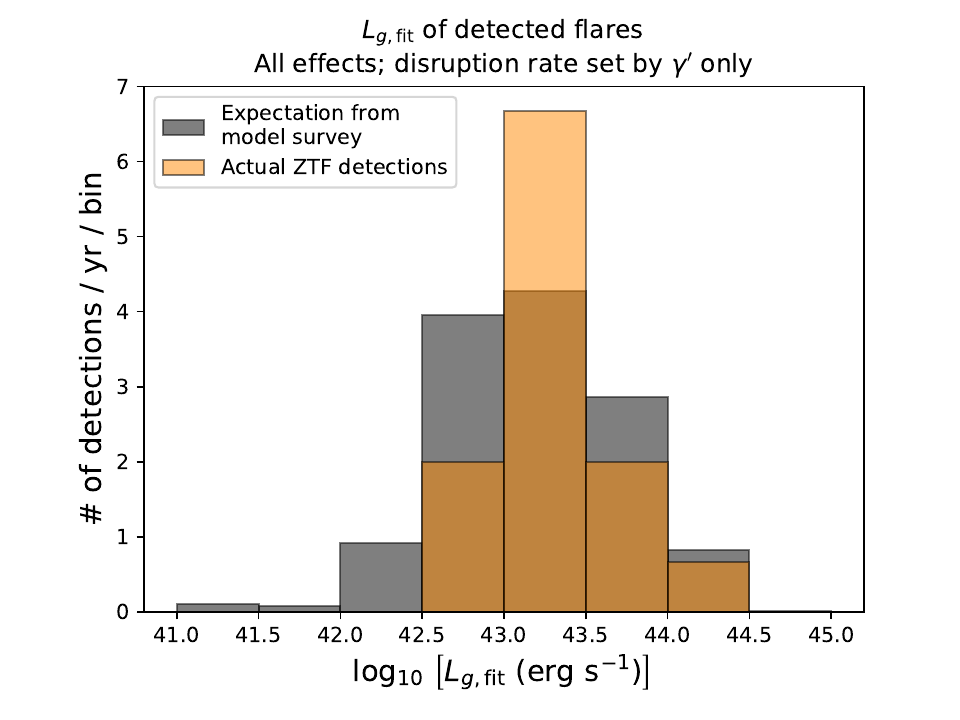}
\caption{Distribution of $L_{g, \rm fit}$ expected from the fiducial model survey, in gray, compared to the distribution of $L_{\rm g, fit}$ of TDFs discovered by ZTF, in orange. The two distributions peak at a similar value, but the model distribution is wider than the ZTF distribution. A K-S test indicates that the two distributions are marginally inconsistent with each other: the $p$-value for rejecting the hypothesis that these distributions are the same is 0.047.   }
\label{fig:Lg_ZTF_comp}
\end{figure}

Figure~\ref{fig:Lg_ZTF_comp} compares the fiducial model's expected distribution of $L_{g,\rm fit}$ to the ZTF sample. The two distributions peak at similar values, but the model distribution is wider than the inferred from the data. A K-S test leads to a p-value of 0.047 for rejecting the hypothesis that the underlying distributions are the same. This suggests that there is more work to be done in designing the joint probability distribution for $L_{\rm bb}$ and $T_{\rm bb}$ in the model so that the detected flare distribution it generates may be consistent with the data.

\begin{figure}[htb!]
\includegraphics[width=0.5\textwidth]{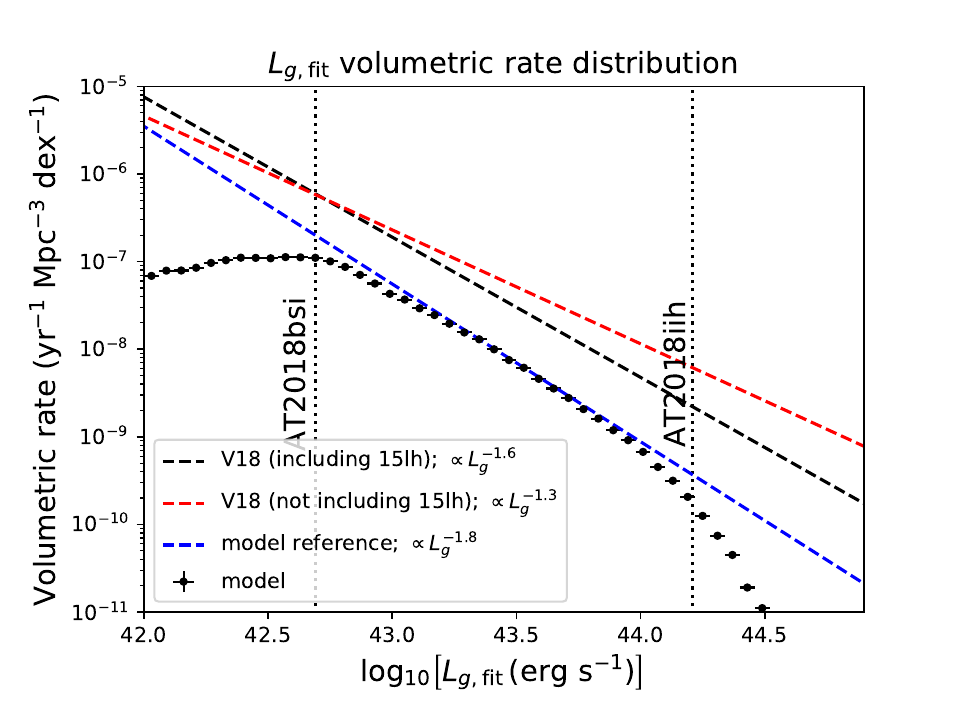}
\caption{The volumetric, rest-frame $g$-band LF for the fiducial model is plotted with black markers. The dashed vertical lines represent the minimum and maximum $L_{g,\rm fit}$ values in the ZTF sample. A reference power-law $\propto L_g^{-1.8}$ is drawn in blue, and this roughly approximates the model distribution for the relevant range of luminosities. The dashed black and red lines are the LFs from \citet{van-Velzen2018-1} when ASASSN-15lh was and was not included in the sample, respectively.} 
\label{fig:LgVol}
\end{figure}

The volumetric $L_{g,\rm fit}$ rate distribution for the fiducial model is compared to the empirical LF in Figure~\ref{fig:LgVol}. Over the range of $L_{g,\rm fit}$ values of the 17 ZTF flares published in \citet{van-Velzen2020}, the volumetric LF for the model roughly follows $dN/d\log(L_g) \propto L_g^{-1.8}$. This is slightly steeper than the dependencies measured in \citet{van-Velzen2018-1}, which ranged from roughly $L_g^{-1.6}$ if ASASSN-15lh was included, and $L_g^{-1.3}$ if that flare was not included in the fit. The figure also indicates that, over the range of luminosities in the ZTF sample, the overall normalization of the volumetric $g$-band LF in the fiducial model is about a factor of 4 lower than the LF function from \citet{van-Velzen2018-1} that includes ASASSN-15lh. However, it should be kept in mind that the expected number of flares per year in the fiducial model survey ($\approx$ 13 flares per year) is quite close to the flare detection rate in ZTF (17 flares in $\approx$ 1.5 years, or approximately 11 flares per year). While we do not fully understand yet how the volumetric LF inferred from the model can be lower that what has been measured previously, while at the same time the model expectation matches the ZTF detection rate, this might result from a cancellation of errors. That is, some of our assumptions about $\psi$ and the flare temperature distributions might lead to a higher detection rate than is realistic, or we are assuming a higher detection efficiency than is realistic, but at the same time we are using a lower overall volumetric TDE rate than is realistic.

The $L_g^{-1.8}$ dependence that is evident in Figure~\ref{fig:LgVol} is also steeper than the $L_{\rm bb}^{-1.5}$ dependence built into $\psi(L_{\rm bb})$. There are several possible reasons for this. The $g$-band LF in the model depends on many details, including the manner in which $T_{\rm bb}$ is sampled, the minimum and maximum luminosities that govern where $\psi(L_{\rm bb})$ is nonzero, how those limiting luminosities depend on black hole mass, how dust shapes the observed spectrum, and how black hole mass correlates with the disruption rate and dust content of each galaxy. Of all of these possible explanations, we suspect that our handling of $T_{\rm bb}$ is most likely responsible for the steeper $L_g$ dependence, although we have not verified this yet. These details also are crucial for understanding where the model LF departs from a simple power-law at the lowest and highest values of $L_g$ in Figure~\ref{fig:LgVol}.

\begin{figure}[htb!]
\gridline{\fig{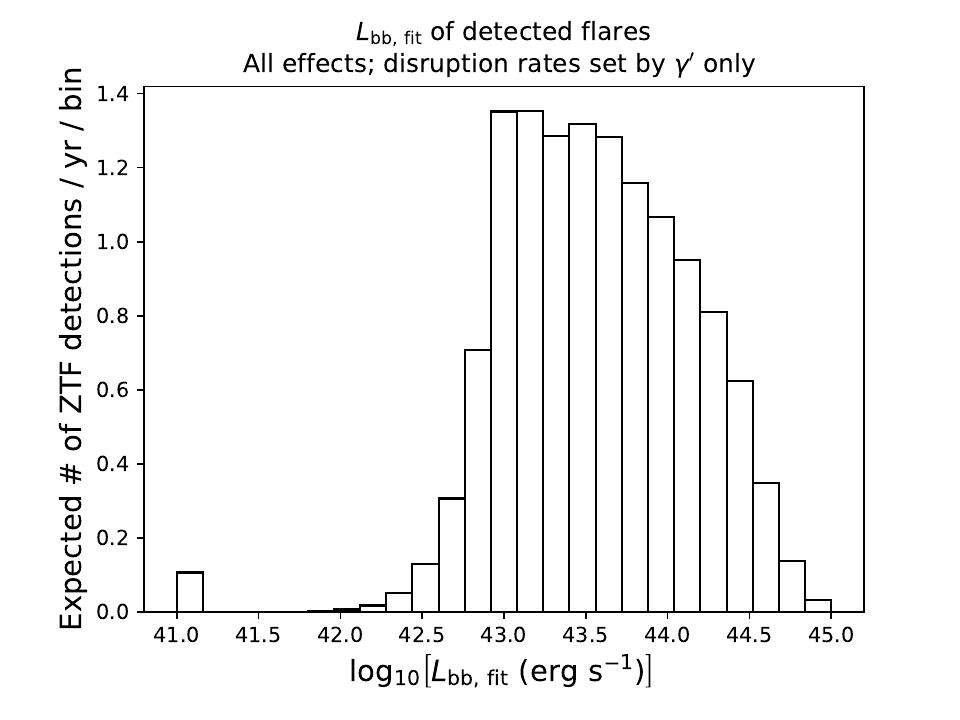}{0.4 \textwidth}{(a)}}
\gridline{\fig{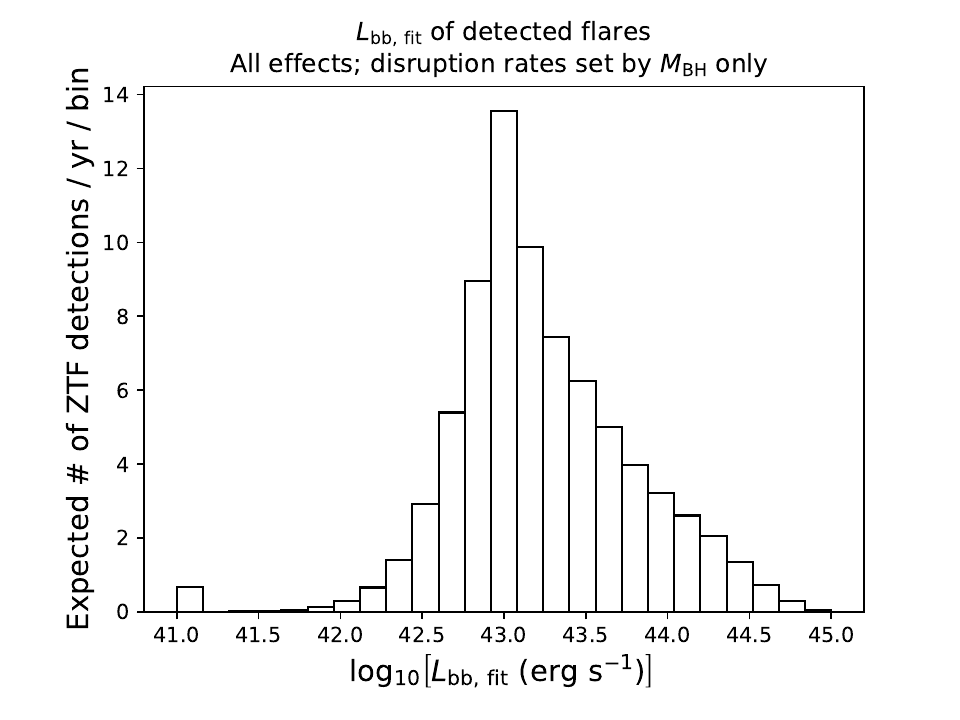}{0.4\textwidth}{(b)}}
\gridline{\fig{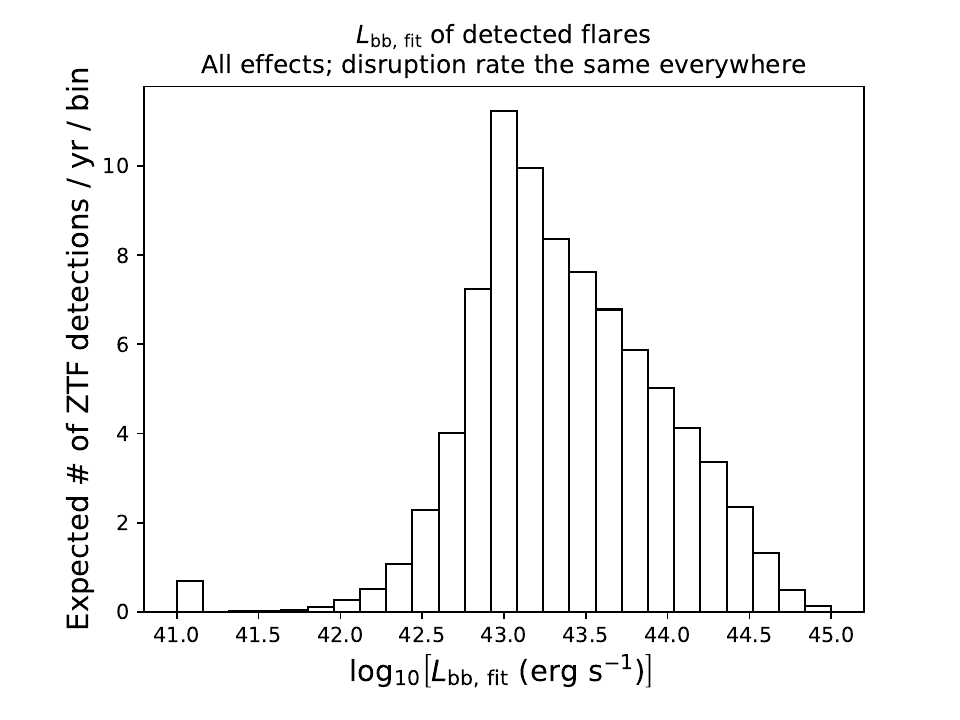}{0.4\textwidth}{(c)}}
\caption{The distributions of $L_{\rm bb, fit}$ corresponding to detected flares, for model surveys where the TDE rate is set in three separate ways. The top-most panel corresponds to the fiducial model.}
\label{fig:LoptFit1D3Panel}
\end{figure}

Now we move on to the distribution of $L_{\rm bb, fit}$, shown in Figure~\ref{fig:LoptFit1D3Panel}. For the fiducial case where disruption rates are set by $\gamma^\prime$, the distribution of $L_{\rm bb, fit}$ is relatively flat between $10^{43}$ and $10^{44}$ \ergspersec. Even though the model is only generating flares with $L_{\rm bb, fit} > 10^{43}$ \ergspersec, the distribution of $L_{\rm bb, fit}$ does extend a bit fainter than $10^{43}$ \ergspersec, mostly because of the effects of dust, which tend to make flares appear dimmer than their unobscured values. Still, the fraction of flares from the model with fitted luminosities below $10^{43}$ \ergspersec\ is small. The fact that the distribution is not prominently peaked at $10^{43}$ \ergspersec, where $\psi(L_{\rm bb})$ is largest, means that a large fraction of these flares are coming from galaxies sufficiently distant that only flares generated at sufficiently high luminosity are visible. 

For the other two cases, the distribution peaks at $10^{43}$ \ergspersec. This is because these flares are largely coming from a population of more nearby galaxies, where a larger fraction of the faint events sampled from $\psi(L_{\rm bb})$ is visible. These distributions also have a larger number of flares in the tail that extends fainter than $10^{43}$ \ergspersec. This reflects the fact that a larger fraction of these flares are coming from dustier galaxies, which lowers the value of $L_{\rm bb, fit}$.

\begin{figure}[htb!]
    \includegraphics[width=0.5\textwidth]{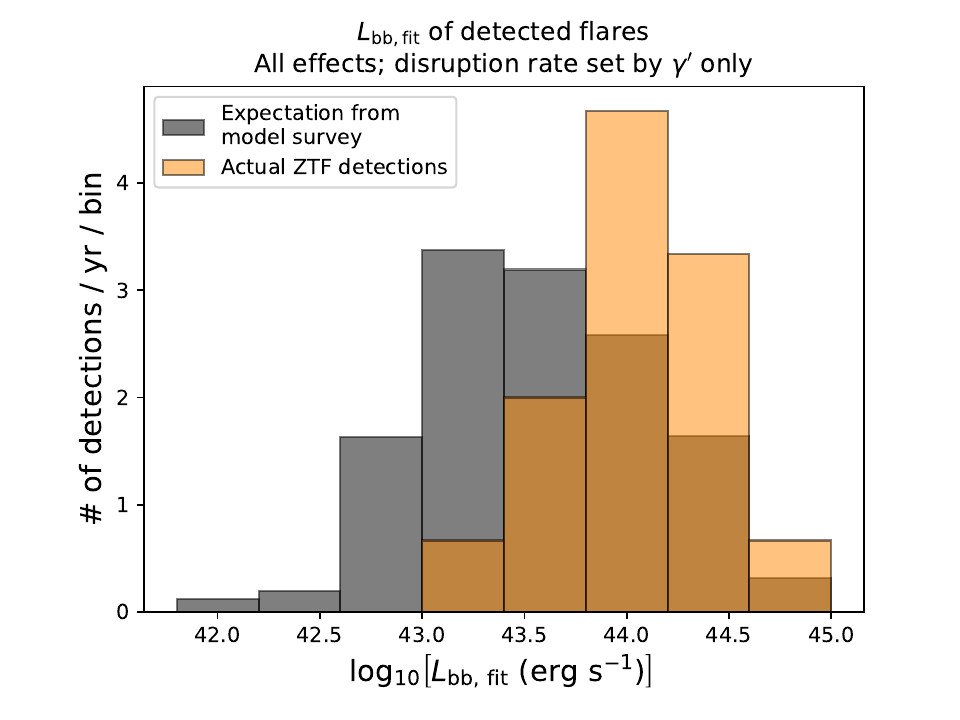}
    \caption{Comparison of expected distribution of $L_{\rm bb,fit}$ from the fiducial model survey with the distribution from ZTF TDF detections. A K-S test confirms that these two distributions are not consistent ($p = 0.0004$).} 
    \label{fig:LoptFit_ZTF_Comp}
\end{figure}

In Figure~\ref{fig:LoptFit_ZTF_Comp} we compare the distribution of $L_{\rm bb, fit}$ for the fiducial model survey with the real ZTF detections (the outlier flares at $L_{\rm bb, fit} \le 10^{42}$ \ergspersec\ from the model survey are not included in the lowest luminosity bin in this figure). Here we find a clear disagreement between the model and the data. The model expects the flare luminosity distribution to be weighted toward fainter values than the data indicate. This suggests that there is still room to improve our understanding of the intrinsic flare luminosity distribution $\psi(L_{\rm bb})$. One possibility is that the true distribution does not continue to follow the $dN/d\log L \propto L^{-1.5}$ distribution to values of $L_{\rm bb}$ as low as we have assumed for $\psi(L_{\rm bb})$. However, making such a change would affect the match to the overall flare detection rate and the distribution of host redshifts, which are currently quite good. Another likely area for improvement is the distribution of $T_{\rm bb}$ and its potential correlation with $L_{\rm bb}$ --- this will be discussed in section~\ref{sec:TemperatureResults}.

To summarize, our construction of $\psi(L_{\rm bb})$ based on the empirical $L_g$ LF, along with our decision to sample $T_{\rm bb}$ uniformly between 10,000 and 50,000 Kelvin, leads the model to expect a $g$-band luminosity distribution that peaks at the same value as the ZTF distribution. However, the modeled $L_g$ distribution is wider than is seen in ZTF. Also, the modeled distribution of $L_{\rm bb, fit}$ is significantly different than the ZTF distribution, expecting lower values of $L_{\rm bb, fit}$ than are observed.

This failure to match $L_{\rm bb,fit}$ cannot be remedied simply by using a different flux limit for the survey detection criteria. For example, if we were to use a flux cutoff at apparent magnitude 18, instead of 19 as we assumed in the fiducial model, the primary effect is to shift the population of flares to more nearby galaxies, without significantly changing the expected distribution of $L_{\rm bb,fit}$ and $T_{\rm bb,fit}$.

\subsubsection{Combined distribution of \mbh\ and $L_{\rm bb, fit}$ of detected flares}

\begin{figure}[htb!]
    \includegraphics[width=0.5\textwidth]{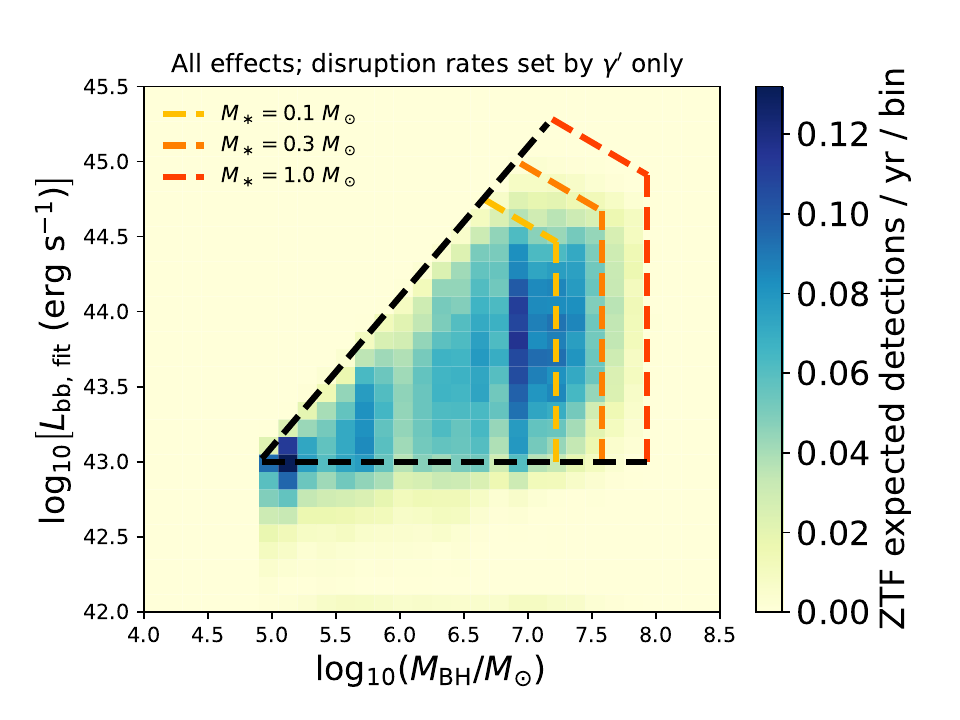}
    \caption{Combined distribution of \mbh\ and $L_{\rm bb, fit}$ of detected flares. The black dashed lines indicate restrictions on the unobscured flare luminosity based on the black hole mass and our assumptions about the flare LF. The colored dashed lines represent restrictions based on the Hills mass and the maximum rate of mass fallback after disruption, both of which depend on the mass of the disrupted star.} 
    \label{fig:LFitMbh2D}
\end{figure}

In Figure~\ref{fig:LFitMbh2D} we look at the 2D histogram for $L_{\rm bb, fit}$ and \mbh. In addition to presenting the results of the previous two subsections in a unified manner, this plot also illustrates how a number of our assumptions relating \mbh, the mass $M_\ast$ of the disrupted star, and $L_{\rm bb}$, are used. The diagonal dashed black line represents the Eddington limit. The horizontal dashed black line represents our choice for $L_{\rm bb, min}$. Some values of $L_{\rm bb, fit}$ fall below this value, especially at lower \mbh, because of our treatment of how host dust obscuration alters the fitted luminosity. Again we point out that the value of \mbh\ at which these two black dashed curves intersect effectively defines a lower limit on \mbh\ for detected flares.

The three colored dashed lines relate to the role that stellar mass plays in setting allowable flare luminosities and values for \mbh. The vertical portion of these curves corresponds to the Hills mass for the given stellar mass (see equation~\ref{eq:Hills}). The diagonal portion of these curves represents the luminosity limit for the maximum mass fallback rate at this stellar mass and \mbh, assuming $\beta = 1$ and $0.1$ radiative efficiency.

When flares are randomly generated they are sampled from $\psi(L_{\rm bb})$, with the appropriate upper and lower limiting values that depends on \mbh\ and $M_\ast$, as depicted in this plot (see also section~\ref{sec:FlareLF} and Appendix~\ref{sec:ImplementLF}).

\subsubsection{Distribution of Eddington ratios of detected flares}

\begin{figure}[htb!]
\gridline{\fig{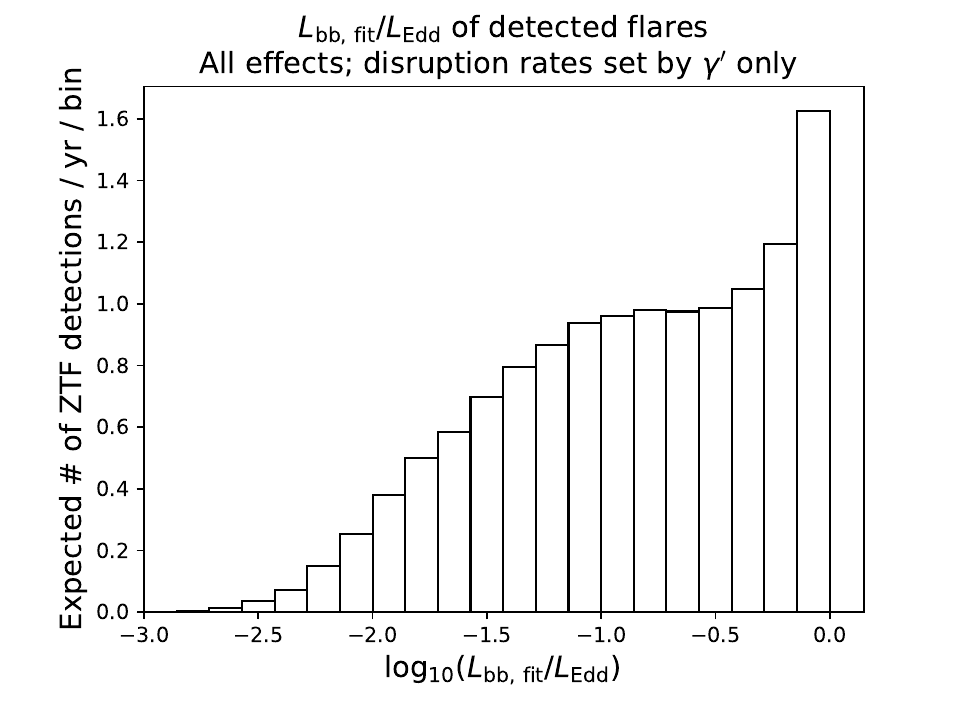}{0.4 \textwidth}{(a)}}
\gridline{\fig{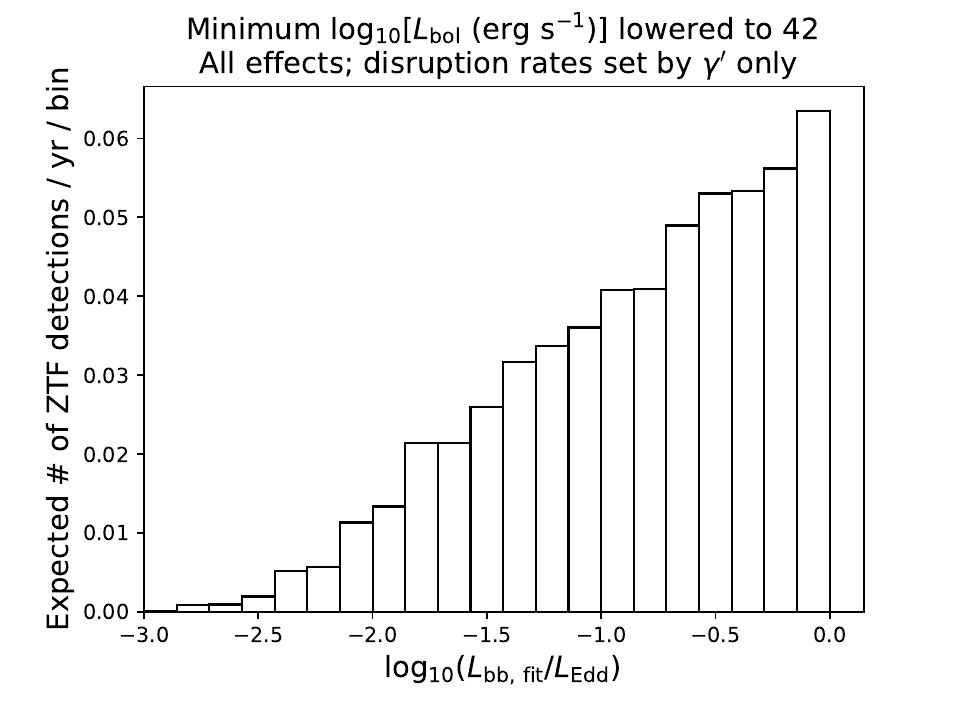}{0.4\textwidth}{(b)}}
\gridline{\fig{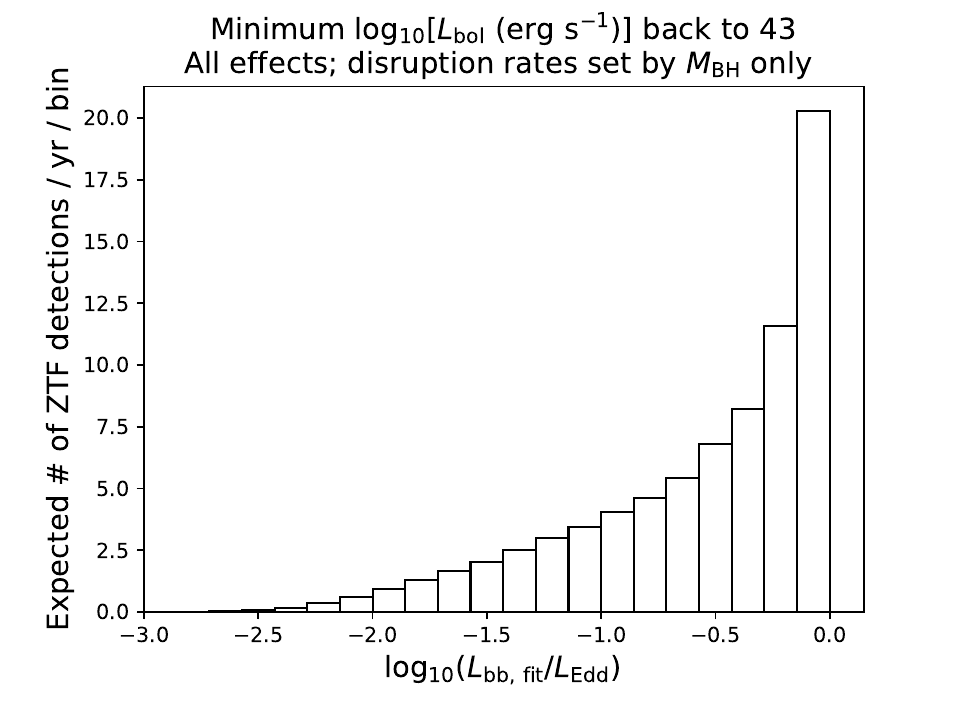}{0.4\textwidth}{(c)}}
\caption{The distributions of $L_{\rm bb, fit} / L_{\rm Edd}$ corresponding to detected flares, for three different model scenarios. The top-most case corresponds to the fiducial model.}
\label{fig:EddRatio3Panel}
\end{figure}

Figure~\ref{fig:EddRatio3Panel} shows distributions of Eddington ratios for detected flares in several model surveys. For the fiducial case with rates set by $\gamma^\prime$, and using fitted luminosities, the model expects a large fraction of detected flares to be at the Eddington limit. The fact that the Eddington limit is playing such an important role becomes important when trying to understand anything related to the \mbh\ distribution for detected flares. Another important feature of this distribution is that it is mostly flat between $\log_{10}(L_{\rm bb, fit}/L_{\rm edd})$ of $-0.5$ to $-1.5$. This is partly a property of the $\psi(L_{\rm bb})$ power-law, $dN/d\log(L) \propto L^{-n^\prime}$ for $n^\prime = 1.5$, but as we will discuss shortly it also depends on our choice of $L_{\rm bb, min}$. Decreasing $n^\prime$ (making $\psi(L_{\rm bb})$ less weighted toward fainter events) without adjusting $L_{\rm bb, min}$ would cause this histogram to increase more steadily toward higher Eddington ratios. Increasing $n^\prime$ (making $\psi(L_{\rm bb})$ weighted even more toward fainter events) would lead to a peak in the distribution at a low Eddington ratio. 

Meanwhile, when we lower $L_{\rm bb, min}$ to $10^{42}$ \ergspersec , keeping $n^\prime =1.5$, the distribution steadily increases for larger values of $\log_{10}(L_{\rm bb, fit}/L_{\rm edd})$, without any flattening at intermediate values. Keep in mind that the overall number of cases has dropped by about a factor of 24, so the change in distribution shape is attributable to losing flares, evidently with the largest loss at intermediate luminosity values. 

Finally we consider the case where the disruption rate is set by \mbh\ rather than $\gamma^\prime$, but move $L_{\rm bb, min}$ back to $10^{43}$ \ergspersec. This causes the Eddington ratio to be even more highly weighted toward high Eddington ratios. This can be largely understood as a consequence of the disruption rate being higher for lower values of \mbh. 

For the above cases, using the true unobscured value of $L_{\rm bb}$ rather than the fitted value to the obscured flare does not qualitatively change the shape of the distributions. This fact might change if the fraction of detected flares with large amounts of dust obscuration becomes even larger.

A key takeaway from this section is that the observed Eddington ratio distribution is sensitive to not only the details of the flare probability distribution, but also the way that \mbh\ is related to the disruption rate. It also seems to be generally true that, even when $\psi(L_{\rm bb})$ is highly weighted toward fainter events, the observed Eddington ratio distribution will tend to be weighted toward higher Eddington ratios. Therefore caution must be exercised when attempting to back out the population-averaged accretion conditions (specifically, how the typical accretion rate or generation of radiation via shock-heating compares to the critical value set by the Eddington limit and a typical radiative efficiency) compares to the observed distribution of Eddington ratios. 

\subsection{Further exploring the role of dust in setting properties of host galaxy detections}

\begin{figure}[htb!]
\gridline{\fig{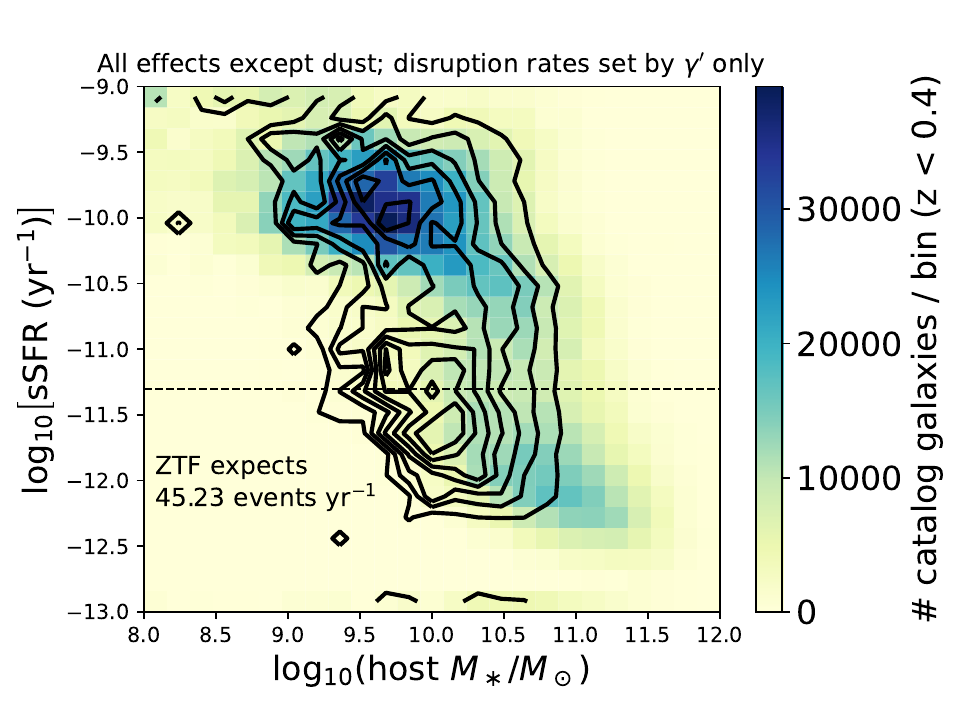}{0.5 \textwidth}{(a)}}
\gridline{\fig{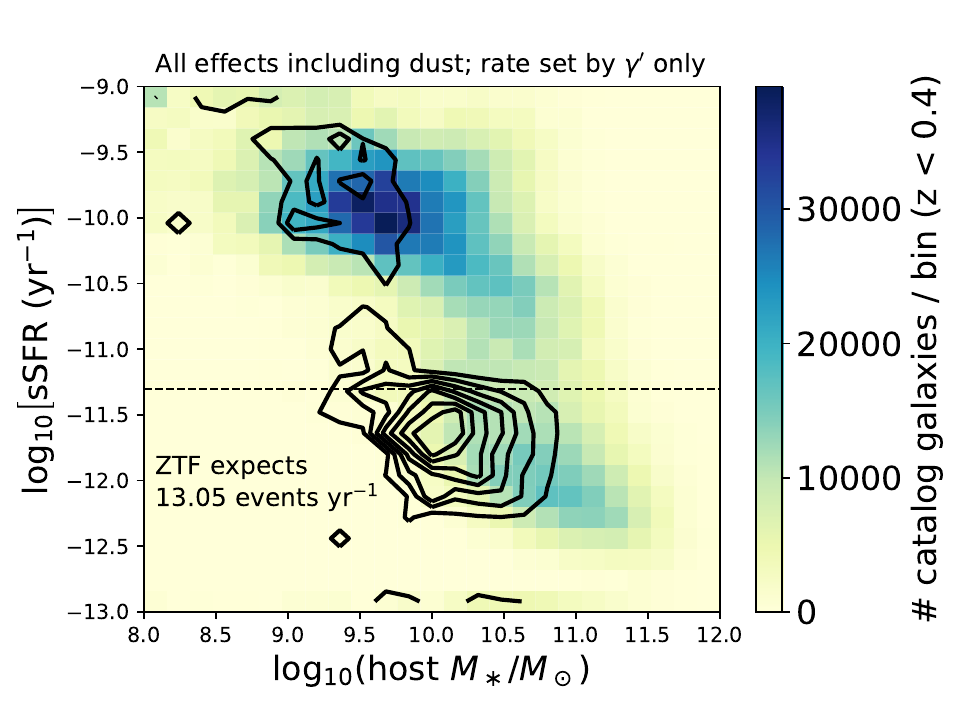}{0.5\textwidth}{(b)}}
\caption{Two-dimensional histogram of distribution of host galaxy total \mstar\ and sSFR for detected flares. The top panel corresponds to a model survey that is identical to the fiducial survey in every respect except that it does not account for obscuration by host dust. The bottom panel corresponds to the fiducial model survey, which does account for dust obscuration.  The horizontal dashed line represents our cut at $\log_{10}[{\rm sSFR \,\, (yr}^{-1}{\rm )}] = -11.3 $, which we use to separate star-forming galaxies with more dust from quiescent galaxies with less dust. Galaxies with properties that fall outside of the ranges of the plotted bin values are included in the edge bins.}
\label{fig:MstarsSFR_Fiducial}
\end{figure}

In Figure~\ref{fig:MstarsSFR_Fiducial} we explore the distributions of hosts of detected flares binned based on sSFR and total host \mstar, both for the fiducial model survey, and for a model survey which does not account for dust obscruation but that is otherwise identical. Including dust does not have much of an effect on the quiescent galaxies, but wipes out most of the flares from star-forming galaxies, as expected. The peak of the distribution of flares detected in quiescent host galaxies is offset from the peak of the number distribution of those galaxies in the catalog. More specifically, the flares come from a lower-stellar mass population, which can largely be understood as a consequence of the Hills mass suppressing visible disruptions in galaxies that have higher stellar mass, because those galaxies tend to have higher \mbh. 

\begin{figure}[htb!]
\gridline{\fig{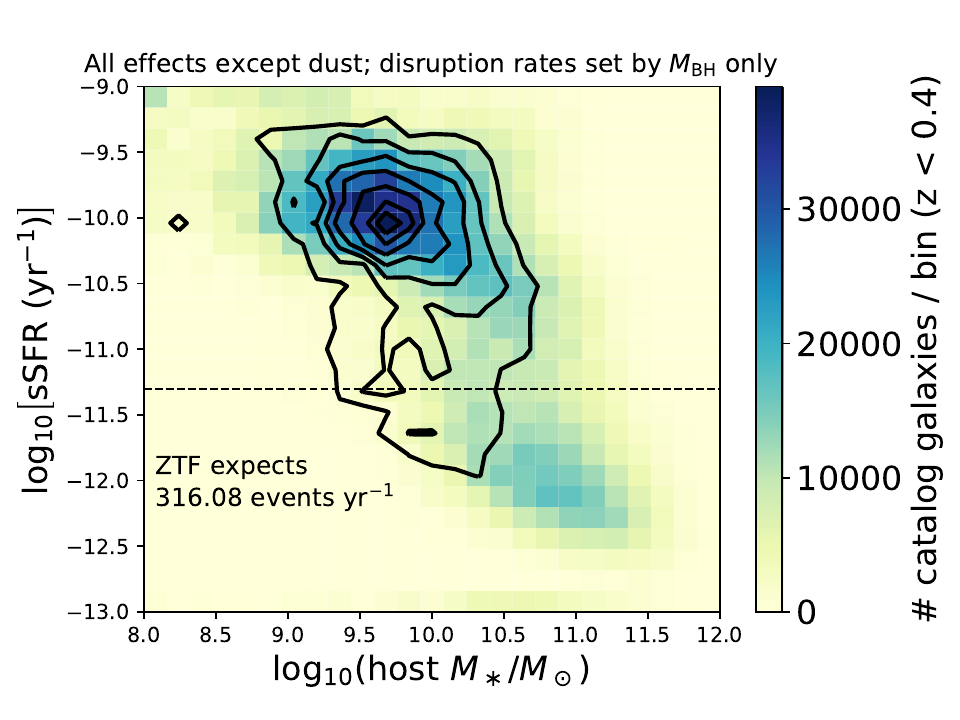}{0.5 \textwidth}{(a)}}
\gridline{\fig{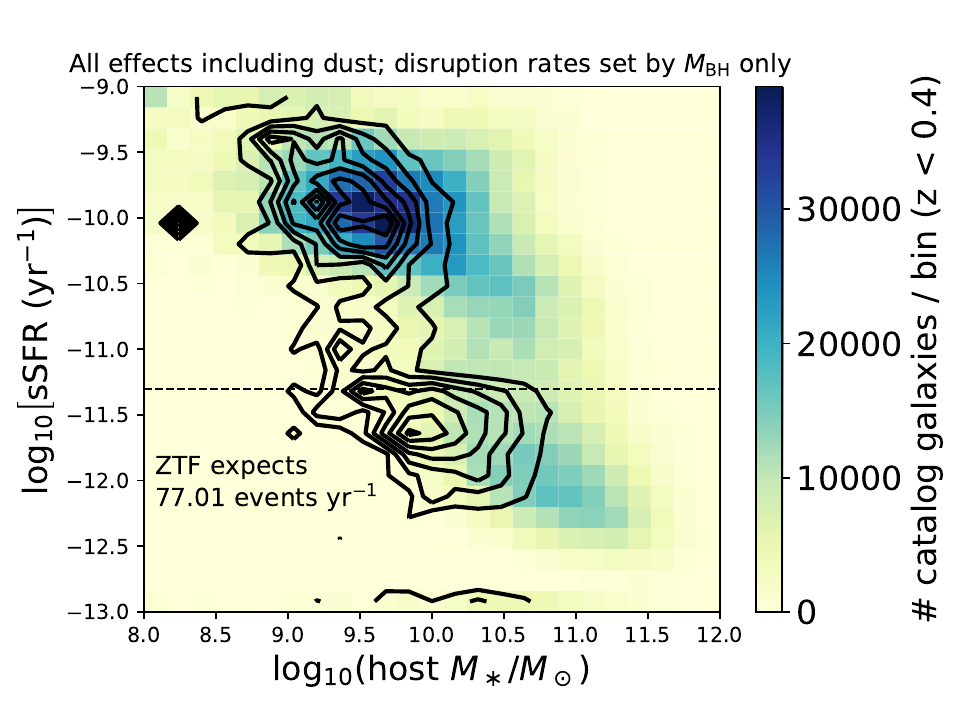}{0.5\textwidth}{(b)}}
\caption{The same as Figure~\ref{fig:MstarsSFR_Fiducial}, but for model surveys in which the TDE rate is set by \mbh. Without any effects of dust, and if the disruption rate increases with decreasing \mbh as in equation~\ref{eq:MbhRate}, then a survey like ZTF should have seen the vast majority of its flares in star-forming galaxies (top panel). Even when accounting for dust, using the disruption rate from equation~\ref{eq:MbhRate} expects that over half of flares will be detected in star-forming galaxies (bottom panel).}
\label{fig:MstarsSFR_MbhRate}
\end{figure}

In Figure~\ref{fig:MstarsSFR_MbhRate} we show the same plots, but for model surveys in which the TDE rate is set by \mbh. In this case, without dust, detected flares come overwhelmingly from star-forming galaxies. This is mostly because those galaxies tend to have lower \mbh, and here the rate is higher in galaxies with lower \mbh, similar to the preliminary calculation in section~\ref{sec:preliminary}. The shape of the detected flare distribution in this case largely follows the distribution of the number of galaxies in these bins at $z < 0.4$ in the catalog, although the peak of the model detected flare distribution is offset very slightly to higher host \mstar\ and lower sSFR when compared to the catalog distribution. When obscuration by host dust is accounted for, most of the flares in the star-forming galaxies are lost, allowing the flare contribution from the quiescent galaxies to contribute an identifiable peak. However, as before, the peak of the distribution of flares detected in quiescent host galaxies is offset from the peak of the number distribution of those galaxies in the catalog.

\begin{figure}[htb!]
\gridline{\fig{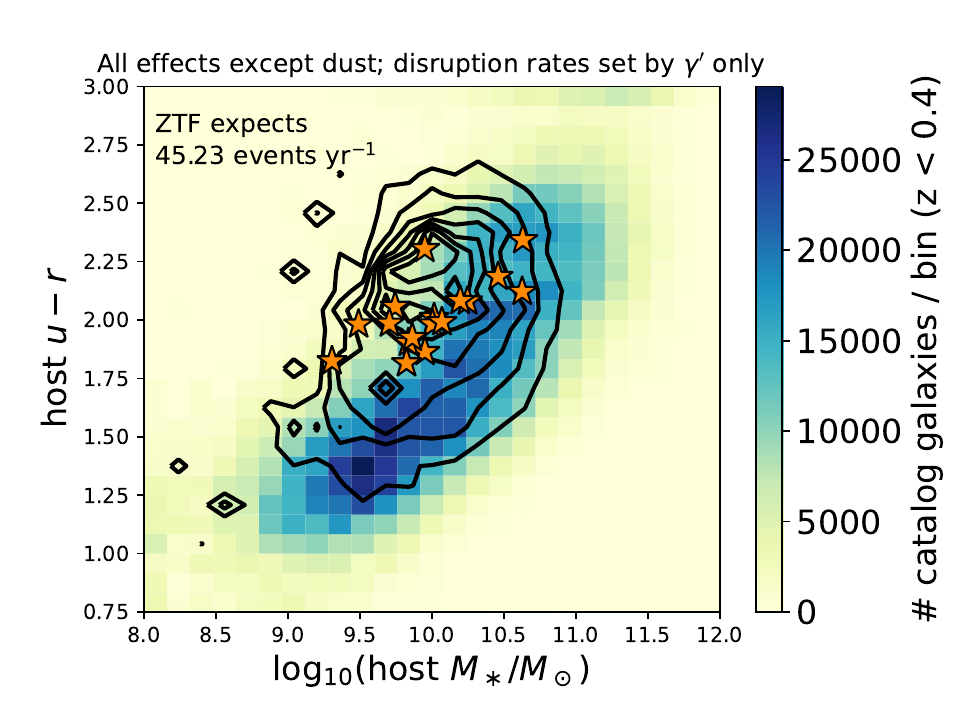}{0.5 \textwidth}{(a)}}
\gridline{\fig{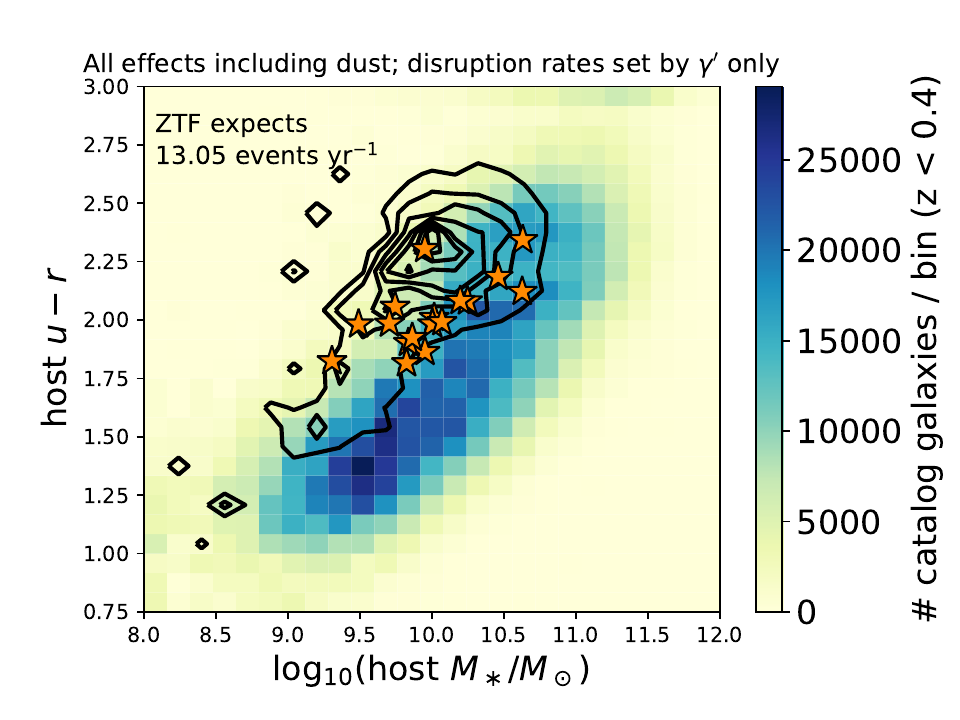}{0.5\textwidth}{(b)}}
\caption{Two-dimensional histogram of distribution of host galaxy total \mstar\ and $u - r$ for detected flares. The top panel corresponds to a model survey that is identical to the fiducial survey in every respect except that it does not account for obscuration by host dust.  The bottom panel corresponds to the fiducial model survey, which does account for dust obscuration. The orange stars represent ZTF TDF detections. Galaxies with properties that fall outside of the ranges of the plotted bin values are included in the edge bins.}
\label{fig:MstarsUminusR_Fiducial}
\end{figure}

In Figure~\ref{fig:MstarsUminusR_Fiducial} we look at detections in the space of host $u-r$ versus host \mstar\ for the fiducial model survey, because it is in this space where the green valley can be identified. A prominent result from the ZTF sample \citep[and hinted at by earlier samples from various surveys, as noted by][]{Law-Smith2017} is that TDF hosts seem to be found unusually often in green valley galaxies. For the flux-limited galaxy sample we are using, the bi-modality in this space is not as immediately discernible. The situation is also complicated because the galaxy distribution in this space evolves with redshift. Nevertheless, it is still clear that the ZTF detections fall in the intermediate range of $u -r$ in this plot. 

For the fiducial model, where rates are set by $\gamma^\prime$, we have already seen that there is good agreement with the expected distribution of host stellar masses (Figure~\ref{fig:MstarZ_ZTF_Comp}. The modeled $u - r$ is distribution is shifted to redder values than the observed distribution, by about 0.25 mags (panel b). It is not entirely clear why this difference exists compared to the real detections. It might be related to the assumptions that go into how the model treats dust obscuration. Or, it might be because the true disruption rate and/or flare LF is further enhanced in the green valley galaxies in a way that is not included in the model. Even when dust is not included, the peak of the distribution is too red in host $u - r$ space compared to the ZTF detections, although it extends to cover much more of the bluer galaxies. The distribution with dust included has smaller dispersion along both of these axes, better matching the dispersion of the ZTF detections than the model without dust, which results in a higher dispersion in host $u - r$.

\begin{figure}[htb!]
\gridline{\fig{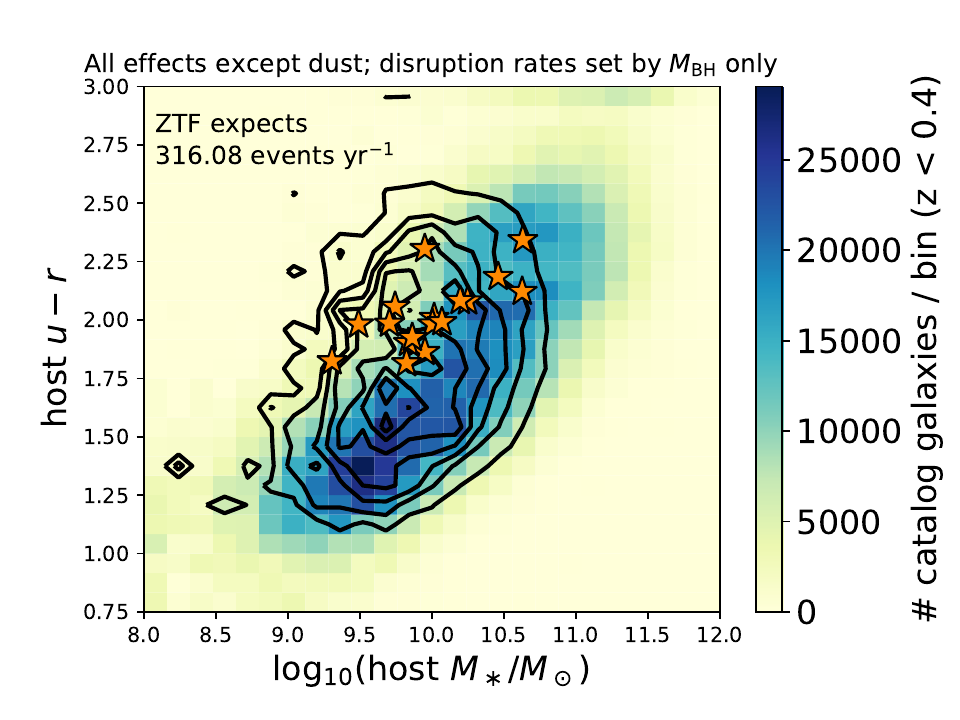}{0.5\textwidth}{(a)}}
\gridline{\fig{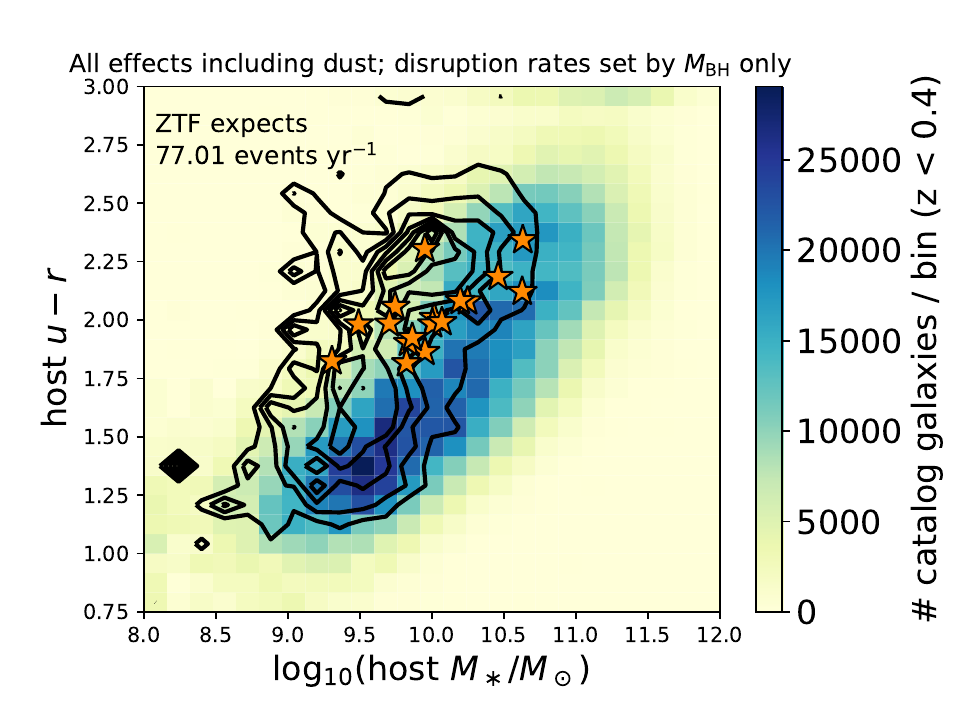}{0.5\textwidth}{(b)}}
\caption{Like figure~\ref{fig:MstarsUminusR_Fiducial}, but for model surveys where the TDE rate is set by \mbh. Host dust obscuration is accounted for in the bottom panel, but not in the top panel.}
\label{fig:MstarsUminusR_MbhRate}
\end{figure}

In Figure~\ref{fig:MstarsUminusR_MbhRate} we look at the same plots for \mbh-dependent rates. The model distribution of hosts of detected flares is more multi-modal. It favors bluer galaxies (lower $u - r$) more than the case when the rates were set by $\gamma^\prime$. The inclusion of dust cuts out flares with higher host stellar mass, and with bluer colors. The final distribution with dust is still multi-modal, close to bi-modal, and this is the same bi-modality that we saw when the vertical axis was sSFR. Both of the models with and without dust have large dispersion in $u - r$, larger than the real ZTF sample.

To summarize, when the effect of host dust obscuration is not included in a model survey that is otherwise identical to the fiducial model, roughly equal numbers of flares are expected to be detected in star-forming galaxies as in quiescent galaxies. In the fiducial model survey, which does account for host dust obscuration, most of the formerly detectable flares from TDEs occurring in star-forming galaxies are rendered undetectable. In the space of total host $M_\ast$ and host $u-r$ for detected flares, the fiducial model shows broad overlap with the distribution of the ZTF flares (panel b of Figure~\ref{fig:MstarsUminusR_Fiducial}). However, the model distribution is centered at a higher value of $u-r$ compared to the observed distribution. While the fiducial survey does not fully explain the preference of observed TDFs to have hosts in the green valley, it does help to quantify the size of the TDE rate enhancement that is required for those green valley galaxies in order to match the data. The expected distribution of flares in this parameter space is also affected by the details of how we account for dust obscuration in the survey. 

\subsection{Central surface brightness distributions of TDF hosts}

\citet{French2020-0} pointed out that the host galaxies of TDFs tend to have brighter centers than a matched population of nearby (co-moving distance $< 40$ Mpc) elliptical galaxies at similar stellar masses \citep[taken from the \mbox{ATLAS$^{3{\rm D}}$} sample,][]{Cappellari2011} , but these hosts have similar central surface brightnesses (SBs) to those of post-starburst galaxies out to $z \sim 0.12$ \citep[as in the sample described in][]{Yang2008}. We can look for similar patterns in the central SBs of mock host galaxies for detected flares in our model survey. The central SBs of the mock catalog galaxies are determined based on the pure Sersic model parameter entries from the catalog, evaluated at 0.1 kpc angular diameter distance from the galaxy center, and corrected for cosmological dimming. 

\begin{figure}[htb!]
\gridline{\fig{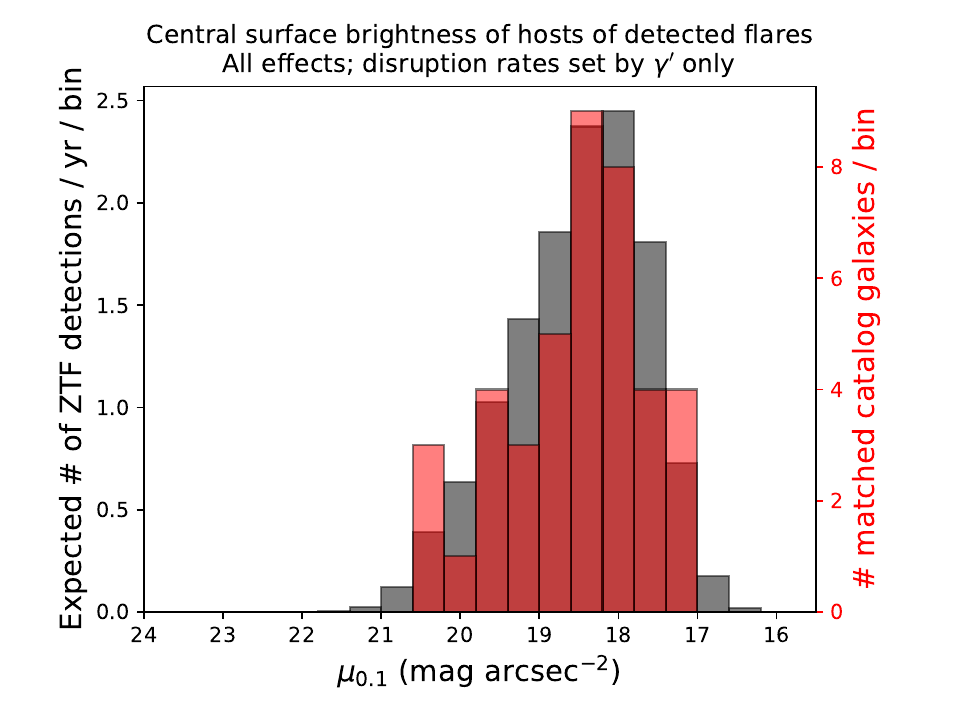}{0.5 \textwidth}{(a)}}
\gridline{\fig{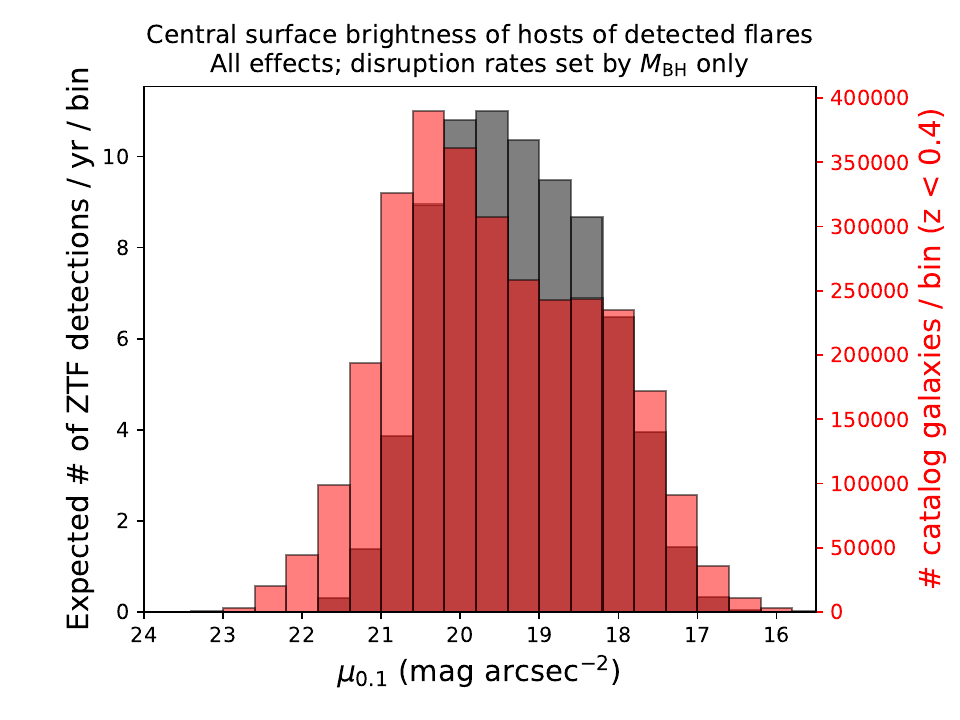}{0.5\textwidth}{(b)}}
\caption{Distributions of central surface brightnesses of TDF hosts compared to relevant subsets of the mock catalog.}
\label{fig:Mu100_catalog}
\end{figure}

The black histogram in the first panel of Figure~\ref{fig:Mu100_catalog} shows the central SBs of the hosts of detected flares for our fiducial model survey. This is compared to a subset of the mock catalog that roughly corresponds to the \mbox{ATLAS$^{3{\rm D}}$} sample of nearby elliptical galaxies at the mass range of interest to us. Specifically, the red population satisfies $9.7 < \log_{10}({\rm host} \,\,\mstar / \msun) < 10.2$, $z < 0.02$, and $u - r > 0.25 \, \log_{10}({\rm host}\,\,\mstar / \msun) - 0.4$ . This final cut on color is meant to roughly select for mock catalog galaxies that represent elliptical galaxies; it is a substitution for a cut on morphological classification, which is not in the catalog. 

The red and black distributions are similar. So, this panel is telling us is that the TDF hosts in our model survey, which span a cosmological volume out to $z \sim 0.4$, tend to resemble the galaxies in the mass-matched \mbox{ATLAS$^{3{\rm D}}$} sample of nearby elliptical galaxies. This comes about for two reasons: first, in the fiducial model survey the stellar disruption rates in galaxies are set by $\gamma^\prime$, which is highest in the catalog for red, quiescent galaxies that tend to be elliptical. Second, these galaxies are also less dusty on average, which further enhances their contribution to detected flares compared to contributions from dustier star-forming galaxies. The peak of the red distribution is at about 18 mags arcsec$^{-2}$, which is slightly fainter than the values inferred for the mass-matched \mbox{ATLAS$^{3{\rm D}}$} sample, which peaks at around 17 at $0.1$ kpc \citep{French2020-0}. This mismatch is likely due to the fact that the mock catalog uses pure Sersic fits, whereas \citet{French2020-0} use Sersic-plus-disk models. 

We must explain why the central SBs of TDF hosts in our model survey resemble the nearby ellipticals quite well, whereas real TDF hosts have even brighter central SBs that are characteristic of some post-starburst galaxies. Again, this is likely attributable to the lack of detailed surface brightness models down to $0.1$ kpc angular diameter distance resolution for the galaxies that make up the mock catalog. By design, the catalog contains galaxies out to distances much larger than the distance for which such SB measurements are commonly made for large galaxy samples, so it is difficult to accurately populate the catalog with SB profiles at the required resolution. Relatedly, the catalog only stores parameters for pure-Sersic fits. The result is that the Sersic fits in the catalog often underestimate central galaxy SBs, and so galaxies with central SBs brighter than 16 mags arcsec$^{-2}$ are under-represented (at least using the pure Sersic fit parameters), whereas such bright SBs are common in real TDF hosts. Further work is required to improve the mock catalog to allow it to clearly identify galaxies that possess the SB profiles found in real TDF hosts. 

In the second panel, the black histogram now represents the same central surface brightness distribution, but for the model survey in which disruption rates are set by \mbh\ rather than $\gamma^\prime$. We see that this distribution is now weighted more toward galaxies with fainter centers than was the case when the rates were set by $\gamma^\prime$. In fact, the TDF hosts in this model survey now have central SBs distributed in a manner very similar to the distribution of the all the galaxies in the mock catalog at $z < 0.4$ (red histogram). However, the red and black distributions do exhibit separation, with the mean of the black histogram lying at a brighter value than for the catalog galaxies. In other words, when we specify that the galaxy disruption rate be set by \mbh\ rather than $\gamma^\prime$, the resulting distribution of TDF hosts does not exhibit as strong a preference for galaxies with bright centers. However, for the case with rates set by \mbh, the host galaxies still exhibit a slight preference for galaxies with brighter centers when compared to the population of all galaxies in a flux-limited survey out to $z$ of $0.4$. This must come about because of correlations that exist between \mbh, host \mstar, and dust extinction (the galaxy quantities involved in setting the disruption rate and flare visibility) with the galaxy central SB. 

\begin{figure}[htb!]
    \includegraphics[width=0.5\textwidth]{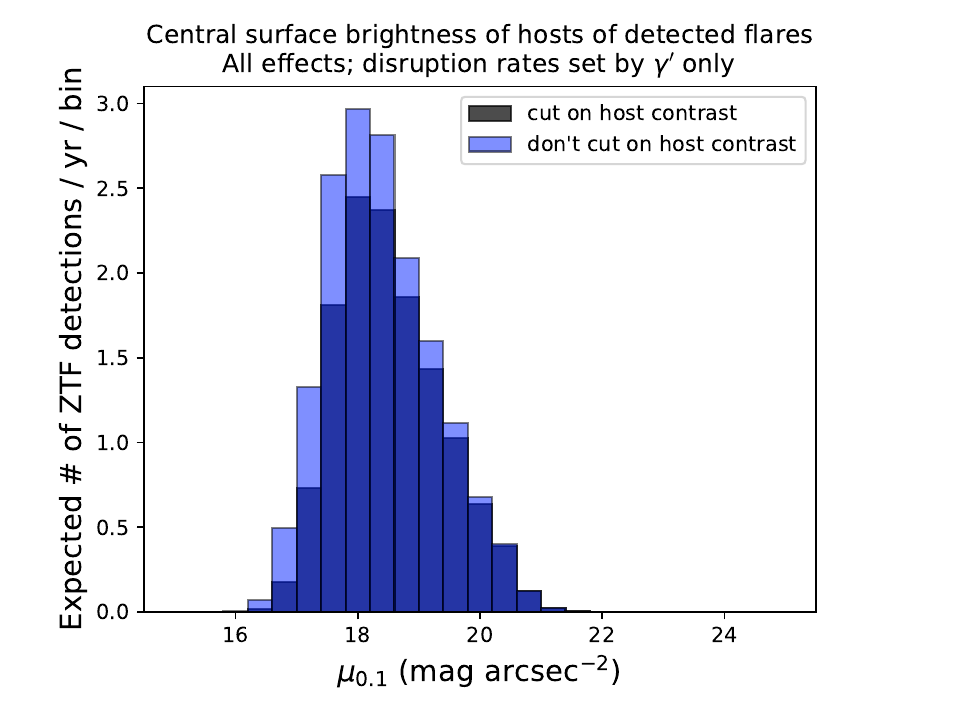}
    \caption{Distributions of host central SBs in the fiducial model survey, with and without applying the cut on contrast with host PSF light.} 
    \label{fig:Mu100_InclusionTest}
\end{figure}

Next we explore the effect of the requirement in our model survey that the peak flux from the transient be sufficiently bright compared to the PSF light from the host. In Figure~\ref{fig:Mu100_InclusionTest} we see that removing this requirement from our survey detection criteria would lead to an increased detection rate for galaxies with brighter centers, as expected. The detection rate in galaxies with fainter centers is not affected very much. The overall detection rate would increase from 13.05 events per year to 16.29 events per year, roughly a 25\% increase. As discussed above, post-starburst galaxies with especially bright centers are under-represented by the pure Sersic fit parameters in the mock catalog. This suggests that the true effect of requiring transients to sufficiently contrast with host PSF light might be even more important than the current models suggest - we may be missing an even larger fraction of flares (greater than 25\%) coming from such galaxies which are not able to clearly outshine their host’s central light.

To summarize, the  TDF  hosts  in the fiducial  model  tend to resemble nearby elliptical galaxies in terms of their surface brightness profiles. For this aspect of the modeling, we are limited by detailed surface brightness profiles for all the host galaxies in our mock catalog, which extends out to higher redshifts. Our modeling also suggests that a non-negligible number of TDFs ($\gtrsim 25\%$ of the detected number) are being missed because they cannot outshine their host's central light, and this effect might become even more significant when more realistic surface brightness profiles are used in the mock galaxy catalog.
\subsection{The effect of host dust on fitted blackbody parameters}
\label{sec:ResultDustEffects}
\begin{figure}[htb!]
\gridline{\fig{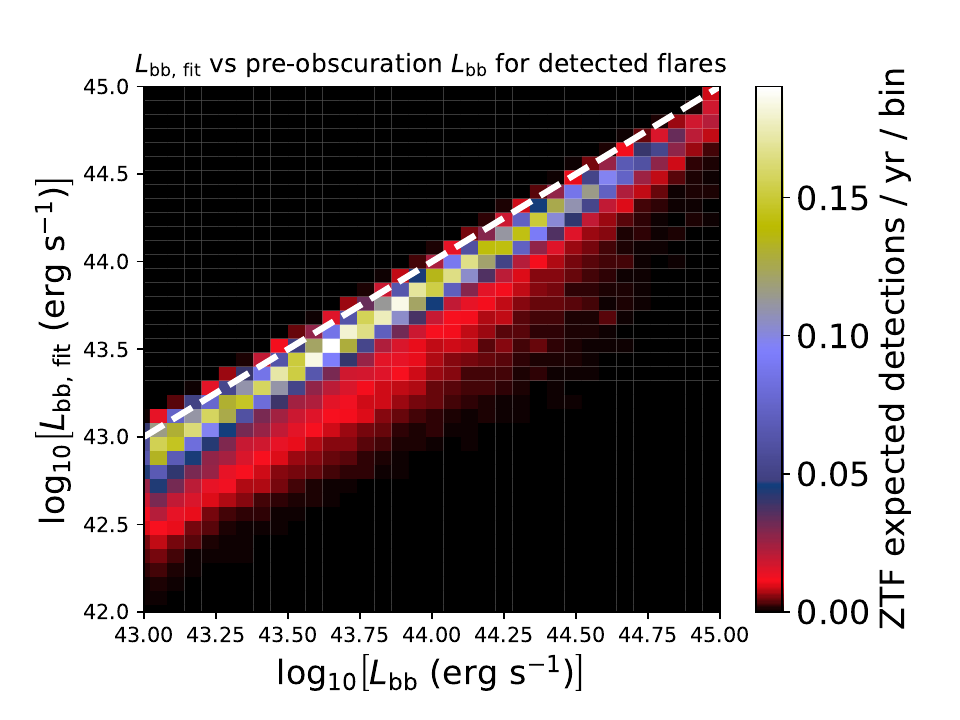}{0.5 \textwidth}{(a)}}
\gridline{\fig{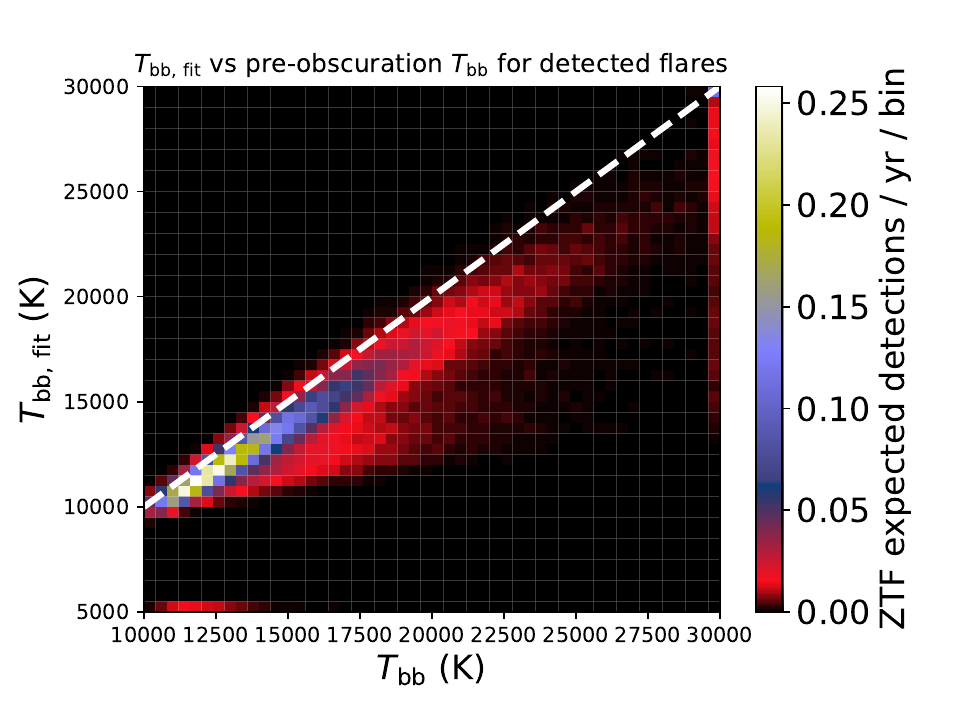}{0.5\textwidth}{(b)}}
\caption{Comparison of how the fitted, dust-obscured values for flare luminosity and temperature in detected flares in the model survey compare to the input, unosbscured values. The flare temperature distribution is weighted toward lower temperatures because these tend to have higher optical flux and are easier to detect when the temperature is sampled independently of the luminosity. }
\label{fig:DustEffectFit}
\end{figure}

When there is no dust, a fitting procedure that relies on photometric measurements in the ZTF $r$- and $g$-bands, along with Swift UVW1, UVM2, and UVW2, leads to excellent agreement between the fitted blackbody parameters and the true values (not shown). The presence of dust leads to fitted luminosities and temperatures that are both smaller than would be the case without dust, because of its extinction and reddening properties. We compare the fitted parameters with their true, unobscured values for the fiducial model in Figure~\ref{fig:DustEffectFit}, and we have chosen a color scheme that allows the tail of values near zero to be more easily visible. The white dashed line indicates a fitted value precisely equal to the unobscured value. In the absence of dust, all flares would cluster tightly around that line. Here, most of the detections are from galaxies with little dust. Thus, the effects on the fitted blackbody parameters are relatively minor. For the luminosity, there is increased spread in the fitted luminosity for lower values of the unobscured luminosity. This spread decreases at higher luminosities. This is because, for a flare to have a higher luminosity, the Eddington limit has to accommodate that luminosity. This can only be the case for the higher mass black holes, which generally tend to live in galaxies with less dust.

For the temperature comparison, in addition to the primary trend where the fitted temperature is a small factor lower than the true unobscured value because of small amounts of dust that we have assumed are present even in quiescent galaxies, one can faintly see a second spoke where the fitted temperature is reduced by a larger factor. That second population represents the detections in star-forming galaxies that have a substantial amount of dust. As expected, in this more highly dust-obscured population, the amount by which the fitted temperature disagrees from the intrinsic temperature becomes larger as the intrinsic temperature grows larger, because the dust opacity increases at shorter wavelengths. Since temperature and luminosity are fitted as independent parameters, it is not the case that flares with temperature fits of a factor of, say, $1.5$ too low compared to the intrinsic value correspond to luminosity fits that are a factor of $(1.5)^4$ too low --- that would only be the case if the fit were requiring a pre-specified blackbody radius that was held constant for the fit.

We emphasize that the number of model TDF detections that are highly dust obscured (the cluster of red-colored pixels at lower values of $T_{\rm bb, fit}$ in panel (b) of figure \ref{fig:DustEffectFit}) is small compared to the number of detections in galaxies that have much less dust obscuration. In other words, in this model, if one were to focus only on the detected flares, one might conclude that dust is not playing a large role. This conclusion would be incorrect, however, because in fact roughly twice as many flares as detected flares are being missed because of dust obscuration in the model, as we saw in panels $(e)$ and $(f)$ of Figure~\ref{fig:SixPanel}. In this sense we say that the dust has a ``guillotine'' effect, removing a large number of otherwise detectable flares, while not leaving very many highly obscured flares in the detected sample.

Our assumptions about host dust content, and its dependence on total host $M_{\ast}$ and sSFR (Section~\ref{sec:HostDust}), do have some effect on how this guillotine effect operates. To explore this, we considered a model in which the same dust obscuration parameterization (Equation~\eqref{eq:GarnBestLaw}) was used in all galaxies, regardless of their specific star formation rate. In this model, there is a higher amount of dust obscuration on average in galaxies with sSFR $< 10^{-11.3}$ yr$^{-1}$ than was the case for the fiducial model. Thus, the overall expected detection rate in this model drops to approximately 5.3 flares per year for this new model, compared to 13 events per year in the fiducial model. Among detected flares for this new model, 4.9\% of them should have $A_V > 1.0$, compared to 1.4\% of flares in the fiducial case. These fractions can be compared to the fractions of galaxies in the mock catalog with median $A_V > 1.0$ and with $8.5 < \log_{10}(M_\ast/M_\odot) < 11.5$ and $z < 0.4$. These fractions are 59.7\% and 36.4\% for this new model and the fiducial model, respectively. So for flares with $A_V > 1.0$, the approximate under-representation factors of observed flares compared to relative numbers of matched catalog galaxies are 12 and 26 for the new model and the fiducial model, respectively. If we instead consider all flares with $A_V > 0.5$, these under-representation factors change to 2.2 and 5.2 for the new model and the fiducial model, respectively.

\subsection{Flare temperature distributions}
\label{sec:TemperatureResults}

\begin{figure}[htb!]
\gridline{\fig{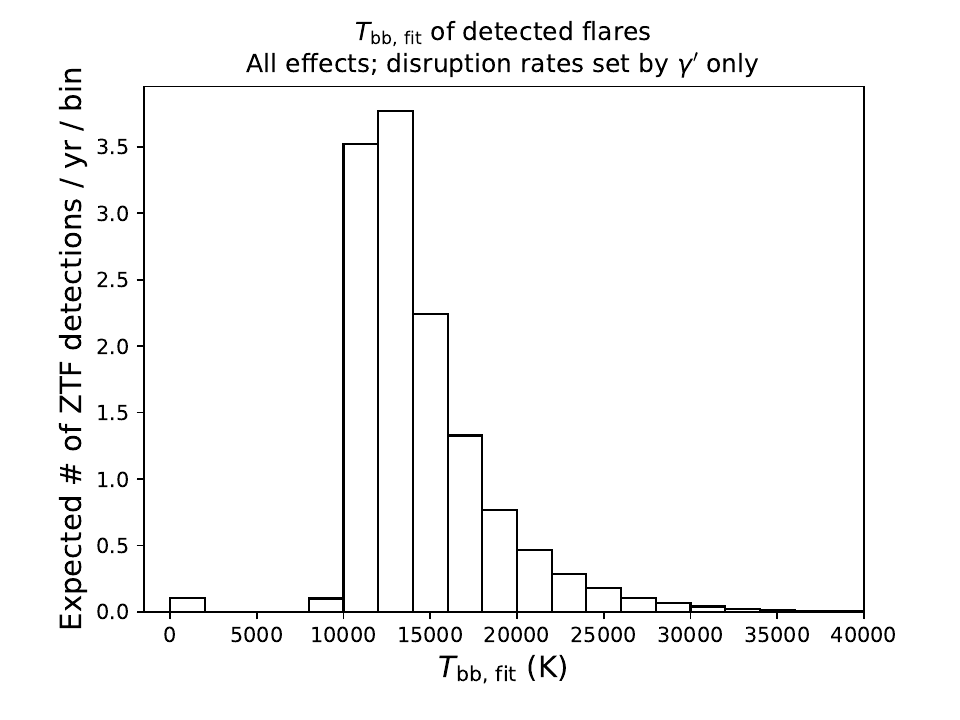}{0.5 \textwidth}{(a)}}
\gridline{\fig{Rfit1d.pdf}{0.5\textwidth}{(b)}}
\caption{Distributions of $T_{\rm bb, fit}$ and $R_{\rm bb, fit}$ for detected flares in the fiducial model survey. These will be compared to the ZTF TDF fits in Figure~\ref{fig:TandR_ZTF_Comp}.}
\label{fig:TandRFits}
\end{figure}

A striking result of this modeling exercise is that, for an assumed prior distribution that is flat for $T_{\rm bb}$, the distribution of temperatures for detected flares in the model survey is highly weighted toward low temperature. This can be seen in the second panel of Figure~\ref{fig:DustEffectFit}, and more directly in the projected 1d histogram for fitted temperatures shown in Figure~\ref{fig:TandRFits}. Recall that the true, unobscured temperatures were drawn uniformly from 10,000 to 50,000~K. The fact that some detected flares in the model have fitted temperatures lower than 10,000~K is a result of host dust obscuration. The small numbers of flares with fitted temperatures at zero are cases where the automated fit procedure did not succeed.

This preference for flare detections at lower temperature can be understood because, for a fixed overall luminosity budget, the flux at optical $r$ and $g$ bands increases steadily as temperature is lowered from 50,000 to 10,000~K.

\begin{figure}[htb!]
    \gridline{\fig{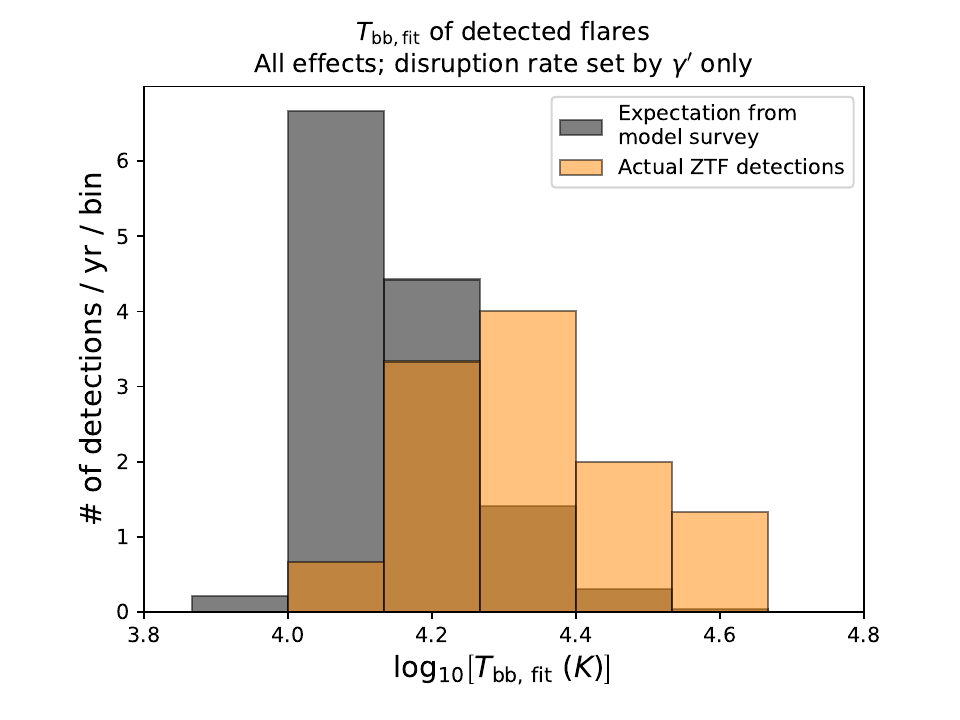}{0.5 \textwidth}{(a)}}
    \gridline{\fig{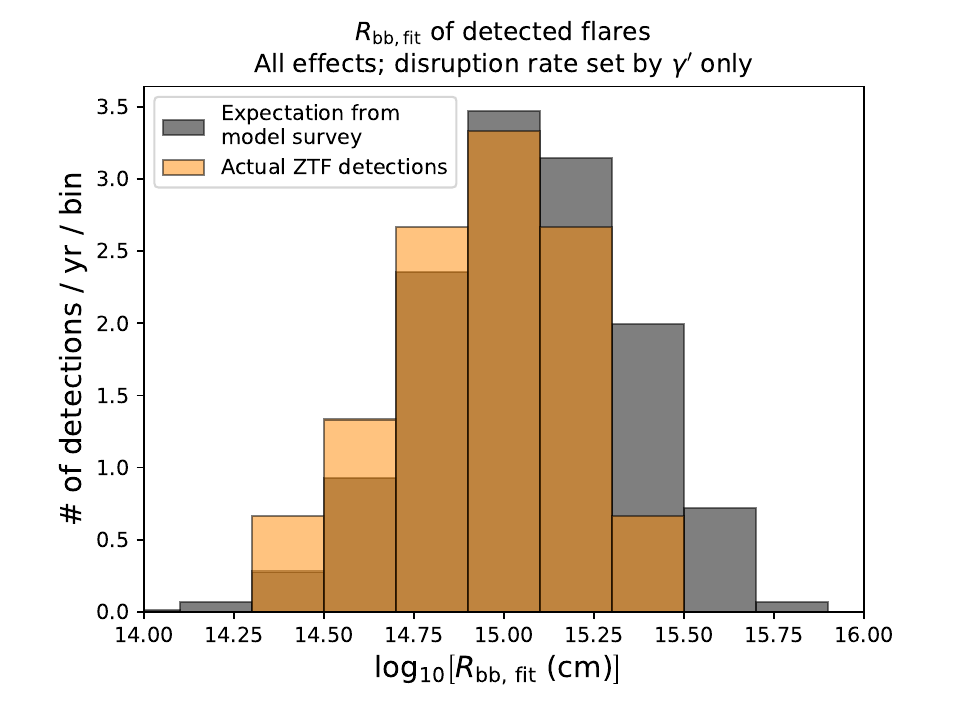}{0.5 \textwidth}{(b)}}
    \caption{Comparison of distributions of $T_{\rm bb, fit}$ and $R_{\rm bb, fit}$ for detected flares in the fiducial model survey to the distributions of flares detected by ZTF. The model expectation for $T_{\rm bb, fit}$ is inconsistent with the data ($p = 10^{-5}$), favoring values that are too low. The model expectation for $R_{\rm bb, fit}$ favors values that are slightly high compared to the data, but the distributions are roughly consistent (rejection $p$-value 0.14).  } 
    \label{fig:TandR_ZTF_Comp}
\end{figure}

However, the distribution of observed flares in real surveys, including ZTF, favors higher temperatures, as can be seen in Figure~\ref{fig:TandR_ZTF_Comp}, and a K-S test confirms that the model distribution for $T_{\rm bb,fit}$ is not consistent with the ZTF distribution ($p < 0.001$). It is important to remember that we have allowed the luminosity and temperatures of the flares to be sampled independently, which amounts to letting the photospheric radius of the corresponding spherical blackbody emission model to vary freely to accommodate these values. If we look at the distribution for the fitted values of $\log_{10}[R_{\rm bb, fit} \,\,({\rm cm})]$ in the model survey, we see the values fall mostly between 14.5 and 15.5, with a peak at about 15.1. This is similar to the real ZTF survey results, although the real detections tend to favor slightly smaller fitted values, with a median at around 14.8. A K-S test suggests that the model distribution for $R_{\rm bb,fit}$ is still roughly consistent with the ZTF distribution, with a rejection $p$-value of 0.14. To summarize, the model is favoring lower temperatures, lower luminosities, and slightly larger photospheric radii for detected flares, in such a way the overall detection rates are similar to the real survey rates.

\begin{figure*}[htb!]
    \plottwo{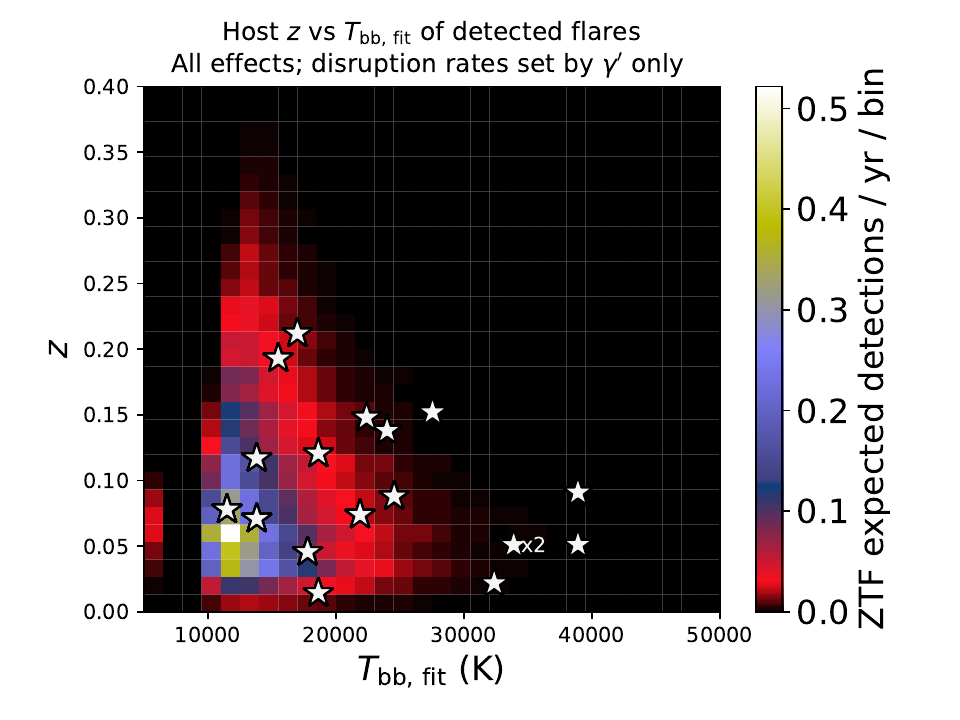}{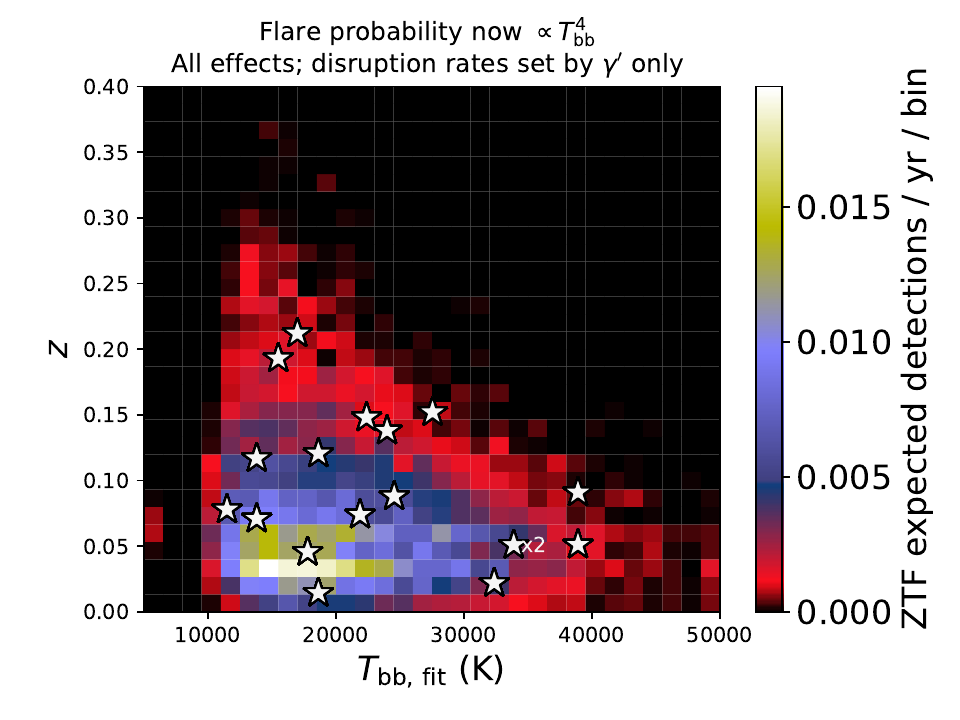}
    \caption{Two-dimensional histogram of host $z$ and detected flare $T_{\rm bb, fit}$ for the fiducial model survey in which input unboscured flare temperature is uniformly distributed between 10,000 and 50,000 K, and a model in which flare probability is weighted by $T_{\rm bb}^4$ (first and second panels, respectively). The white stars represent ZTF TDF detections. Flares with fitted temperatures that fall outside of the ranges of the plotted bin values are included in the edge bins.} 
    \label{fig:ZvsT}
\end{figure*}

Another factor at play in determining which flares at given temperatures and luminosities are detected is the distance to the host, which in turn corresponds to the host redshift $z$. Redshift also has a lower-order effect via the changing rest frame of the emission, which acts to increase flare optical visibility at higher redshift. In the first panel of Figure~\ref{fig:ZvsT} we consider how the fitted blackbody temperature correlates with host redshift for detected flares. When interpreting this figure one must also consider how the fraction of dust-obscured TDFs correlates with the redshift. But overall the pattern is that, for fitted flare temperatures above approximately 13,000 Kelvin, the model indicates that flares at higher $z$ become less detectable with increasing $T_{\rm bb, fit}$. In the model, since luminosity and temperature are considered independent variables, the temperature is playing a large role in setting the optical flux of the flares, so that higher temperature flares must have higher luminosities in order to be visible at a given luminosity distance. It remains to be seen whether evidence for such an effect can be found in real surveys. If this effect is not found, this would be an argument against the modeling strategy used here where flare temperature and luminosity are sampled independently.

Struck by the fiducial model’s strong preference for low temperatures for detected flares, we considered what would happen if the assumed probability distribution for $T_{\rm bb}$ were not uniformly but distributed, but power-law distributed in favor of higher flare temperatures. We tried a sequence of increasingly steep power-laws, eventually stopping at $dN/dT \propto T^4$ (note that this $T^4$ rate dependence is separate from the formula for blackbody luminosity given a fixed photospheric area). Overall this process resulted in a lower model detection rate, with the $T^4$ case yielding an expected ZTF detection rate of less than one flare per year. The model distribution of $T_{\rm bb, fit}$ for detected flares is still noticeably weighted toward faint flares, with the peak in the distribution below 20,000 Kelvin. The second panel of Figure~\ref{fig:ZvsT} shows the histogram of $T_{\rm bb, fit}$ versus $z$ for this case weighted toward higher flare temperatures. This makes the coupling between $z$ and $T_{\rm bb, fit}$ in the distribution of detections even more prominent. Again, this is largely a consequence of our choice to independently sample luminosity and temperature. If we instead chose to sample $R_{\rm bb}$ and $T_{\rm bb}$, or $R_{\rm bb}$ and $L_{\rm bb}$, instead of $L_{\rm bb}$ and $T_{\rm bb}$, or more directly tied these quantities to disruption parameters, it is possible this pattern in the $z$ vs $T_{\rm bb, fit}$ histogram would not appear as prominently. 

In both panels of Figure~\ref{fig:ZvsT}, flares with fitted temperatures that fall outside of the ranges of the plotted bin values are included in the edge bins, and we are again using the color scheme that lets the tail of bins with low counts be more easily visible.

To summarize, the fiducial model survey expects a distribution of $T_{\rm bb, fit}$ that is weighted too much toward low temperatures compared to the ZTF distribution. This is likely a consequence of our decision to sample $T_{\rm bb}$ uniformly and independently of $L_{\rm bb}$. Attempting to remedy this discrepancy by only adjusting the $T_{\rm bb}$ distribution, and keeping it independent of $L_{\rm bb}$, worsens the model agreement with the data in many other respects. It is clear that what is required is a better model for the joint distribution of $T_{\rm bb}$ and $L_{\rm bb}$. Even better would be a formalism for determining $T_{\rm bb}$ and $L_{\rm bb}$ based on disruption parameters which themselves could be sampled from theoretically derived distributions. 

\subsection{Predictions for VRO/LSST}
\label{sec:VROLSST}
We can use this modeling framework to anticipate the distributions of properties for TDFs to be detected with VRO/LSST. The single-visit limiting magnitude for the VRO $r$-band is expected to be $\approx 24.5$ \citep{Ivezic2019-0}, so for this model we require $m < 23$ in both $r$ and $g$, which provides a buffer similar to the one we used for ZTF so as to see rise and fall of the light curve in both bands. We also adjust the $r$ and $g$ filter wavelengths to match the VRO filters, but we still approximate the filters as delta functions peaked at the mean filter wavelength. To compute the final rate expectation we again assume 15,000 square degrees of effective sky coverage at these depths, but survey strategy might affect the appropriate value to use in future modeling. 

VRO reference images are expected to be as faint as $m_r \approx 27$ \citep{Ivezic2019-0}. This poses a limitation for our modeling, as the mock galaxy catalog we are using only includes galaxies as faint as $m_r = 22$. We partially account for this by eliminating  the host contrast requirement from the survey selection criteria. However, we caution that this model will not account for flares coming from galaxies with $m_r > 22$, which might be present in real VRO detections.

\begin{figure}[htb!]
\gridline{\fig{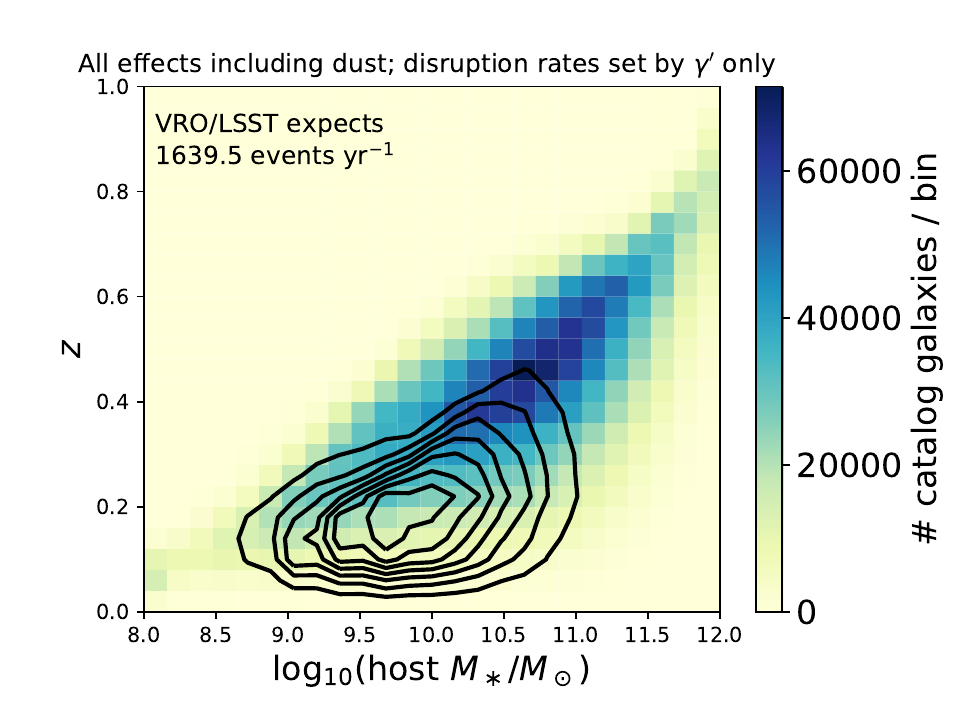}{0.5 \textwidth}{(a)}}
\gridline{\fig{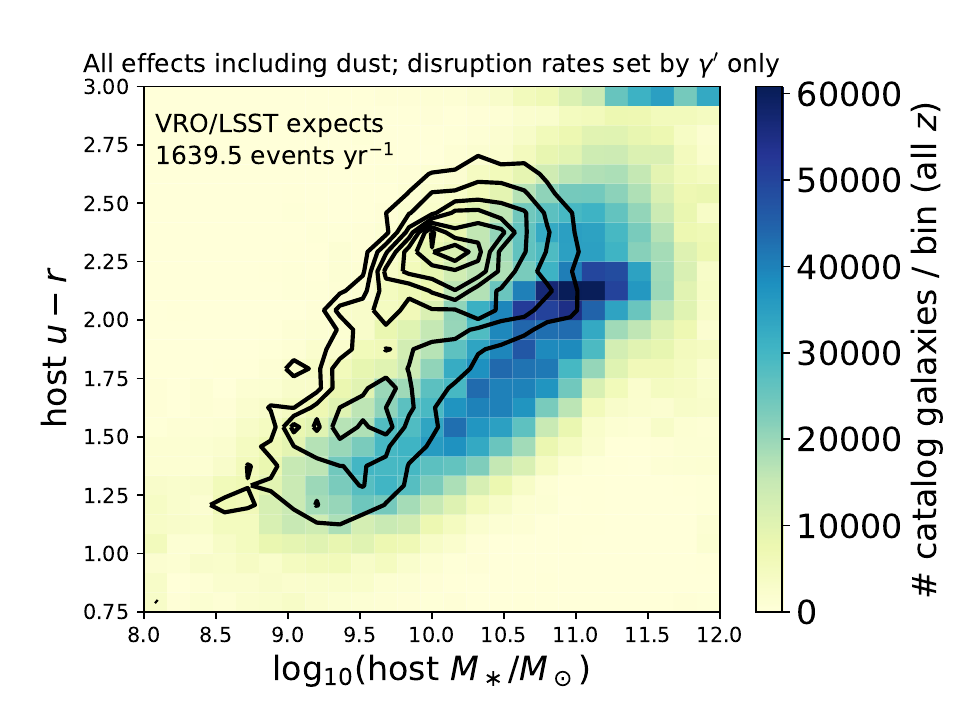}{0.5\textwidth}{(b)}}
\caption{Distributions of host-galaxy properties for detected flares in a model TDF survey that incorporates details based on the upcoming VRO/LSST. To compute the expected number of TDF detections per year, the survey sky coverage was assumed to be 15,000 square degrees, with perfect efficiency over that region, subject to the requirement that the peak transient flux is brighter than 23rd magnitude in both the $g$ and $r$ bands. This model does not account for details of the survey field selection or cadence. This model also uses the same galaxy catalog as was used for the ZTF models, which only includes galaxies with $m_r < 22$.}
\label{fig:VROPredictions}
\end{figure}

The first panel of Figure~\ref{fig:VROPredictions} shows a 2D histogram for the host galaxies of the expected TDF detections for this VRO-like model survey, binned by host total \mstar\ with $z$. The overall expected detection rate is $\approx 125$ times higher than the fiducial model for ZTF. Note that the distribution now extends to higher $z$, but in such a way that traces the overall correlation of host \mstar\ with $z$ (the contours are stretched diagonally up and to the right), which is related to the flux limit assumed in the survey. It also seems that VRO should be more sensitive to flares from galaxies with lower total \mstar\, so the effective \mbh\ cutoff imposed by $L_{\rm bb, min}$ and the Eddington limit might become even more important. 

The second panel of Figure~\ref{fig:VROPredictions} shows that in host $u-r$ versus host \mstar\ space, the peak of the distribution for detected flares remains largely unchanged compared to the ZTF model. However, the fractional contribution from bluer, low-mass galaxies becomes larger. 

This model's expectation of $\approx 1600$ TDF detections per year for VRO is somewhat low compared to previous estimates of this rate in the literature. \citet{van-Velzen2011} estimated that VRO would find $\approx 4000$ flares per year, by scaling the flare detection rate in SDSS stripe 82 imaging data based on its flux limit and sky coverage. Specifically, this method assumes the detection rate is proportional to $\Omega \, F_{\rm lim}^{-1.5}$ where $\Omega$ is sky coverage and $F_{\rm lim}$ is the limiting flux for transient detection. We can apply the same scaling argument to the ZTF detection rate, and if VRO is able to detect flares up to 4 magnitudes fainter than ZTF, and has similar sky coverage to ZTF, we then expect approximately 3000 flares per year. Since the fiducial model presented in this paper expects a number of flares per year that is similar to to the actual ZTF detections, one might then expect that the model presented in this section would also expect roughly 3000 TDF detections per year. One reason why the present model predicts a lower rate is because, as already mentioned, the mock galaxy catalog has not been extended to include fainter hosts. Another reason is that the fraction of flares that are affected by the model's treatment of dust obscuration has increased. 

Meanwhile, this model is neglecting several other details that will affect the real VRO TDF detection rate. For example, it does not account for the details of how the survey will select its observing fields, and the cadence of these observations in various bands. Along these lines, this model can be compared with the predictions of \citet{Bricman2020-0}, who used the VRO simulation framework to incorporate details concerning survey strategy, and used a simulation framework built specifically for VRO to catalog potential host galaxies. They predicted between approximately 3,000 to 8,000 TDF detections per year, depending on the SMBH mass distribution and the operable flux cutoffs for TDF detection. Their prediction of a higher detection rate might be due to a larger catalog of potential host galaxies, but is also affected by a number of other detailed differences between their simulation methodology and ours. Among these differences, they assumed a rate of $10^{-5}$ disruptions per year in all galaxies. They also used the {\tt MOSFiT} transient modeling package to generate TDF multi-band light curves \citep{Guillochon2018,Mockler2019}, and they based these light curves on stars of 1 solar mass in all cases.

Both the model presented in this paper and the \citep{Bricman2020-0} predictions are subject to uncertainty due to the effect of sample contamination by interlopers, and the possibility of incorrectly classifying TDFs as other types of transients. For example, the predicted detection rates for the models developed in this paper refer to expected rates of TDFs that pass selection criteria based solely on survey photometry in the $g$ and $r$ bands, but realistically only a small fraction of these will have their detection status strengthened with follow-up observations, and these follow-ups will be biased toward more nearby events \citep{Bricman2020-0}. Additionally, depending on survey strategy, flares might not have enough detections in multiple bands, which can be crucial for confident classification of TDFs \citep{Gezari2018WP}.

\section{Discussion}
\label{sec:Discussion}
Our main results are
\begin{itemize}
    \item Current optical surveys for TDFs are probably missing a population of flares in star-forming galaxies, because of dust. Based on our model assumptions, the TDFs in surveys such as ZTF might account for only 30\% of the flares that would be detectable if no dust were present in the host galaxies. However, this fraction is subject to substantial uncertainty related to the distribution of dust in galaxies. This selection effect should be accounted for in rate estimates and constraints on galaxy occupation fraction of SMBHs.
    \item Dust obscuration acts as a ``guillotine,'' preventing many flares from being detectable while leaving relatively few highly obscured flares detectable. Consequently, dust may seem to play a small role in obscuring the flares that are detected, even if it is playing a large role in setting the number of flares that are detected.
     \item While selection effects cannot totally explain the observed over-representation of observed TDFs in E+A galaxies (or more broadly the green valley), selection effects do seem to have an important role in setting the magnitude of the required rate enhancement in those galaxies.
    \item Our fiducial model survey, which uses a TDE rate description from \citet{Stone2016-1} that depends only on galaxy surface brightness information and an overall disruption rate normalization  (quoted here as Equation~\ref{eq:GammaPrimeRate}), along with our other assumptions about the flare luminosity distribution, can match the overall flare detection rate and the distributions of host redshift and total stellar mass of the hosts of TDFs detected by ZTF fairly well. However, no model survey explored in this study was able to produce a good match to the distributions of all observed quantities. A more sophisticated treatment is required for the rate prescription and/or the luminosity and temperature distributions of the flares, and how these functions correlate with galaxy properties. The absolute flare detection rate in our model survey is also subject to any uncertainty that affects the normalization of Equation~\ref{eq:GammaPrimeRate}.
    \item Surveys such as ZTF might also be missing a population of TDFs at a rate of $\gtrsim 25\%$ of the observed detection rate because the flares are not sufficiently bright when compared to the central light of their host.
    \item Connecting the observed TDF sample to the underlying volumetric rate of disruptions is very sensitive to how faint the flare luminosity distribution extends with non-zero probability. As such, measured rates can only provide lower limits on the volumetric rate, and the limits are less constraining the fainter the luminosity distribution extends.
    \item As a corollary of the last point, a more solid theoretical understanding of the luminosity distribution is therefore highly desirable. This would not only sharpen rate estimates, but would also allow future models to account for correlations between flare luminosity and other galaxy or disruption properties that are not accounted for here.
    \item The framework developed here can be adapted to TDF surveys at X-ray wavelengths, and to surveys of other astronomical transients, provided that the method is updated to make use of survey cadence and to account for light curve rise times.
    \item The framework here can be adapted to make use of more sophisticated models for stellar disruption rates and flare properties. The code used to generate the results in this paper has been made publicly available at \href{url}{https://github.com/nroth/tdegb}.
\end{itemize}

Future work must focus on, among other things:

\begin{itemize}
    \item A better understanding of how $L_{\rm bb}$ and $T_{\rm bb}$ are set by \mbh, $M_\ast$, stellar composition, and orbital parameters of the TDE. 
    \item The role of BH spin in determining when disruptions take place outside the SMBH event horizon, and possibly in setting $L_{\rm bb}$ and $T_{\rm bb}$.
    \item Extracting the entries in the mock galaxy catalog that correspond to E+A galaxies and further quantifying how the expected flare detection rate in those galaxies in the model compares to the observed rate.
    \item Updating the mock catalog to include more detailed stellar surface brightness profile information, and to account for the presence or absence of nuclear star clusters.
    \item Incorporating the latest results in calculation of stellar dynamical disruption rates, accounting for details such as nucler star clusters and non-isotropic velocity distributions \citep[e.g.][]{Loubser2020-0}, and how these rates correlate with other host galaxy properties.
    \item Accounting for how the dust extinction toward the SMBH at the galaxy center might deviate, in a statistical sense, from the extinctions measured from Balmer decrements of distributed star-forming regions in the galaxy. This might be related to the mass-dependent prevalence of dust lanes \citep{Dalcanton2004-0}.
    \item Incorporating light curve rise and fall information, not just peak luminosity and temperature, on detection rates in surveys with varying cadences.
    \item Incorporating the effects of detections of TDF interlopers such as nuclear SNe and other types of AGN variability, and the effects of incorrectly classifying TDFs as other types of transients.
    \item Incorporating possible deviations from the standard \mbh--$\sigma$ relation, which might arise at the low-mass end.

\end{itemize}

\section*{Acknowledgments}
We thank Suvi Gezari, Erica Hammerstein, Nicholas Stone, Michael Kesden, Julian Krolik, and Hugo Pfister for helpful conversations.
NR acknowledges the support from the University of Maryland  through the Joint Space Science Institute Prize Postdoctoral Fellowship. 

ZTF is supported by the National Science Foundation under Grant No. AST-1440341 and a collaboration including Caltech, IPAC, the Weizmann Institute for Science, the Oskar Klein Center at Stockholm University, the University of Maryland, the University of Washington, Deutsches Elektronen-Synchrotron and Humboldt University, Los Alamos National Laboratories, the TANGO Consortium of Taiwan, the University of Wisconsin at Milwaukee, and Lawrence Berkeley National Laboratories. Operations are conducted by COO, IPAC, and UW.

The authors thank the Yukawa Institute for Theoretical Physics at Kyoto University. Discussions during the YITP workshop YITP-T-19-07 on International Molecule-type Workshop ``Tidal Disruption Events: General Relativistic Transients'' were useful to complete this work.

An early version of this work was presented at the Aspen Center for Physics winter workshop ``Using Tidal Disruption Events to Study Super-Massive Black Holes,'' where helpful feedb\
ack was provided. The Aspen Center for Physics is supported by National Science Foundation grant PHY-1607611.

Funding for the SDSS and SDSS-II has been provided by the Alfred P. Sloan Foundation, the Participating Institutions, the National Science Foundation, the U.S. Department of Energy, the National Aeronautics and Space Administration, the Japanese Monbukagakusho, the Max Planck Society, and the Higher Education Funding Council for England. The SDSS is managed by the Astrophysical Research Consortium for the Participating Institutions: the American Museum of Natural History, Astrophysical Institute Potsdam, University of Basel, University of Cambridge, Case Western Reserve University, University of Chicago, Drexel University, Fermilab, the Institute for Advanced Study, the Japan Participation Group, Johns Hopkins University, the Joint Institute for Nuclear Astrophysics, the Kavli Institute for Particle Astrophysics and Cosmology, the Korean Scientist Group, the Chinese Academy of Sciences (LAMOST), Los Alamos National Laboratory, the Max-Planck-Institute for Astronomy (MPIA), the Max-Planck-Institute for Astrophysics (MPA), New Mexico State University, Ohio State University, University of Pittsburgh, University of Portsmouth, Princeton University, the United States Naval Observatory, and the University of Washington.

Lawrence Livermore National Laboratory is operated by Lawrence Livermore National Security, LLC, for the U.S. Department of Energy, National Nuclear Security Administration under Contract DE-AC52-07NA27344.

This document was prepared as an account of work sponsored by an agency of the United States government. Neither the United States government nor Lawrence Livermore National Security, LLC, nor any of their employees makes any warranty, expressed or implied, or assumes any legal liability or responsibility for the accuracy, completeness, or usefulness of any information, apparatus, product, or process disclosed, or represents that its use would not infringe privately owned rights. Reference herein to any specific commercial product, process, or service by trade name, trademark, manufacturer, or otherwise does not necessarily constitute or imply its endorsement, recommendation, or favoring by the United States government or Lawrence Livermore National Security, LLC. The views and opinions of authors expressed herein do not necessarily state or reflect those of the United States government or Lawrence Livermore National Security, LLC, and shall not be used for advertising or product endorsement purposes.

\appendix

\section{Applying dust obscuration and K-correction}
\label{sec:ImplementDustKcorrection}

In the rest frame of the host galaxy (which we will refer to as the ``emitted'' frame, labeled with the subscript ``emit''), an \emph{unextincted} source emitting as a blackbody will have specific luminosity given by
\begin{equation}                  
\label{eq:BlackbodyEmissionHost}            L_{\nu, \rm emit,unextincted} = L_{\rm bb} \frac{ \pi B_{\nu, \rm emit}(T_{\rm bb}) }{\sigma_{SB}T_{\rm bb}^4} \, \, ,
\end{equation}
where $B_{\rm \nu}(T)$ is the Planck function and $\sigma_{SB}$ is the Stefan-Boltzmann constant. Extinction will lower this specific luminosity by a factor $f_{\nu,\rm emit}$:
\begin{equation}                                                            \label{eq:FluxReductionFactor}              f_{\nu,emit} \equiv \frac{ L_{\nu,\rm emit,extincted}}{ L_{\nu,\rm emit,unextincted}} \,\, ,
\end{equation}
and the fact that $f_{\nu}$ is larger for higher optical and UV frequencies gives rise to reddening. From the definition of extinction $A_\nu$ and the relation between magnitude and flux we have
\begin{equation}
\label{eq:AnuDefinition}
A_{\nu, \rm emit} = -2.5 \log_{10}(f_{\nu,\rm emit}) \,\, .
\end{equation}

The extinction law from \citet{Calzetti2000} returns a flux-reduction function which we will label as ${\cal C}_{\nu}$, which is a function of the emitted frequency and a parameter $R_V$. This function is defined via the relation
\begin{equation}
\label{eq:CardelliFactorDefinition}
A_{\nu,\rm emit} = A_{V,\rm emit} {\cal C}_{\nu,\rm emit} \, \, ,
\end{equation}
where $V$ refers to $V$-band. 
Our models must specify how we allow $A_V$ to vary across different galaxies. For quiescent galaxies, which we have defined as those galaxies whose catalog entry indicates  sSFR $< 10^{-11.3}$ ~yr$^{-1}$, we assume a median $A_V$ of 0.2 and we further assume that the distribution about this median is Gaussian, with a floor at $0$. For galaxies with higher sSFR entries in the catalog, we follow the \citet{Garn2010-1} fit for how $A_{{\rm H} \alpha}$ depends on host galaxy mass (equation~\ref{eq:GarnBestLaw}). That extinction is in terms of $A_{{\rm H} \alpha}$, so in these cases we first use equation~(\ref{eq:CardelliFactorDefinition}) to solve for $A_V$, which then allows us to write
\begin{equation}                                                        \label{eq:AnuEmitFinalSF}                   A_{\nu,\rm emit} = \frac{A_{{\rm H}\alpha}}{{\cal C}_{{\rm H}\alpha}} {\cal C}_{\nu,\rm emit} \, \, .                            
\end{equation}
If the band of observation is approximated with a delta-function filter, we have \citep[e.g.][]{Hogg1999}:
\begin{equation}                                                            \label{eq:ObservedFlux}                     \nu_{\rm obs} F_{\nu, \rm obs} = \frac{\nu_{\rm emit} L_{\nu,\rm emit,extincted}}{4 \pi D^2_L(z)} \, \, ,                           \end{equation}
where $D_L(z)$ is the cosmological luminosity distance, and
\begin{equation}                                                            \label{eq:NuEmitObserved}                   \nu_{\rm emit} = (1 + z) \nu_{\rm obs} \, \, .
\end{equation}
Putting everything together,
\begin{equation}                  \label{eq:ObservedExtinctedFlux}            F_{\nu,\rm obs} = \frac{ 1 + z}{4 \pi D_L^2(z)} \, 10^{- \frac{A_{\nu, \rm emit}}{2.5}} \, L_{\rm bb} \frac{ \pi B_{\nu, \rm emit}(T_{\rm bb}) }{\sigma_{\rm SB}T_{\rm bb}^4} \, \, .
\end{equation}
This flux can then be converted into an AB magnitude for the purpose of determining whether it passes survey flux limits or sufficiently contrasts with the host galaxy light contained in the PSF.

\section{Temperature and luminosity fitting procedure}
\label{sec:Tfit}
As mentioned in the main text, we fit blackbody spectra to the flux measurements at the the mean filter wavelengths for the ZTF $r$- and $g$-bands, along with the UVW1, UVM2 and UVW2 bands for the UV-optical telescope aboard the Neil Gehrels Swift Observatory. These bands were chosen in order to replicate the actual measurements used to fit the properties of the ZTF TDFs. 

When performing the fit, the fluxes in each of the five bands are K-corrected such that the fit is finding the values of $L_{\rm bb}$ and $T_{\rm bb}$ as would be measured in the galaxy rest-frame, although the extinction and reddening from the dust has been applied in models that include the effects of dust. It is therefore important to note that even if the actual extincted spectrum is no longer a blackbody, the fit is still treating the spectrum as a blackbody in order to determine the fitted values of $L_{\rm bb}$ and $T_{\rm bb}$, because in practice all the spectra of the ZTF TDFs are fit to what are assumed to be unobscured blackbody spectra in the galaxy rest-frame. Also as mentioned in the main text, in order to approximate statistical error for the measurement, we apply a Gaussian random error, with $\sigma$ = 10\% of the original value, to the flux measurement (in magnitudes) in each band before performing the fit. 

Explicitly, each of the extincted, noise-adjusted specific luminosities in the five bands is compared to equation~\ref{eq:BlackbodyEmissionHost}, where equation~\ref{eq:NuEmitObserved} is used to relate the galaxy rest-frame frequencies to the instrumental values that correspond to the mean filter wavelengths. The simultaneous fit for $T_{\rm bb}$ and $L_{\rm bb}$ is performed using the GNU Scientific Library, specifically the nonlinear least-squares fitting functions with weights, as defined by the ``gsl\_multifit\_nlinear'' header file. Once fitted values $T_{\rm bb,fit}$ and $L_{\rm bb,fit}$ have been determined, we define $L_{g, \rm fit}$ as 
\begin{equation}
L_{g, \rm fit} \equiv \nu_{g,\rm emit} L_{\rm bb,fit} \frac{ \pi B_{\nu, \rm emit}(T_{\rm bb,fit}) }{\sigma_{SB}T_{\rm bb,fit}^4} \, \, .
\end{equation}
By including the leading factor of $ \nu_{g,\rm emit}$, $L_{g, \rm fit}$ has units of \ergspersec. 

\section{Treatment of flare luminosity probability distribution}
\label{sec:ImplementLF}
As discussed in section~\ref{sec:FlareLF}, we assume that all flares are at least as bright as some chosen minimum value $L_{\rm bb, min}$. Each combination of \mbh\ and \mstar\ then sets a maximum possible luminosity $L_{\rm bb, max}$, which is tied to the Eddington limit, or at sufficiently high stellar mass, tied to the maximum mass fallback rate (see also Figure~\ref{fig:LFitMbh2D}). We further assume that the probability distribution for flares to take on value of $L_{\rm bb}$ between these values is a power-law weighted toward fainter events. 

So, generally speaking, when we randomly draw a value for $L_{\rm bb}$ for our simulated flares, we use a probability distribution of the form 
\begin{equation}                            \label{eq:FlareLF}                          \psi(L_{\rm bb}) = \left (\frac{-n +1}{L_{\rm bb, max}^{-n + 1} - L_{\rm bb, min}^{-n + 1}} \right ) L_{\rm bb}^{-n} \,\, ,
\end{equation}
whenever $n \ne 1$ and where the factor in parentheses ensures that the probability integrates to 1 between $L_{\rm bb, min}$ and $L_{\rm bb, max}$. All models in this paper used $n = 2.5$, although it may be useful to explore other values in future work.

For numerical implementation, it is useful to recast this in non-dimensional form. When $n \ne 1$ we can proceed as follows:
\begin{eqnarray}
 n^\prime &\equiv& n - 1 \,\, , \qquad  y \equiv \left(\frac{L_{\rm bb}}{L_{\rm bb,max}} \right)^{n^\prime} \,\, , \qquad  y_{0} \equiv \left(\frac{L_{\rm bb, min}}{L_{\rm bb, max}} \right)^{n^\prime} \,\, , \nonumber \\   \psi(y) &=& 
 \begin{cases}
 \frac{y_{0}}{1 - y_{0}} y ^{-2}  \qquad &{\rm for} \,\, y \,\, {\rm given \,\, by} \,\, L_{\rm bb, min}< L_{\rm bb} < L_{\rm bb, max} \,\, , \\ 0 \qquad &{\rm otherwise}.
 
 \end{cases}
\end{eqnarray}

The sampling is performed via inverse transform sampling: a uniform random deviate $u$ is drawn, which is then used to produce a $y$ (and hence an $L_{\rm bb}$):
\begin{equation}                                                            \label{eq:InverseTransformLF}
y = \frac{y_{0}}{1 - u(1 - y_{0})} \,\, .
\end{equation}

The special case of $n = 1$ is handled differently. In this case:
\begin{eqnarray}
 y &\equiv& \left(\frac{L_{\rm bb,max}}{L_{\rm bb}} \right) \,\, , \qquad  y_{0} \equiv \left(\frac{L_{\rm bb, max}}{L_{\rm bb, min}} \right)\,\, , \nonumber \\   \psi(y) &=& 
 \begin{cases}
 \frac{1}{\ln(y_{0})} \frac{1}{y}  \qquad &{\rm for} \,\, 1 < y < y_{0} \,\, , \\ 0 \qquad &{\rm otherwise}.
 \end{cases}
\end{eqnarray}
This time, $y$ (and hence $L_{\rm bb}$) is produced in terms of a random uniform deviate $u$ via the relation
\begin{equation}
y = \exp[u \ln(y_0)].
\end{equation}

\section{Monte Carlo calculation of observed flare rates and distributions}
\label{sec:MonteCarloExplanation}
We can determine the combined flare detection rate for all the galaxies in the mock galaxy catalog by performing a weighted sum over all galaxies:
\begin{equation}
\frac{dN_{\rm TDF,obs}}{dt_{\rm obs} d\Omega} = \frac{1}{\Omega_{\rm cat}} \sum_i \dot{N}_{{\rm TDE},i} \frac{1}{1 + z_i} f_i \, \, ,
\end{equation}
where $\Omega_{\rm cat}$ is the sky area that corresponds to the mock catalog, the sum is over $i$ is for all galaxies in the catalog, $\dot{N}_{{\rm TDE},i}$ is the per-galaxy disruption rate for each galaxy (see section~\ref{sec:PerGalaxyRates}), the $1/(1+z)$ factor accounts for cosmological time dilation, and $f_i$ is the fraction of all disruptions in galaxy $i$ that lead to detectable flares.

To estimate $f_i$, we randomly generate 100 ($N_{\rm trials})$ trial disruptions in each galaxy, and determine the fraction of these that lead to flares that pass the survey selection criteria, given our assumptions about the flare luminosity and temperature distributions and the effects of host obscuration. In more detail, for each trial disruption, we first sample the mass $M_\ast$ of the disrupted star from the present-day mass function in the galaxy, which is a truncated Kroupa IMF (see section~\ref{sec:PerGalaxyRates}). If \mbh\ for the galaxy is larger than the Hills mass for that given stellar mass, the disruption does not lead to a detectable flare. Otherwise we sample $L_{\rm bb}$, $T_{\rm bb}$, and $A_V$ for the flare, as described in section~\ref{sec:Method} and Appendix~\ref{sec:ImplementLF}. We then determine the observed flux in each of the ZTF $r$- and $g$- bands using equation~\ref{eq:ObservedExtinctedFlux}. We can then determine whether these fluxes pass our requirements for overall detectability, host contrast, and observed color. For the fiducial model, these requirements are: peak ZTF $m_r$ and $m_g$ both $< 19$, the host contrast requirement $m_{\rm TDF} < m_{\rm PSF} + 1$ (equation~\ref{eq:HostContrast2}), and peak $m_g - m_r < 0$.

Once this is done, the expected number of flares detected in the survey per time interval $\Delta t _{\rm surv}$ and sky area $\Omega_{\rm surv}$ is given by 
\begin{equation}
N_{\rm TDF, obs} = \Delta t_{\rm surv} \Omega_{\rm surv} \frac{dN_{\rm TDF,obs}}{dt_{\rm obs} d\Omega} \, \, .
\end{equation}

To determine the expected distributions of flares with various properties, we store all the attributes of each detected flare, along with a weight value $w_j$:
\begin{equation}
\label{eq:FlareWeights}
w_j = \left( \frac{\Delta t_{\rm surv} \Omega_{\rm surv}}{N_{\rm trials} \Omega_{\rm cat}} \right) \, \dot{N}_{{\rm TDE},i} \frac{1}{1 + z_i} \, \, ,
\end{equation}
where $\dot{N}_{{\rm TDE},i}$ and $z_i$ are the overall TDE rate and redshift, respectively, for the host galaxy for trial flare $j$. Distributions of flare properties can then be created by binning detected flares based on their attributes and using these weights. These weighted histograms are then already normalized in the sense that summing the weighted values in all the bins is equal to $N_{\rm TDF, obs}$.

\bibliographystyle{apj}

\end{document}